\documentclass[12pt]{article}
\usepackage{arxiv}
\usepackage{setspace}
\renewcommand{\shorttitle}{Human-AI Collaboration: \textit{Partial vs. Full Automation}}
\renewcommand{\headeright}{}
\renewcommand{\undertitle}{}

\usepackage{amsmath}
\usepackage{bm}
\usepackage{newtxtext}
\usepackage{newtxmath}
\usepackage{pdfpages}

\usepackage{booktabs}
\usepackage{colortbl}
\usepackage{array}
\usepackage{multirow}
\usepackage{makecell}
\usepackage{tabularx}
\usepackage{longtable}
\usepackage{threeparttable}
\usepackage{adjustbox}

\usepackage{graphicx}
\usepackage{subcaption}
\usepackage{float}
\usepackage{tikz}

\usepackage{algorithm}
\usepackage{algpseudocode}

\usepackage{xcolor}
\usepackage[most]{tcolorbox}
\definecolor{teal_header}{RGB}{0, 95, 115}

\usepackage{enumitem}
\usepackage{appendix}
\usepackage{siunitx}
\usepackage{chngcntr}
\usepackage{afterpage} 
\usepackage{needspace}

\usepackage{listings}
\usepackage{caption}
\renewcommand{\footnoterule}{%
  \kern -3pt
  \hrule width \textwidth
  \kern 2pt
}
\usepackage{natbib}
\bibpunct[, ]{(}{)}{,}{a}{}{,}

\usepackage{hyperref}
\usepackage{pdfpages}

\AtBeginDocument{
  \setlength{\abovedisplayskip}{14pt}
  \setlength{\belowdisplayskip}{14pt}
  \setlength{\abovedisplayshortskip}{10pt}
  \setlength{\belowdisplayshortskip}{10pt}
}

\begin{document}

\title{Economics of Human and AI Collaboration:\\[4pt]
When is Partial Automation More Attractive than Full Automation?
}

\author{
  \textbf{Wensu Li$^{a}$ \quad
  Atin Aboutorabi$^{b}$ \quad
  Harry Lyu$^{a}$ \quad
  Kaizhi Qian$^{c}$} \\[4pt]
  \textbf{Martin Fleming$^{a}$ \quad
  Brian C.\ Goehring$^{d}$ \quad
  Neil Thompson$^{a}$}\thanks{Corresponding author: \texttt{neil\_t@mit.edu}} \\[8pt]
  $^{a}$Massachusetts Institute of Technology \quad
  $^{b}$\'{E}cole Polytechnique F\'{e}d\'{e}rale de Lausanne \\[2pt]
  $^{c}$IBM Research \quad
  $^{d}$IBM's Institute for Business Value \\[6pt]
  \texttt{\{wensu, hlyu, marti264, neil\_t\}@mit.edu} \\[2pt]
  \texttt{atin.aboutorabi@epfl.ch} \quad
  \texttt{kqian@ibm.com} \quad
  \texttt{goehring@us.ibm.com}
}

\date{}

\maketitle
\vspace{-3mm}
\begin{abstract}
This paper develops a unified framework for evaluating the optimal degree of task automation. Moving beyond binary automate-or-not assessments, we model automation intensity as a continuous choice in which firms minimize costs by selecting an AI accuracy level, with outcomes ranging from labor-only production (no automation) to partial automation (human--AI collaboration) to full automation. Our framework has two components. On the supply side, we estimate an AI production function through fine-tuning scaling-law experiments that link model performance to data, training steps, and model size. Because large language models, computer vision systems, and foundation-model-based AI more generally exhibit predictable but diminishing returns to these inputs, the cost of achieving higher accuracy is convex: reaching good performance may be relatively inexpensive, but pushing toward near-perfect accuracy becomes disproportionately costly. Full automation is therefore often not cost-minimizing; instead, partial automation -- an interior solution in which firms choose an intermediate level of AI deployment and retain human workers for the residual workload -- frequently emerges as the cost-minimizing equilibrium. On the demand side, we take an information-theoretic approach, introducing an entropy-based measure of task complexity that maps model accuracy into a labor substitution ratio, quantifying how much human work time AI displaces at each accuracy level and formalizing the division of work between AI and human collaborators.

We calibrate the framework with O*NET task data, a survey of 3,778 domain experts, and GPT-4o-derived task decompositions, and implement it empirically in computer vision -- a domain where abundant scaling-law data exists. We find that task complexity shapes labor substitution: tasks with few subtasks and low complexity see high substitution rates, while tasks with many subtasks and high complexity favor limited partial automation. Scale of deployment is a fundamental determinant of the automation frontier: AI-as-a-Service and AI agents spread fixed development costs across many users, sharply expanding the set of economically viable tasks.

At the firm level, cost-effective automation captures approximately 11\% of computer-vision-exposed labor compensation; under economy-wide deployment, the economically viable share rises sharply. Since large language models and other AI systems exhibit similar scaling-law economics, the mechanisms we identify extend beyond computer vision. Our findings reinforce that partial automation -- where AI assists rather than replaces human judgment -- is often the economically rational long-run outcome, not merely a transitional phase on the path to full automation.
\vspace{2mm}

\end{abstract}
\setstretch{1.2}

\newpage
\section{Introduction}\label{sec:Intro}
The central strategic question facing firms in the AI era is no longer whether tasks can be automated, but whether they should be automated, and, critically, to what extent. Emerging AI systems have fueled expectations of large productivity gains while simultaneously raising concerns about job loss and the pace of labor market adjustment. Much of the early economics of automation framed this problem as a binary choice: a task is either automated or not, and exposure is measured by whether a technology is technically capable of performing the required task. Yet for most real-world tasks, automation is better understood as a continuum.\footnote{In contrast to the economics literature, the computer science literature has long recognized automation as a continuum of human-machine interaction. See \citet{sheridan1978human}; \citet{parasuraman2000model}, who propose a multi-dimensional continuum; and \citet{endsley1999level}, who propose a taxonomy of 10 levels of automation for dynamic control tasks.} Firms can adopt AI systems that fully replace human effort, or they can deploy AI as a tool that handles part of the work while humans perform the remainder. Understanding when partial automation is the optimal solution -- rather than full automation or no automation -- is central to predicting how AI will reshape production and employment.

Why does partial automation deserve this central role? The answer lies in the cost structure of modern AI systems. Large language models, computer vision systems, and foundation-model-based AI more generally exhibit scaling laws: model performance improves predictably as training data, model size, and compute are scaled up, but with sharply diminishing returns at higher accuracy levels. As a result, the marginal cost of improving AI performance rises steeply. Achieving ``good'' accuracy on a task may be relatively inexpensive, but pushing from good to near-perfect accuracy -- the level often required for full automation -- can be orders of magnitude more costly. When the marginal cost of further accuracy improvement exceeds the marginal labor saving it enables, firms optimally stop short of full automation. The AI system handles the portion of the task it can perform cost-effectively, and human workers resolve the remaining uncertainty. Partial automation is therefore not merely a transitional state on the path to full automation; it is frequently the long-run cost-minimizing equilibrium.

This cost structure implies three possible outcomes for any given task: no automation, partial automation, and full automation. Which outcome is optimal depends on whether the labor savings from improved AI performance justify the sharply rising cost of achieving that performance. Our framework formalizes these three cases and shows that partial automation is often the most prevalent outcome, precisely because the jump from partial to full automation is disproportionately expensive for many tasks.

\begin{figure}[]
    \centering
    \includegraphics[width=\linewidth]{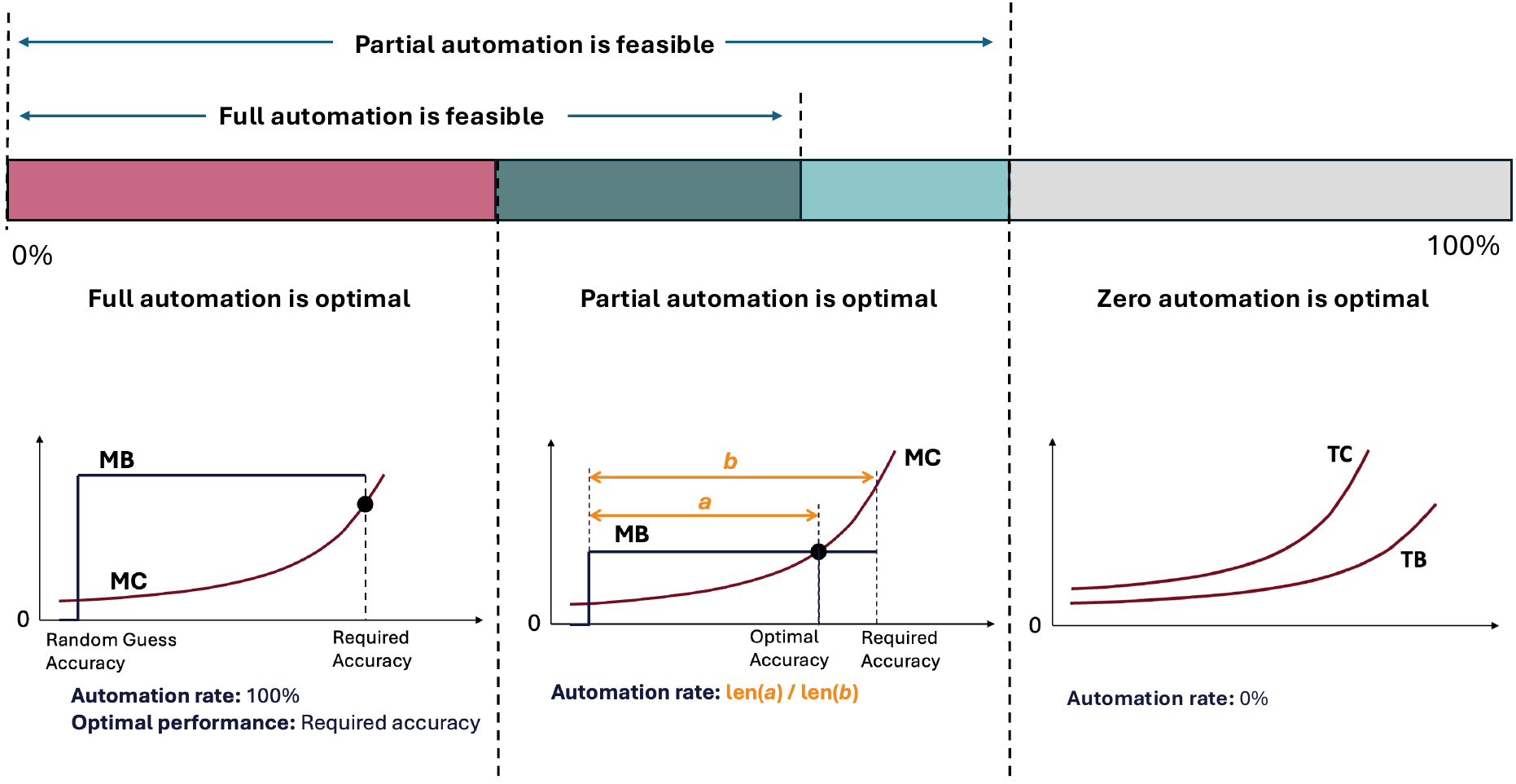}
    \caption{Breakdown of occupation compensation by task-level automation type: different cases of partial equilibrium, ranked from most to least benefit from AI automation.
    }
    \label{fig:partial_equilibrium} 
\end{figure}

Figure~\ref{fig:partial_equilibrium} illustrates this logic schematically. The horizontal axis traces the automation rate from zero to full automation. When the convex AI cost function lies below the labor-saving benefit throughout the relevant performance range, full automation is both feasible and optimal. When the steeply rising marginal cost curve intersects the marginal benefit curve before the required accuracy level is reached, the firm optimally stops at an interior solution -- partial automation -- and human workers complete the residual workload. When fixed development costs alone exceed the potential labor savings, no automation is warranted. The key insight is that the middle region -- partial automation -- occupies the largest share of the task space, precisely because the convexity of scaling-law cost functions makes the jump from partial to full automation disproportionately expensive for most tasks.

Firms increasingly begin with foundation models or off-the-shelf AI systems that already deliver some baseline level of performance. For some tasks, that baseline is sufficient. For many others, achieving the accuracy required for reliable deployment requires additional investment in task-specific data, training compute, or larger models. The economic problem is therefore not simply whether AI can perform a task, but whether it is worth paying to move far enough up the performance frontier.

In contemporary task models \citep{autor2003skill, acemoglu2011skills, acemoglu2018race, acemoglu2018modeling}, automation results in the reallocation of tasks from labor to capital. Over the past decade, with the advent of large language models and deep learning models, the literature has moved beyond labor-capital substitution. Automation research has focused on exposure to tasks potentially automated through these technologies or robotics \citep{frey2017future, brynjolfsson2017machinelearning, eloundou2024gpts}. A growing literature now examines the conditions under which technically feasible automation is economically viable. One strand emphasizes the composition of tasks and the expertise required for execution: whether AI substitutes for low-expertise or high-expertise tasks shapes both wage and employment outcomes \citep{autor2025expertise}. Another strand studies human--AI collaboration, showing that many tasks are most productively executed when AI and workers share responsibility \citep{NBERw33949, brynjolfsson2025generative, shao2025futurework}. A third line of work investigates AI scaling laws, documenting how performance improves -- and costs escalate -- with sharply diminishing returns at high accuracy levels \citep{kaplan2020scaling, rosenfeld2021scaling, hoffmann2022training, thompson2022computationallimitsdeeplearning}. Macroeconomic analyses link these micro-level choices to aggregate employment outcomes, emphasizing that the net impact of AI depends on the balance between automation of existing tasks and the creation of new, complementary activities \citep{acemoglu2025simple, hampole2025ailabor}.

These strands offer important insights but are typically studied in isolation. What remains absent is a unified framework that connects (i) the technical feasibility of task automation, (ii) the costs of achieving different levels of AI performance, (iii) the required accuracy for economically acceptable task performance, and (iv) the scale at which AI systems are deployed. In particular, we lack a microeconomic model that treats full automation, partial automation, and no automation as competing options within a single optimization problem, and the data to quantify how much labor is optimally automated with current technologies. Without such a framework, it is difficult to answer basic questions: When is it optimal for firms to invest in automation? When they do, is it optimal to fully replace workers on a task, or to retain them as collaborators? And how does the scale of deployment -- within a firm, an industry, or the entire economy -- shift the boundary between these choices?

In this paper, we develop such a framework. We reconceptualize automation for the era of large language models and deep learning. Prior automation models were shaped by robotics, where capital directly replaces labor. Modern AI systems introduce a fundamentally different economic structure: the binding constraint is accuracy rather than physical capability, and achieving the accuracy required for reliable deployment means navigating the steep cost curves imposed by scaling laws. We model each task as a bundle of discrete classification decisions with a required accuracy threshold and decompose it into automatable and non-automatable components. On the supply side, we estimate an AI production function from fine-tuning scaling-law experiments that link performance to data, training steps, and model size. On the demand side, we introduce an entropy-based mapping from model accuracy to labor substitution, which quantifies how much human work time AI displaces at each level of performance. Together, these components yield a task-level optimization problem in which firms choose among labor-only production, partial automation, and full automation.

The supply side generates the convex cost structure that drives the partial-automation result: because each input to the AI production function exhibits diminishing returns, the cost of achieving progressively higher accuracy rises steeply. The demand side provides a quantitative account of how work is divided between AI and human collaborators along the automation continuum. Our information-theoretic approach leverages the well-established relationship between entropy and human processing time: higher model accuracy reduces the residual uncertainty that humans must resolve, and this reduction in uncertainty translates into labor savings. This is what allows us to model partial automation as a precise, quantifiable outcome rather than a vague intermediate category.

We implement this framework empirically in computer vision -- one of the most developed areas of AI and one where abundant scaling-law data exists \citep{svanberg2024beyond}. We bring the framework to data by combining several new empirical inputs with existing economic statistics. First, we use O*NET to identify 420 computer-vision-exposed tasks across 263 occupations and measure how much worker time in each occupation is allocated to each task.\footnote{An O*NET task refers to the standardized task descriptors in the U.S.\ Department of Labor's Occupational Information Network (O*NET), which provides a detailed taxonomy of job tasks for each occupation. We later discuss how these task definitions are incorporated into our data construction in Section \ref{sec:results}. Across the entire O*NET database, there are 1,016 occupations and 18,796 tasks, for an average of 18.50 tasks per occupation. There are also 73,308 direct work activities -- or subtasks -- for an average of 3.81 subtasks per task.} Second, we use a large survey of workers and domain experts to elicit task-specific required accuracy, which we translate into entropy-based complexity and performance targets. Third, we use GPT-4o to extract for each task the number of vision subtasks, the number of classes per subtask, and the share of the task that is inherently visual, with manual validation by human coders. Finally, we integrate wage, employment, and firm-size distributions from U.S.\ statistical agencies to scale task-level decisions up to the occupation, firm, industry, and economy-wide levels.

Calibrated to these data, our model delivers four main findings:

First, at the task level, the convexity of the AI cost function and the entropy-based mapping from accuracy to labor savings imply that marginal costs of improving model performance eventually exceed marginal labor savings for many tasks. As a result, partial automation is often cost-minimizing, even when full automation is technically feasible. This occurs both when AI falls short of the accuracy required for full task replacement and when only a subset of subtasks is economical to automate.

Second, task complexity shapes labor substitution. For tasks with fewer subtasks and low complexity, high labor substitution is often optimal. Holding complexity fixed, moving from small-scale deployment to medium-scale deployment increases labor substitution. For more complex tasks, labor substitution is reduced at both small- and medium-scale deployment.

Third, at the firm level, we find that approximately 11\% of the labor compensation allocated to computer-vision-exposed tasks is economically attractive to automate. Of this, most labor saving comes from partial rather than full automation, with AI systems handling a share of computer-vision-related work while humans retain the residual workload. These estimates pertain to a single AI modality; the economically viable share of automation would be considerably larger when the framework is extended to large language models and other foundation models that cover additional task types.

Fourth, the scale of deployment fundamentally shapes automation incentives. When a single AI-as-a-Service offering or AI agent can be shared across all firms performing the same task, fixed development costs are distributed over a much larger user base. This substantially expands the range of tasks for which adoption is economically viable and shifts the optimum toward higher-quality models and higher automation rates. Conditional on adoption, firms typically deploy systems that automate the large majority of the targeted task, leaving only a small residual share to human workers. Under economy-wide deployment, the economically viable share of computer-vision-exposed labor compensation rises substantially. Because large language models and other foundation models exhibit similar scaling-law cost structures, the analytical framework and the qualitative mechanisms we develop are designed to extend to AI-based automation more broadly.

Our contributions are fivefold. First, we provide a unified microeconomic framework that endogenizes the choice among full automation, partial automation, and no automation at the task level, explicitly linking these choices to task complexity, required accuracy, and AI scaling laws. Second, we introduce an entropy-based mapping from model performance to labor saving that provides a quantitative framework for modeling human--AI collaboration and defining feasibility and optimality along the automation continuum. Third, we show that scaling-law diminishing returns create a fundamental asymmetry: the cost of moving from partial to full automation can be orders of magnitude larger than the cost of partial automation itself, so that full automation is often not cost-minimizing, while partial automation frequently emerges as the interior optimum. Fourth, we combine new experimental and survey evidence with large administrative datasets to calibrate this framework to real-world computer vision tasks across the U.S.\ economy, yielding quantitative estimates of how much labor is optimally automated under current technologies and prices. Fifth, by comparing firm-level deployment, AI-as-a-Service, and AI agents, we show how economies of scale in AI development reshape the automation frontier and help explain why early adoption is concentrated in large firms and standardized tasks, rather than in small firms and complex, high-variability activities.

More broadly, automation potential depends jointly on technical feasibility and economic feasibility. Occupations with standardized, low-complexity tasks are more likely to satisfy both conditions, while occupations involving high variability, tacit knowledge, and complex classification demands remain less attractive targets for automation. Although we calibrate the framework to computer vision -- where abundant scaling-law data permits precise empirical implementation -- the underlying cost structures and optimization logic apply to any AI system exhibiting diminishing returns to scale, including large language models and multi-modal foundation models.

Our work highlights that partial automation is not merely a transitional state on the path to full automation, but often the long-run cost-minimizing outcome. As costs fall and delivery models such as AI-as-a-Service expand access, the frontier of economically viable automation will continue to widen -- but the fundamental asymmetry between partial and full automation costs implies that human--AI collaboration will remain a central and durable mode of adoption across a broad range of tasks.

The remainder of the paper proceeds as follows. Section \ref{sec:related_work} reviews related work, covering the transition from technical feasibility to economic viability, the reconceptualization of automation as a continuum, scaling-law regularities, and the connections between micro-level decisions and macro-level employment outcomes. Section \ref{sec:theory} presents the theory of cost optimization, including the supply-side AI cost function, the demand-side entropy-based mapping from accuracy to labor substitution, and the firm's optimization problem across the three automation regimes. Section \ref{sec:data} estimates the AI production function from fine-tuning scaling-law experiments and characterizes its properties, including performance elasticities and input substitutability. Section \ref{sec:results} quantifies the automation frontier by bringing the framework to data --calibrating it to computer vision tasks using O*NET task data, a domain expert survey, and GPT-4o-derived task decompositions -- and presents results at the task, firm, industry, and economy-wide levels. Section \ref{sec:conclusion} concludes.

\section{Related Work}
\label{sec:related_work}
The fundamental question facing firms in the AI era is not whether tasks can be automated, but whether they should be automated, and if so, to what extent. This question has driven the economic analysis of AI automation to evolve from early exposure-based assessments toward frameworks that evaluate the feasibility and optimality of human--AI collaboration. This evolution rests on three key insights that directly motivate our microeconomic approach: (1) technical possibility does not guarantee economic viability; (2) automation exists on a spectrum rather than as a binary choice; and (3) the economics of AI deployment depend fundamentally on scaling laws and cost structures that vary across tasks and required accuracy levels.

Building on these insights, the related literature can be grouped into four strands that collectively inform our framework. Section~\ref{sec:lit_feasibility} reviews the transition from technical feasibility to economic viability. Section~\ref{sec:lit_recon} surveys evidence on human--AI collaboration and the reconceptualization of automation as a continuum. Section~\ref{sec:lit_scaling} links scaling-law regularities to the cost structure of AI systems. Section~\ref{sec:lit_macro} connects these micro foundations to macro-level employment outcomes. Section~\ref{sec:lit_positioning} positions our contribution within this broader literature.

\subsection{From Technical Feasibility to Economic Viability}
\label{sec:lit_feasibility}
Early task-based automation theories trace back to \citet{autor2003skill}, who built on \citet{zeira1998workers}.
\citet{acemoglu2011skills} combine elements of prior work and consider a continuum of tasks, with technologies augmenting factors of production through a range of possible outcomes: increasing worker productivity, increasing capital productivity, automating work, or creating new tasks \citep{acemoglu2024task}. With the combination of automation and new tasks, this literature established that automation induces the contraction of the range of tasks performed by labor.

In recent years, initial AI automation research focused primarily on exposure---the technical possibility of tasks being automated through machine learning or robotics \citep{frey2017future, brynjolfsson2017machinelearning, eloundou2024gpts}. However, recent scholarship has pivoted toward the more economically relevant question: when does technical feasibility translate into profitable automation decisions?

This shift toward economic viability is exemplified by \citet{autor2025expertise}'s expertise framework, which demonstrates that the wage-and-employment impact of automation depends critically on whether AI eliminates low-expertise or high-expertise subtasks. When automation removes inexpert work, it raises average task expertise, bidding up wages while reducing employment among workers whose skills the automated tasks replaced; eliminating expert work produces the opposite effect. This expertise perspective strengthens the argument that task composition---not just exposure---determines economic viability and directly affects the cost-benefit computation that firms face. This is precisely the margin our cost-minimization model addresses by incorporating both required accuracy levels and the complexity structure of tasks.

Complementing this task-composition insight, \citet{hampole2025ailabor} provide firm-level evidence that labor-demand effects depend on the distribution of AI exposure across tasks within firms---that is, within-firm heterogeneity. They find that while mean exposure reduces labor demand, concentrated exposure in specific tasks can increase demand through labor reallocation toward remaining work. This finding validates our framework's emphasis on partial automation: when AI handles concentrated, high-exposure subtasks while workers focus on complementary activities, overall productivity can increase without proportional job displacement.

Beyond technical feasibility, adoption frequently requires substantial supporting investments in managerial know-how, organizational restructuring, and process redesign. \citet{Brynjolfsson2021productivity} characterize this pattern through a ``productivity J-curve,'' in which productivity gains materialize only after firms make costly intangible and organizational investments. These studies highlight that adoption costs extend beyond model development to include the broader investments needed to integrate AI systems into production.

\subsection{Reconceptualizing Automation}
\label{sec:lit_recon}
The recognition that automation operates along a spectrum---from full automation to partial automation---has motivated a growing body of research on how workers and AI jointly contribute to task execution. However, much of the economic literature has assumed that automation yields full task replacement of workers with machines. With the advent of generative AI and deep learning, the wholesale replacement of labor with capital is no longer the only choice. Partial automation has become a possibility: AI models are able to perform tasks and subtasks while augmenting the skills that workers bring, making the choice more than binary. At the occupation level, some roles can be automated while others continue to be performed by co-workers. At the task level, automated tasks can free up workers for tasks requiring human skills. At the subtask level, automation requires workers to augment subtasks with new or existing capabilities. The choice of automation versus augmentation can thus be conceptualized as partial versus full automation.

Human involvement for task completion and quality is the focus of \citet{shao2025futurework}, who introduce the Human Agency Scale (HAS), a five-level audit framework (H1--H5). The scale centers on the ability of workers to act independently while adopting AI agents and provides a shared language to capture the spectrum between automation and augmentation. HAS is based on assessments from 1,500 workers across 104 occupations and annotations from 52 AI experts, covering 844 occupational tasks. Their WORKBank dataset shows that although roughly 46\% of tasks fall into an ``Automation Green-Light'' category, a comparably large share lies in ``Red-Light'' or ``R\&D Opportunity'' zones where some degree of human involvement is preferred. This evidence highlights that many tasks are not suitable for full automation and instead benefit from configurations in which AI systems provide assistance while workers continue to guide or refine outputs.

Field evidence also demonstrates the potential value of partial automation. \citet{brynjolfsson2025generative} and \citet{handa2025economic} show that partially automated workflows can substantially increase productivity without proportional reductions in labor. In a large-scale field experiment with radiologists, \citet{NBERw31422} find that simply providing AI predictions does not always improve diagnostic accuracy; gains arise only when experts are able to integrate AI outputs with contextual information. This supports the broader insight that human and machine inputs are often complementary, particularly when tasks require nuanced judgment or contextual understanding.

On the theoretical side, \citet{NBERw33949} develop a sufficient-statistic approach for allocating decisions between people and predictive systems in classification settings. They show that optimal policies direct cases with high-confidence AI predictions to automated decision-making and assign uncertain cases to humans. Importantly, they find that once such allocation is in place, providing AI predictions to humans yields limited additional benefit: humans tend to under-respond to model output and reduce effort when shown confident AI predictions. These results highlight that the primary efficiency gains come from assigning cases to the agent---human or AI---best suited to handle them, rather than from simultaneous joint decision-making. However, the most suitable agent could change with learning.

Our paper takes a novel information-theoretic approach to quantifying human--AI collaboration. We leverage entropy---a measure of uncertainty and information content---to map AI model accuracy into effective labor savings. Higher model accuracy reduces the residual uncertainty that humans must resolve, and this reduction translates proportionally into less required labor time. This connection between entropy and processing time is well supported empirically: \citet{hick1952rate} and \citet{hyman1953stimulus} establish that reaction time in choice tasks increases linearly with task entropy, and subsequent work extends this relationship to reading time and binary decision tasks \citep{lowder2018lexical, hu2022human}. By embedding these information-theoretic relationships into an economic framework, we provide a rigorous quantitative bridge between AI model performance and labor market outcomes.

\vspace{-2mm}
\subsection{Scaling Laws and the Economics of AI Quality}
\label{sec:lit_scaling}
The economic feasibility of both full and partial automation depends not only on technical feasibility but also on the cost structure required to achieve a given level of performance. Recent advances in understanding how AI performance scales with data, compute, and model size provide an empirical basis for assessing these costs. These relationships---commonly referred to as AI scaling laws---describe how increases in training data, computational resources, and model parameters translate into performance improvements \citep{kaplan2020scaling, rosenfeld2021scaling, hoffmann2022training, thompson2022computationallimitsdeeplearning}. The relationship typically follows a power law with diminishing returns as systems approach fundamental limits, thereby quantifying the cost of improving model quality.

For computer vision specifically, \citet{svanberg2024beyond} demonstrate that transfer-learning curves exhibit consistent diminishing-return exponents across classification tasks, implying sharply rising compute costs at very high accuracy levels. Economically, moving from ``good'' to ``near-perfect'' accuracy can be vastly more expensive, yielding limited incremental benefit. This convex cost structure explains why firms often select partial automation: when the marginal cost of accuracy exceeds its marginal benefit, retaining human oversight becomes the cost-minimizing choice. Our framework embeds these empirical scaling relationships into a formal cost function linking model quality, required accuracy, and task complexity, thereby quantifying when automation remains economically feasible and when human collaboration dominates.

These firm-level cost boundaries naturally aggregate into broader labor-market outcomes, providing a bridge to macroeconomic analyses of AI adoption.

\subsection{Integrating Micro-Foundations with Macro-Employment Effects}
\label{sec:lit_macro}
The broader employment implications of AI adoption depend fundamentally on the micro-level decisions that determine how tasks are allocated between workers and automated systems. Recent macro-level research emphasizes that aggregate effects arise from the interplay of task-level substitution, within-firm reallocation, and economy-wide productivity gains. At the aggregate level, employment outcomes emerge from how task-level automation decisions propagate through firms and industries. \citet{hampole2025ailabor} use firm-level variation in AI adoption to show that task-level substitution is largely offset by productivity gains and the reallocation of labor toward complementary activities, yielding modest net employment effects. This finding aligns with \citet{acemoglu2025simple}, who argues that employment outcomes depend on the relative pace of new-task creation versus automation-driven substitution. In this view, labor-market effects depend not only on technological capability, but also on how firms reorganize production and reallocate tasks in response to it.

\subsection{Positioning Our Contribution}
\label{sec:lit_positioning}

Existing research has advanced understanding along several dimensions of the automation process---ranging from the expertise composition of tasks and the role of human agency to the scaling-law cost structures that govern model performance and the macro employment effects of technological adoption. Yet these strands remain largely separate. What is missing is a unified perspective that links these insights to the firm's choice among full automation, partial automation, and no automation---the margin through which technical progress ultimately shapes economic outcomes.

Our framework addresses this gap by connecting the technical foundations of AI performance with the economic determinants of adoption. We draw on the expertise-based view of task composition to capture heterogeneity in automation potential; on evidence of human--AI complementarity to motivate partial automation; on scaling-law research to characterize the cost of achieving required accuracy levels; and on macroeconomic findings to interpret how such micro decisions aggregate into economy-wide outcomes. When full automation is costly or produces unreliable performance---conditions that both micro and macro evidence shows to be common---firms rationally choose partial automation and reallocate labor toward complementary activities rather than eliminating it. This mechanism provides a micro-founded rationale for why AI tends to augment, rather than displace, labor in practice.

\section{Theory}
\label{sec:theory}
The starting point for our framework is that firms today can access foundation models -- pre-trained AI systems that deliver some baseline level of performance across a broad class of tasks without task-specific investment. For some tasks, this baseline performance is sufficient for deployment. For many others, however, the accuracy required for reliable task execution exceeds what the foundation model provides out of the box. In such cases, the firm must invest in task-specific fine-tuning: additional training data, more training steps, and potentially larger model architectures. These investments follow the scaling laws documented in the AI literature: each additional unit of data, compute, or model capacity yields diminishing returns in accuracy, so that the cost of improving AI performance is convex and steeply increasing at high accuracy levels.

This cost structure gives rise to the central economic trade-off of our model. The labor savings from higher AI accuracy are approximately proportional to the reduction in uncertainty that the AI system achieves (as formalized in Section~\ref{sec:benefit}), while the costs of achieving that improvement are convex (as formalized in Section~\ref{sec:cost}). When the benefit exceeds the cost throughout the relevant performance range, full automation is optimal. When the convex cost curve overtakes the benefit before the required accuracy is reached, partial automation -- deploying AI at an interior accuracy level and retaining human workers for the residual -- is cost-minimizing. When even modest AI investment is uneconomical, no automation is warranted. The formal model below makes these three cases precise.

\subsection{Model Set Up}
\label{subsec:model}
To fix ideas, it is helpful to begin with a concrete example, though our framework is not specific to this setting. Consider radiology: each diagnostic case (e.g., reading an X-ray or CT scan) involves a sequence of discrete classification decisions such as determining whether an image is normal or abnormal, identifying the likely condition, and, when applicable, assessing severity. This example simply illustrates that many tasks can be decomposed into a number of discrete decisions and that different decisions may require different accuracy levels. Motivated by this general structure, we model each task $i$ as requiring $Y_i$ discrete decisions and define $a_i$ as the accuracy level at which those decisions must be performed. These primitives apply broadly---to any setting in which AI can potentially automate part of a multi-step decision process---and they will allow us to formalize the economics of full and partial automation in the sections that follow.

Consider a firm that must complete task $i$ as an integral component of its production process. Previously, this task has been entirely performed by human labor. The firm is now assessing the potential adoption of AI models to either partially or fully automate task $i$.

In the second stage, conditional on adoption ($s_i = 1$), the firm chooses the accuracy level $a_i$. Throughout, we interpret $a_i$ as the quality of the AI system, expressed in terms of its operating accuracy, the accuracy level at which the firm chooses to operate its AI system. Higher $a_i$ corresponds to deploying a more capable and better-performing model.

Assume that the total output of the task is given exogenously as $Y_i$ (the number of discrete decisions the firm must make for task $i$ over the period). In our context, ``output'' refers to the total number of completed decisions required for task $i$, which must achieve the target accuracy $a_i$. Denote the total cost of performing the task $i$ as $C_i(s_i, a_i, Y_i)$.

If the firm chooses labor-based production ($s_i = 0$), it will not face any further decision regarding the quality of AI, and the cost does not depend on $a_i$. In this case, the total cost with a human-only platform is given by:
\begin{equation}
    C_i(0,a_i, Y_i) =  C_i(0, Y_i)=w_{i} \tau_i Y_i.
    \label{eq:human_cost_coef}
\end{equation}
$w_{i}$ represents the exogenous labor wage, and $\tau_i$ represents the labor time it takes to produce one unit of the output. 

Here, we assume that each task consists of two proportions: a vision component and a non-vision component.\footnote{In this paper, we focus on computer vision AI. For other technologies, the ``vision/non-vision'' split can be replaced with ``AI/non-AI'' or a technology-appropriate decomposition.}  When the firm completes the task using labor only, workers spend a $\delta_i$ fraction of their time on the vision component which a human relies on vision but could be replaced by computer vision AI to some extent and $(1-\delta_i)$ on the non-vision component which requires other cognitive or physical abilities and cannot be substituted by computer vision AI.\footnote{For example, consider one of gambling managers' tasks that requires them to circulate among gaming tables to ensure that operations are conducted properly, that dealers follow house rules, and that players are not cheating. The aspects that can be replaced by surveillance camera and computer vision include analyzing the state of the game at each table, detecting abnormal betting patterns such as consistently high winnings, and identifying unusual hand movements of customers. Computer vision cannot detect audio-based cheating, such as whispered communication or coded language.}  We interpret $\delta_i \in [0,1]$ as the fraction of task i that is technically automatable by computer vision AI. When $\delta_i = 0$, the task cannot be automated at all; when $\delta_i = 1$, the entire task's input is technically automatable.

If the firm chooses to introduce computer vision AI, i.e. $s^*_i = 1$, then AI and human labor collaborate to accomplish the task, the total cost of producing $Y_i$ is

\begin{equation}
    C_i(1,a_i, Y_i) = \big[\delta_{i} (1-r_{i}(a_{i})) + (1-\delta_{i})\big] w_{i}\tau_i Y_i +\kappa_i(a_i, Y_i).
    \label{eq:ai_cost_coef}
\end{equation}

Equation~\eqref{eq:ai_cost_coef} consists of two terms. The first term corresponds to the labor cost, and the second term is the cost associated with training and deploying the AI system. For the labor cost, recall that workers originally spent $(1-\delta_i)$ of their time on the non-vision portion of the task, which is not automatable by the AI system and must be completed by human labor. For the computer vision portion, we assume that the AI system saves a proportion $r_i$ of the required labor, so human workers still need to perform the remaining $(1-r_i)$ proportion. $r_i=1$ corresponds to the case of full automation, where all the vision part is performed by the AI system. $0<r_i(a_i)<1$ corresponds to the case of partial automation, where human labor still needs to participate in a portion of the production process; $r_i=0$ corresponds to no automation. The degree of automation, $r_i$, depends on the quality of the computer vision system chosen by the firm, described as accuracy $a_i$. If the firm chooses a high-accuracy AI system, more human labor will be saved. Thus $r_i(a_i)$ is a monotonically increasing function. Section~\ref{sec:auto_rate_collab} describes our approach to obtain $r_i(a_i)$. 

In the radiology example, partial automation corresponds to natural coarse-to-fine workflows: an AI system may triage cases (normal vs. abnormal), or narrow a large label set down to a short list of plausible findings, while the radiologists perform the remaining fine-grained decisions among the narrowed options. We formalize this division of work in Section~\ref{sec:auto_rate_collab}.

The second term $\kappa_{i}(a_{i}, Y_i)$ represents the cost associated with the building and use of the AI system to produce the output of the task. It is increasing monotonically with respect to system quality $a_{i}$ and task output quantity $Y_i$. Section~\ref{sec:ai_cost_min} describes our approach to derive $\kappa_{i}(a_{i}, Y_i)$. 

Equation~\eqref{eq:ai_cost_coef} highlights the trade-off the firm faces in choosing different quality levels of the AI-system. The higher the accuracy, the more human labor will be saved, but the AI-related costs will also become higher.

Essentially, $\kappa_{i}(a_{i}, Y_i)$ describes the supply of AI technology in terms of quality; $r_{i}(a_{i})$ describes the demand of AI technology in terms of quality. Section \ref{sec:cost} and Section \ref{sec:benefit} will detail our modeling of the supply and demand aspects of AI technology, respectively. Appendix~\ref{app:firm_backward_induction} describes the backward induction approach to solve the firm's optimization problem and summarizes different scenarios.

For each task, the firm's optimization problem is summarized as follows
\begin{equation}
    \min_{a_i, s_{i}} C_{i}(s_i, a_i, Y_i).
    \label{eq:cost_min}
\end{equation}

Assume that the total output of the task is given exogenously as $Y_i$ (the number of discrete decisions the firm must make for task $i$ over the period). In our context, ``output'' refers to the total number of completed decisions required for task $i$, which must achieve the target accuracy $a_i$. Denote the total cost of performing the task $i$ as $C_i=C_i(s_i,a_i,Y_i)$.

\subsection{Assumptions and Scope}
\label{sec:assumptions}
Our analysis is conducted within a partial equilibrium framework. Specifically, we take the wage of labor $w_i$, the cost of computing and data used by AI systems, and the output level $Y_i$ of each task as given. These variables are not endogenously determined within the model. Additionally, we assume that the set of tasks required in the economy is fixed and exogenously specified. That is, we do not consider the possibility of new tasks emerging due to advances in AI capabilities. The task composition within each occupation is also assumed to remain constant.

Moreover, when evaluating the benefits of AI adoption, we focus exclusively on the labor-saving aspect - i.e., the cost reduction achieved through substitution of human labor with AI in the vision component of the task. We do not account for potential additional gains from AI systems that could arise from improved performance or accuracy levels beyond human levels, which might enhance the quality or output of a task.  

For more capable systems, the question is not whether systems will be created that have better capabilities than human workers. Rather, when and if more capable systems are economically attractive to build, so are systems with capabilities equal to human workers. Consequently, our modeling approach will correctly identify the extent and timing of automation. The challenge to our approach would occur if building a more capable system becomes economically attractive before the equal-capabilities system. We argue that this is unlikely to be a common occurrence because improving the capability of AI systems results in an enormously rapid increase in the cost of these systems, as shown by \citet{thompson2022computationallimitsdeeplearning} and as is consistent with foundational computer science work in this area (\citet{kaplan2020scaling}, \citet{henighan2020generative}, \citet{mikami2021scalinglawsynthetictorealtransfer}). Since less capable systems are unlikely to be able to substitute for human workers, and more capable ones are likely to become economically attractive only later, the modeling that will best predict the automation of human labor is the model with equivalent capabilities. And, because such a system provides similar benefits to the human doing that task, one can compare the economic attractiveness of these systems by comparing their costs.

Similarly, any economic benefits stemming from the creation of entirely new products or services enabled by AI technologies are not considered.

\subsection{Supply of AI: In Terms of Both Quality and Quantity}
\label{sec:cost}
In this section, we discuss how to derive the function $\kappa_i(a_i, Y_i)$. Recall that $\kappa_i(a_i, Y_i)$ characterizes the cost of deploying an AI system with a given quality requirement $a_i$ and usage level $Y_i$ -- that is, the task output produced using the system. Specifically, $\kappa_i(a_i, Y_i)$ represents the cost associated with implementing an AI system that delivers quality level $a_i$ and generates task output $Y_i$.

One of the foundational discoveries in the scaling law literature is that AI models exhibit predictable improvements in performance, $a_i$, when the inputs to training are scaled up in a systematic manner.\footnote{In this context, training specifically refers to fine-tuning a foundation computer vision (CV) model on a task tailored to a specific occupational setting.}  We identify three key input factors that affect the quality of an AI system: data, model size, and training steps. Increasing any of these inputs leads to higher compute requirements, and both data and compute contribute directly to the overall cost of developing the AI system. Once a model with a given quality level has been successfully trained, we can consider the fixed cost associated with achieving that level of quality to be determined. However, increasing the model's usage---that is, increasing the number of inference runs---raises the required amount of compute during deployment. As a result, the cost associated with AI system also increases with the usage level $Y_i$.

\subsubsection{Scaling Law: The Production Function for AI Quality}
\label{subsubsec:scaling_law}

To illustrate our model, return to the radiology example. One core task in this occupation that can be augmented or substituted by a computer vision system is the diagnosis of diseases and abnormalities from medical images such as X-rays or CT scans. In this context, the accuracy of the computer vision system for this diagnostic task,
\begin{equation}
 a_i = Q(D_i, T_i, M_i; m_i, n_i),
\label{eq:accuracy_production}
\end{equation}
can be measured by its diagnostic performance and is modeled as a production function of three critical inputs: data $D_i$, training steps $T_i$, and model size $M_i$, together with task-specific parameters $(m_i,n_i)$.

Data ($D_i$) refers to the amount of data used for training the computer vision system. Specifically, in our example, it consists of paired X-ray images and their corresponding diagnoses. Training steps ($T_i$) represents the number of iterations during the training process where the model's parameters are updated. Multiple training steps constitute an epoch, where the entire dataset is passed through the model once. We use $T_i=D_i \cdot epoch_i$as a proxy for training steps. Model Size ($M_i$) denotes the size of the computer vision system, characterized by the number of parameters within the AI model.\footnote{A commonly included input in traditional production functions---labor---is notably absent from our production function. This omission reflects the idea that labor is not a key driver of improvements in the performance (i.e., quality) of an AI system. While the costs associated with employing AI experts, engineers, and support staff must be accounted for, we incorporate these labor inputs into the fixed cost component (introduced in the following section).}  In equation~\eqref{eq:accuracy_production}, $m_i$, the number of subtasks, and $n_i$, the number of classes per subtask, serve as key parameters that characterize the complexity of a given O*NET task.

In general, as the number of distinct diseases that a diagnostic system is able to detect increases, the value of $m_i$ will increase accordingly. Moreover, if diagnosis becomes more granular, for example, distinguishing between early-stage and advanced-stage illness, or between mild and severe forms, then each subtask will require more output classes, and $n_i$ will exceed 2. In both cases, increases in $m_i$ and $n_i$ reflect a rise in task complexity. Achieving higher diagnostic accuracy in such complex settings will require a greater investment in resources in terms of input data, computational power, or model sophistication.\footnote{In Appendix \ref{app:a_and_m}, we explain how the number of tasks $m_i$ influences the interpretation of required accuracy. Simply put: when two O*NET tasks share the same required accuracy, the one with a higher $m_i$ value will impose stricter requirements on both individual vision task accuracy and overall, AI model performance.}

\subsubsection{Cost Minimization for AI Systems} 
\label{sec:ai_cost_min}

The firm needs to address the cost minimization problem: given a required level of model performance $a_i$ and usage level $Y_i$, how to determine the optimal input bundles in developing and adopting the AI system?

The objective function in this cost minimization problem is:
\begin{equation}
    K(D_i,T_i,M_i,I_i)=c_F+c_D D_i+c_T M_i T_i + c
    _I I_i.
\end{equation}

In our earlier discussion of AI quality, we introduced three types of inputs that affect system performance. We now extend the framework by introducing a fourth input, denoted by $I_i$, which does not affect model quality but captures the computational resources expended for model inference.\footnote{Inference refers to the process of using a trained model to generate outputs.} Specifically, $I_i$ reflects the GPU hours invested to perform inference and accomplish the task over the period. Unlike the other inputs, $I_i$ does not affect the quality of the model but is instead directly linked to the quantity of output, that is, the level of usage. We assume that $I_i = \tau_\text{GPU} M_i Y_i$ is the amount of GPU time needed to produce one unit of task output. $I_i$ is proportional to $M_i$ because larger models have more parameters and more computations, so they require more GPU time to process inputs during inference. $\tau_{\text{GPU}}$ is the number of GPU hours required for a model size $M_i$ used $Y_i$ times. The term $c_I I_i$ captures the variable cost incurred during deployment, which increases with system usage.

We use $c_F$ to denote the fixed cost component, which does not vary with the model's quality or usage level. In the context of this paper, $c_F$ primarily reflects the cost of hiring an engineering team to develop, train and maintain the AI system. 

$c_D D_i+c_T M_i T_i$ represent the variable costs required to train and maintain a model to a given quality level.\footnote{Model maintenance includes monitoring performance, integrating updated data, retraining or fine-tuning as needed, addressing model drift, and ensuring the long-term reliability and safety of the deployed system.} Here, $c_D$ is the cost of increasing data, and $c_T$ is the cost of increasing training computation. Larger model size and more training steps both increase the total amount of training compute. This cost structure arises from the inherent characteristics of training and deploying computer vision models. For instance, increasing model size typically requires more compute per training epoch and greater computational resources per inference. A more detailed definition of each cost term in the formal cost function is provided in Appendix~\ref{app:cost}, and a broader discussion of the economic interpretation of these costs appears in Section~\ref{subsubsec:ai_cost_components}.

We can now formally define the corresponding cost minimization problem as follows:

\begin{equation}
\begin{aligned}
& \min_{D_i, T_i, M_i, I_i} \; \kappa_i(D_i, T_i, M_i, I_i) \\[6pt]
\text{s.t.} \quad & Q(D_i, T_i, M_i; m_i, n_i) \geq a_i \\[6pt]
& \frac{I_i}{\tau_{\text{GPU}} M_i} \geq Y_i
\end{aligned}
\label{eq:define_cost_minimization}
\end{equation}

The objective is to minimize cost while delivering the required accuracy ($a_i$) and supporting the volume of decisions ($Y_i$) necessary. See Appendix~\ref{app:cost_minimization_first} for the first order conditions.

\subsubsection{Cost Components in AI System Development and Deployment}
\label{subsubsec:ai_cost_components}

To interpret the solution to the cost minimization problem, it is helpful to describe how the different terms in the objective function reflect the economic costs of developing and deploying an AI system. The reduced-form cost function $\kappa_i(a_i, Y_i)$ represents the minimum expenditure needed to achieve a system with accuracy level $a_i$ that can support a usage level of $Y_i$. Conceptually, these costs fall into two broad categories: fixed costs associated with building and training the model, and variable costs associated with using it at scale.

The fixed component corresponds to the one-time investment necessary to create a model capable of meeting the target accuracy level. This includes engineering and development labor involved in designing the system, setting up the training pipeline, and maintaining the model throughout its lifecycle.\footnote{For example, maintenance includes managing updates, integrating new training data, addressing model drift, and ensuring the long-run reliability of the deployed system.} It also includes the cost of acquiring and preparing training data, as reflected in the term proportional to $D_i$, since higher accuracy generally requires a larger and more carefully curated dataset. In addition, training a more accurate model requires greater computational resources, which is captured by the component proportional to $M_i T_i$: larger models require more compute per update, and more training steps increase the total computational workload. Together, these fixed elements determine the minimum development cost necessary to produce a model of quality $a_i$.

A second part of the cost arises from deployment and scales with the amount of output the model produces. Once the model is trained, each inference run requires computational resources that depend on the model size. Because larger models involve more parameters and higher per-call compute requirements, the term proportional to $I_i$ captures the expenditure associated with running the system to produce the required $Y_i$ task outputs. This component therefore reflects the variable cost of using the model in practice, and increases with both the model size chosen to achieve accuracy $a_i$ and the usage level $Y_i$ that the firm must support.

These two components---development costs that depend on achieving accuracy $a_i$, and deployment costs that depend on supporting usage $Y_i$---together constitute the overall cost structure summarized in $\kappa_i(a_i, Y_i)$. The structure implies two useful properties. First, the cost is increasing in $a_i$, because achieving higher accuracy requires more data, more computation, or larger models. Second, the cost is increasing in $Y_i$, since each additional model call requires inference compute that grows with the model size. These properties clarify how AI-system costs enter the firm's optimization problem in Section~\ref{sec:equilibrium}, with fixed costs governing whether AI adoption is economically viable at all, and variable costs determining the marginal trade-off between AI usage and human labor for task $i$.

\subsection{Demand for AI Quality}
\label{sec:benefit}
In this section, we explain our modeling of $r_i(a_i)$, which is the proportion of labor time (within the AI-automatable subtask) that the AI system could save given the accuracy of the AI system is $a_i$. $r_i(a_i)$ serves as a core equation characterizing the demand for AI quality. While the introduction of AI can generate economic gains from different channels, e.g. increase in output, better products, we do not attempt to incorporate all benefits into the analysis. To address this, we use the reduction in labor compensation resulting from labor-saving substitution as a proxy for the benefit of AI adoption. We focus specifically on the demand for AI quality.

As we construct a one-to-one correspondence between AI quality and the extent of labor substitution by $r_i(a_i)$, once quality is determined, both the proportion of labor saving and the nature of human-AI collaboration are determined. On the other hand, based on the assumption that the output at task level $Y_i$ is given exogenously in our partial equilibrium framework, we do not require a separate function to explicitly determine the demand for AI quantity or usage. 

To bridge AI quality and labor saving, we draw on two concepts from information theory: entropy and cross-entropy loss. The entropy concept, well suited for computer vision models, is a measure of the amount of missing information before reception. Entropy captures the clarity of vision or the fidelity of sound, quantifying the average level of uncertainty or information associated with the variable's potential states or possible outcomes. Cross-entropy loss is the standard measure for classification problems, such as image recognition. It directly compares the predicted probabilities to the true labels. It provides more informative gradients than functions like mean squared error, which can be slow to converge when predictions are confidently wrong. It also provides a measure of the "distance" between the predicted probability distribution and the true distribution, with the goal of making them as close as possible. Minimizing cross-entropy loss is equivalent to maximizing the log-likelihood of the data. Lower cross-entropy loss increases confidence in correct predictions and generally delivers higher accuracy. Cross-entropy loss penalizes confident incorrect predictions more than uncertain predictions. As a result, minimizing cross-entropy loss during training tends to maximize accuracy, but not with perfect correlation.

A body of literature in psychology provides empirical support for a relationship between entropy and work time. Intuitively, higher model accuracy reduces uncertainty in outcomes, and a reduction in uncertainty corresponds to fewer effective decisions that a human needs to make---thereby translating into less required labor time. The cross-entropy loss commonly used in training AI models---both in the mathematical sense and in the context of our setting---is highly correlated with model accuracy. Together, these relationships allow us to establish a mapping from accuracy to labor saving, linking the technical performance of the AI system to the economic outcome of interest.

\subsubsection{Entropy and Task Complexity}
\label{sec:CE_and_time}

In information theory, classification is essentially a process of narrowing down possibilities. The task entropy, $H_{task}$, measures the number of probabilistically equivalent decisions to make in order to rule out all the possibilities. Formally, denote $p(L=l)$ as the prior probability of each class $l \in \mathcal{L}$. Then, the task entropy is defined as
\begin{equation}
    H_{task} = -\, \sum_{l \in \mathcal{L}} p(L=l) \ln p(L=l).
\end{equation}

We introduced the concept of task complexity in Section \ref{subsubsec:scaling_law}, characterizing it using two parameters, 
$m_i$ (number of subtasks) and $n_i$ (number of classes). Task complexity can be formally captured by its associated task entropy $H_{task}$, with higher values of 
$m_i$ and $n_i$ generally corresponding to higher levels of entropy.

\subsubsection{Cross-Entropy Loss and Classifier Performance}
Cross-entropy loss is a standard metric for evaluating the performance of probabilistic classifiers, encompassing both human and AI decision-makers. It measures the divergence between the true label distribution and the predicted probability distribution over possible outcomes. It characterizes the expected number of additional equivalent decisions required to identify the true class, conditional on the classifier's probabilistic output. A lower cross-entropy loss indicates a closer alignment between predicted beliefs and actual outcomes, signifying a more accurate classifier. Specifically, denote $X$ as an input image, and $p(L=l|X)$ as true class probabilities conditional on the image. In addition, denote $q(L = l | X)$ as the predicted probability of the AI system for each class.  Formally, the cross-entropy loss of a classifier,$\tilde{H}$, is defined as
\begin{equation}
    \tilde{H} = -\, \mathbb{E}_{X} \bigg[\sum_{l \in \mathcal{L}} p(L=l | X) \ln q(L=l|X)\bigg].
    \label{eq:CE_AI}
\end{equation}

Cross-entropy loss $\tilde{H}$ and accuracy $a$ are both commonly used metrics for evaluating classifier performance. When the task complexity dimensions, $m$ and $n$ are held fixed,  $\tilde{H}$ and $a$ approximately follow a monotonic relationship:
\begin{equation}
    \tilde{H} = F(a;m,n)
\label{eq:entropy_accuracy} 
\end{equation}
In several parts of our analysis, we employ this function to transform the accuracy levels into the corresponding cross-entropy loss values. A detailed description of the estimation of this mapping function is provided in Appendix \ref{app:scaling_entropy_acc}. As expected, higher accuracy tends to correspond to lower entropy levels, reflecting improved certainty in classification.

When a computer vision classification task is performed by a human, workers rarely execute classification tasks with complete precision; a certain margin of error is typically accepted as part of routine performance. If we take the typical accuracy achieved by humans on a given classification task as the required accuracy level for any technology performing that task, and denote it by $a_{req}$, then based on Equation~\eqref{eq:entropy_accuracy}, this yields an associated cross-entropy loss of the completed task, denoted by $\tilde{H}_{req}=F(a_{req};m,n)$.

The cross-entropy loss of the AI system, $\tilde{H}_{AI}$, measures how many more probabilistically equivalent decisions to make given the output of AI systems. It is governed by the following inequality:

\begin{equation}
    \tilde{H}_{req}\leq \tilde{H}_{AI} \leq\tilde{H}_{task}.
\end{equation}

\subsubsection{Quantifying AI-Human Collaboration}
\begin{figure}[]
    \centering
    \includegraphics[width=0.6\linewidth]{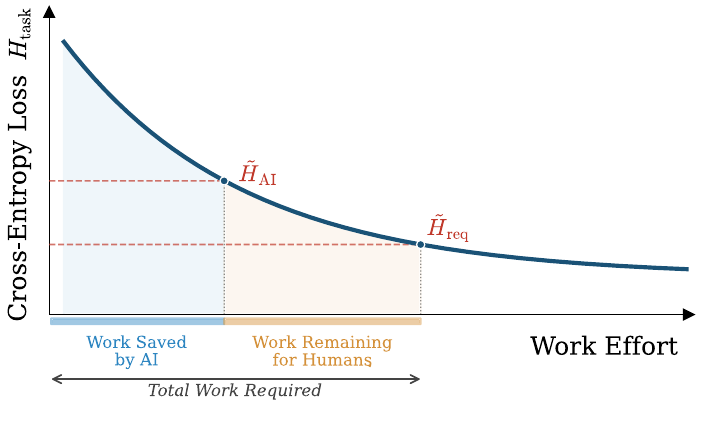}
    \caption{Quantifying Human-AI Work Allocation via Entropy and Cross-Entropy Loss}
    \label{fig:human-ai}
\end{figure}

As established above, completing a vision classification task involves bringing the cross-entropy loss to $\tilde{H}_{req}$. With the introduction of an AI tool, the level of complexity that must be handled by the human worker is reduced. This implies that the AI system effectively completes a portion of the task corresponding to the reduction to  $\tilde{H}_{AI}$. The remaining complexity - from $\tilde{H}_{AI}$ to $\tilde{H}_{req}$ - is resolved by human effort. Figure \ref{fig:human-ai} illustrates this division of the amount of work between the AI system and the human worker.

In this framework, we further assume that human work time is proportional to the number of equivalent decisions that must be made by the human worker. Without AI assistance, human effort corresponds to the reduction in cross-entropy loss from the random guess accuracy to the required accuracy level:
\begin{equation}
\text{Human Work Time} \propto \tilde{H}_{rand} - \tilde{H}_{req} = H_{task} - \tilde{H}_{req}.
\end{equation}

With AI assistance, the human's contribution is reduced, corresponding only to the remaining gap between the AI system's performance and the required accuracy:
\begin{equation}
\text{Human Work Time} \propto \tilde{H}_{AI} - \tilde{H}_{req}.
\end{equation}

As can be seen, our modeling approach implicitly assumes that AI and human labor are substitutes in the production of task-level output. Given the difficulty of directly measuring output at the task level, we restrict our analysis for the purpose of this paper to the benchmark case of perfect substitutability between AI and human inputs.

\subsubsection{Empirical Evidence on the Relationship Between Entropy and Human Work Time}

There is abundant empirical evidence supporting the above assumption that the amount of human labor needed to perform a classification task should be linearly proportional to the number of decisions required, and hence proportional to the reduction in cross-entropy loss/entropy. In particular, \citet{hick1952rate} and \citet{hyman1953stimulus} empirically derived the well-known Hick-Hyman law, which states that reaction time in a choice task increases linearly with task entropy. 

\begin{figure}[]
    \centering
    \includegraphics[width=\linewidth]{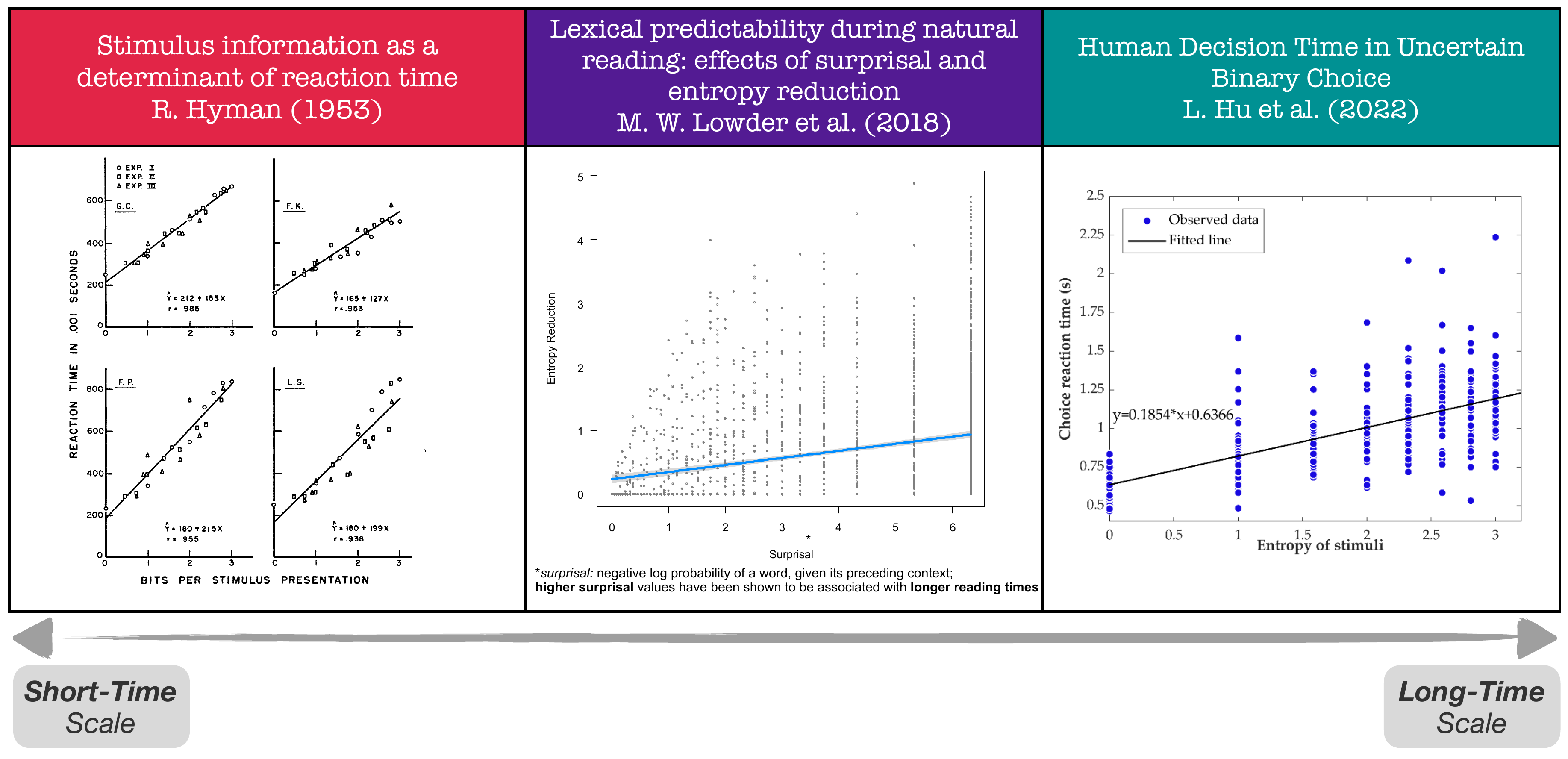}
    \caption{Highlighted empirical evidence linking task entropy to processing time across different time scales.}
    \label{fig:entropy_time_lit}    
\end{figure}

Furthermore, additional studies extend these findings across different domains and time scales. For example, \citet{lowder2018lexical} demonstrate a positive relationship between surprisal and entropy. Since surprisal is defined as the negative log probability of a word given its preceding context, higher surprisal values are associated with longer reading times. Also, \citet{hu2022human} show that reaction times in binary decision tasks escalate with increasing uncertainty. Together, these works and other similar works robustly support the notion that processing time is directly linked to the entropy of the task. Figure~\ref{fig:entropy_time_lit} highlights this line of literature, illustrating how the impact of entropy on processing time spans from rapid perceptual decisions to more extended cognitive tasks.

\subsubsection{Calculating the Labor Substitution Ratio}
\label{sec:auto_rate_collab}

\noindent
$r_i(a_i)$ is the proportion of labor time within the AI-automatable subtask that the AI system could save given the accuracy $a_i$ of the AI system. $r_i(a_i)$ serves as a core equation characterizing the demand for AI quality. We distinguish between two scenarios for the adoption of AI:

\begin{tcolorbox}[colback=gray!5!white, colframe=gray!60!black, title=\textbf{Scenario 1: Full Automation}]
If the AI system is capable of independently meeting the task requirement, i.e., $\tilde{H}_{\text{req}} = \tilde{H}_{\text{AI}}$, then the automation rate is $r_i(a_i) = 1$. In this case, the task is fully executed by the AI, with no human involvement. We refer to this as \emph{full automation}.
\end{tcolorbox}

\begin{tcolorbox}[colback=gray!5!white, colframe=gray!60!black, title=\textbf{Scenario 2: Partial Automation}]
When the required cross-entropy loss is lower than what the AI system can achieve alone, i.e., $\tilde{H}_{\text{req}} < \tilde{H}_{\text{AI}}$, human-AI collaboration is necessary. In this case, the AI is first deployed to reduce the cross-entropy loss from $H_{\text{task}}$ to $\tilde{H}_{\text{AI}}$. Subsequently, human workers contribute to further reduce it to the required level $\tilde{H}_{\text{req}}$. We term this arrangement \emph{partial automation}, and define the corresponding labor substitution ratio for a given task $i$ as:

\[
r_i(a_i) = \frac{H_{\text{task}, i} - \tilde{H}_{\text{AI}}}{H_{\text{task}, i} - \tilde{H}_{\text{req}, i}} = \frac{H_{\text{task}, i} - F(a_i; m_i, n_i)}{H_{\text{task}, i} - \tilde{H}_{\text{req}, i}}.
\]
\end{tcolorbox}

These two cases can be unified under the following expression:

\begin{equation}
    r_i(a_i)=\min\bigg(\frac{H_{task, i}-\tilde{H}_{AI}}{H_{task, i}-\tilde{H}_{req, i}},1\bigg) =\min\bigg(\frac{H_{task, i}-F(a_i; m_i,n_i)}{H_{task, i}-\tilde{H}_{req, i}},1\bigg).
    \label{eq:r}
\end{equation}
It can be easily verified that when the accuracy of the AI model meets the required level, (i.e. $a_i = a_{req,i}$, and thus, $\tilde{H}_{AI,i} = \tilde{H}_{req,i}$), $r_i(a_i) = 1$, which corresponds to the case of full automation. Otherwise, if the performance of the AI system falls short of the required threshold, (i.e., $a_i < a_{req,i}$, and  $\tilde{H}_{AI,i} > \tilde{H}_{req,i}$), $r_i(a_i) < 1$, which corresponds to the case of partial automation.

With $r_i(a_i)$, a measure of demand, and $\kappa_{i}(a_{i}, Y_i)$, a measure of supply, we can solve the firm's cost minimization problem in Equation~\eqref{eq:cost_min}. Recall that the firm's problem is a two-stage decision. In the first stage, the firm chooses to adopt an AI system. If the firm adopts AI, in the second stage, it selects the quality level $a_i$ of the AI system. In Appendix~\ref{app:firm_backward_induction}, we solve this problem using backward induction.

\subsection{Automating Work Through AI Agents}
\label{sec:equilibrium}
In the preceding analysis, we have assumed that each firm operates as an independent decision-making entity, choosing whether to adopt AI for each computer vision task on a firm-specific basis. Under this assumption, any AI system developed is exclusively deployed within the firm that produced it and cannot be shared across organizational boundaries.

However, in many occupations, the same task may be performed in a highly similar manner across different firms. When the production of task-level output is sufficiently standardized and tradable across firm boundaries, a decentralized, firm-by-firm approach to AI deployment may lead to inefficiencies due to the duplication of fixed development costs. In such cases, there is scope for the emergence of intermediaries that supply AI agents or AI-as-a-service, developing and offering AI systems for a given task within an industry to multiple downstream firms. 

Such third-party providers already exist in the LLM space, where firms such as OpenAI, Google, and Anthropic provide AI model services for a known price per use. However, at the industry, subsector, and sector level, domain-specific AI agents and AI-as-a-service providers are generally not yet available. Whether legacy firms or startups innovate such offers, the economics are similar. The adoption principle remains unchanged: buyers will undertake automation whenever expected total benefits exceed total costs. 

What differs under AI agents or AI-as-a-Service is the level at which these quantities are evaluated. At the firm level, the benefit of automation corresponds to the labor compensation saved within an individual organization. Under AI agents or AI-as-a-Service, the same logic applies, but both benefits and costs are assessed jointly across all participating firms. Fixed development and training costs that would otherwise be incurred repeatedly at the firm level are shared instead, while inference costs continue to scale with total usage. Because the benefit pool now reflects the combined labor compensation of all users, the potential gains from automation expand substantially. Conceptually, the optimization problem determining the optimal level of automation remains conceptually identical to the firm-level case; only the scale of the cost--benefit components change, now aggregated over the group of firms rather than defined for a single firm. In Section \ref{sec:Labor Compensation Saving}, we formalize this aggregation by applying the same task-level primitives from the firm analysis while replacing firm-specific quantities with their industry-level counterparts, allowing a direct comparison between firm-level automation and AI agent or AI-as-a-Service within a unified analytical framework.

\section{Scaling Law Experiments: Estimating the Production Function of AI Quality}
\label{sec:data}
In Section ~\ref{sec:theory}, we outlined our approach to analyzing the quality of each computer vision task, with the aim of identifying the optimal level of quality provision. Central to this process is the necessity of defining the production function for computer vision quality. For the majority of industrial applications, we posit that an effective strategy involves leveraging freely available, pre-trained computer vision models and fine-tuning the models for task-specific applications. In other words, the production of computer vision quality, as discussed in this paper, refers to the process of fine-tuning a pre-trained computer vision foundation model.

The AI scaling law literature examines empirical relationships that demonstrate how certain performance metrics of artificial intelligence models, such as model accuracy or loss, improves as a function of increased computational resources, training data, and the number of parameters in AI models. \citet{hestness2017deeplearningscalingpredictable} and \citet{rosenfeld2021scaling}, enable prediction of improvements in model error, as data and compute are scaled. Subsequent research, exemplified by \citet{kaplan2020scaling} and \citet{henighan2020generative}, elucidates optimal input arrangements under constraints on total compute.

However, the direct application of findings from existing research is not feasible for three primary reasons. First, the majority of scaling law studies have concentrated on AI models in the context of large language models, while research specific to scaling laws in computer vision is relatively limited and does not fully meet the requirements of this study. Second, we have identified three critical factors that simultaneously affect both model performance and cost: data, model size, and training steps. For a scaling law to be appropriately utilized as an AI production function, it must account for all three of these factors. However, to our knowledge, the vast majority of existing scaling law studies incorporate only two of these factors. Most commonly, these studies examine the optimal combination of data and model size while holding compute resources constant. Third, existing scaling law literature has largely overlooked the heterogeneity in task difficulty, i.e., the number of classes.

The prevailing computer vision scaling law studies often treat the number of classes as a fixed parameter, commonly using the 1,000 classes found in the ImageNet dataset. However, based on a detailed manual analysis conducted by our AI experts across more than 461 selected computer vision tasks, the number of classes encountered in practical economic activities ranges from as few as two to several thousand, with the majority of tasks involving fewer than ten classes. This variation in the number of classes is markedly different from the assumptions made in existing scaling law studies.

To bridge this gap, we expand the canonical form of scaling law into the following form:

\begin{equation}
\ln(H_{AI}) = \ln\left(\frac{\alpha }{D^a} + \frac{\beta}{T^b} + \frac{\sigma }{M^{c}} +G \right) + k
\label{eq:scalinglaw}
\end{equation}

In the equation above, $H_{AI}$ denotes the cross-entropy loss of the AI model. The logarithmic function on the right side of Equation~\eqref{eq:scalinglaw} includes three fractions, corresponding to the three input factors: data ($D$), training steps ($T$), and model size ($M$). The remaining variables in the equation represent the parameters to be estimated in the scaling law analysis.

The functional form of this scaling law largely follows the classical multi-input scaling laws proposed in prior work by \citet{rosenfeld2021scaling}, \citet{hoffmann2022training}, and others. It clearly deviates from the Constant Elasticity of Substitution (CES) forms commonly used in economic modeling. Instead, it can be viewed as one of the simplest and most general non-CES specifications.

The CES production function is a neoclassical production function that displays a constant percentage change in the factor proportions due to a percentage change in the marginal rate of technical substitution. 
\begin{equation}
    Q = F \cdot \left(\alpha \cdot K^{\rho} + (1-\alpha) \cdot L^{\rho}\right)^{\nu/\rho}
\label{eq:CES production function}
\end{equation}

Where $Q$ = quantity of output, $F$ = total factor productivity, $K$,$L$ = quantities of capital and labor, $\alpha$ = share parameter, $\rho$ = substitution parameter $\rho=((\sigma-1))/\sigma$, and $\sigma$ = elasticity of substitution. Also, $\nu$ is the degree of homogeneity of the production function, when $\nu=1$ there are constant returns to scale, $\nu<1$ there are decreasing returns to scale, and $\nu>1$ there are increasing returns to scale. CES production technology has a constant percentage change in factor proportions (e.g., capital and labor) due to a percentage change in the marginal rate of technical substitution.  In the case of AI, and software more generally, the increase in output from additional data, training steps and/or model size requires little or no increase in labor.\footnote{The two-factor CES production function was introduced by \citet{solow1956econ} and expanded by \citet{arrow1961capital}. The scholarship surrounding the CES production function innovation was well before the economics of software and AI development and production existed.} Further, because the nature of AI production technology is a reduction in loss (increased accuracy and improved decision making), increased factor resources (data, model size and training steps) results in reduced output (cross-entropy loss) as opposed to the traditional increase in produced units. Consequently, a production function in the form of Equation~\eqref{eq:scalinglaw} is used in what follows.

The choice between using a CES versus a non-CES functional form reflects a fundamental trade-off. Adopting a CES form requires imposing the restrictive assumption that the ability to substitute one factor input for another remains constant over different production levels, which appears to contradict the empirical motivation and flexibility underlying the scaling law literature. Conversely, using a non-CES specification sacrifices access to the standard economic analysis tools built upon CES assumptions. Moreover, the non-CES functional form (Equation~\eqref{eq:scalinglaw}) implies that the inputs are complements. Since entropy consistently decreases as any input increases, the parameters $a$, $b$, and $c$ must take positive values to reflect this relationship. As a result, it can be shown that the inputs are pairwise complementary. Interestingly, this also provides a useful observation from an economic perspective: computer scientists appear to implicitly assume that the key inputs in AI model production are fundamentally complementary.

To incorporate the task complexity, $n$, into the scaling law, we assume a log-log dependency of the scaling law parameters on $n$. Thus, the scaling law is expanded as
\begin{equation}
\ln(\tilde{H}_{AI}) = \ln\left(\frac{e^{A_0 + A_1 \ln(n)}}{(\frac{D}{n})^{a_0 + a_1 \ln(n)}} + \frac{e^{B_0 + B_1 \ln(n)}}{T^{b_0 + b_1 \ln(n)}} + \frac{e^{C_0 + C_1 \ln(n)}}{M^{c_0 + c_1 \ln(n)}} +G \right) + K \ln(n)
\label{eq:parametric_scaling_law}
\end{equation}

A lower value of $\tilde{H}_{AI}$ corresponds to improved performance of the AI system, whereas increases in the values of the three input factors---data, training steps, and model size---are expected to enhance the AI system's performance. Consequently, the anticipated outcome is that, within a reasonable range of $n$ values (from 2 to several thousand), the exponents associated with each of the three inputs will be positive, ensuring that $\tilde{H}$ decreases as the levels of these input factors increase.

Equation~\eqref{eq:parametric_scaling_law} defines the parametric relationship between the cross-entropy loss $\tilde{H}_{AI}$ and the input factors, which we abbreviate as $\tilde{H}_{AI} = \tilde{H}_{AI}(D, T, M; n)$. Combining Equations~\eqref{eq:entropy_accuracy} and \eqref{eq:parametric_scaling_law}, the production function $Q(D_i,T_i,M_i; m_i,n_i)$ (Equation~\eqref{eq:accuracy_production}) is derived as:
\begin{equation}
    Q(D_i,T_i,M_i; m_i,n_i)=F^{-1}(\tilde{H}_{AI}(D_i,T_i,M_i; n_i); m_i,n_i)
    \label{eq:q_acc}
\end{equation}
where $F^{-1}(\tilde{H}; m,n)$ represents the inverse function of $F(a;m,n)$ with respect to $a$.

To calibrate this AI quality production function, we fine-tuned a Swin Transformer model under 80 different settings.\footnote{The Swin Transformer is a hierarchical vision transformer that processes images by dividing them into local windows and calculating self-attention within these windows. See \citet{Liu2021swin}.} The computer vision foundation model was pre-trained on 500 randomly selected classes from the ImageNet dataset.\footnote{The ImageNet database contains 1,000 classes. For this study, we randomly selected 500 classes for pre-training and used the remaining 500 classes to simulate applications of varying complexity, spanning from 2-class to 500-class classification tasks.} These 80 settings capture variation across four key dimensions: task complexity, data size, model size, and training steps.

Cross-entropy loss ($H$) measures how closely the model's predicted probability distribution aligns with the true labels. It penalizes cases where the model assigns low probability to the correct class, so lower values indicate better predictive performance. Prediction performance is evaluated as the out-of-sample cross-entropy loss when testing the AI model in an object classification task.

The experiment systematically varies four dimensions of the fine-tuning process.

First, task complexity ($n$) is varied across four levels, measured by the number of distinct outcome classes in the image classification task: 2, 10, 100, and 500 classes, spanning simple binary classification to highly complex multi-class recognition problems. 

Second, training data size ($D$) is varied across five levels, specified as 13, 65, 130, 650, and 1,300 samples per class, corresponding to total training dataset sizes that scale proportionally with the number of classes. For example, a task with 100 classes and 130 samples per class yields a total training set of 13,000 images. 

Third, model size ($M$) is varied across four configurations: 7.3 thousand, 0.4 million, 28.3 million, and 87.8 million parameters, capturing a wide range of model capacities from very small to large-scale architectures. 

Fourth, these three dimensions together yield 80 unique experimental configurations, each independently replicated 50 times ($T$), yielding a total of 4,000 observations. 

The scaling law function related to computer vision fine-tuning is estimated with nonlinear least squares. Following our parametric scaling law introduced in Equation~\eqref{eq:parametric_scaling_law}, to obtain robust estimates for its parameters, we implemented a multi-run optimization strategy. In our approach, we split the data into training ($80\%$) and test ($20\%$) sets, and we performed 20 independent runs with different random initializations for the parametric scaling law. Using SciPy's nonlinear least-squares optimizer (\texttt{curve\_fit}) with a maximum evaluation limit ($500,000$), we fitted the proposed scaling function to the training data by minimizing the residuals between the logarithm of the observed cross entropy loss and the model predictions. For each run, we computed the $R^2$ score on the training data to assess the goodness-of-fit, and the best-performing parameter set (highest training $R^2$) was selected and further evaluated on the test set. In addition to the best parameter estimates, we aggregated the results from all runs to calculate the mean and standard deviation for each parameter, providing a measure of the uncertainty in the estimation process. The initial values of the parameters are randomly drawn from a Gaussian distribution with mean $0$ and variance $0.1$. The resulting parameter estimates and their standard errors are reported in Table~\ref{tab:third_compute2_params}.

\begin{table}[htbp]
\centering
\footnotesize
\caption{Estimated Parameters for the Scaling Law}
\label{tab:third_compute2_params}
\begin{tabular}{lcc}
\toprule
\textbf{Parameter} & \textbf{Estimate} & \textbf{Std. Error} \\
\midrule
$A_0$  & -1.448         & 0.046 \\
$A_1$  & 0.752          & 0.050 \\
$a_0$  & -0.034          & 0.005 \\
$a_1$  & 0.077         & 0.003 \\
$B_0$  & 1.474          & 0.080 \\
$B_1$  & 1.049          & 0.000 \\
$b_0$  & 0.383         & 0.009 \\
$b_1$  & 0.020         & 0.005 \\
$C_0$  & 4.054          & 1.756 \\
$C_1$  & 0.308          & 0.409 \\
$c_0$  & 0.614         & 0.216 \\
$c_1$  & -0.041          & 0.045 \\
$G$    & -0.296         & 0.173 \\
$K$    & -0.150         & 0.044 \\
\midrule
$R^2_{\mathrm{test}}$  & 0.963 & \\
\bottomrule
\end{tabular}
\end{table}
\vspace{3mm}

Our objective is to formulate an empirical scaling law that accurately describes the regime relevant to our study, rather than a universal structural law over all possible combinations of task complexity and resources. Accordingly, Equation~\eqref{eq:parametric_scaling_law} should be interpreted as a fitted relation over the supported regime covered by our experiments and by common computer vision settings. In particular, the monotonicity of the fitted law with respect to task complexity is understood to be verified empirically within this domain---that is, over the range of class counts, data budgets, optimization budgets, and model sizes represented in our experiments and in practically relevant CV configurations. We therefore use the law as a descriptive and comparative model within this regime, and do not claim unrestricted monotonicity or guaranteed validity under arbitrary extrapolation far outside it.

\section{Results: Quantifying the Automation Frontier}
\label{sec:results}
This section describes our approach to quantifying the optimal AI automation decision in the real-world economy, with estimates of cross entropy loss, $\tilde{H}_{AI}$, and labor substitution, $r_i(a_i)$, grounded in the theoretical model developed in Section \ref{sec:theory}. To measure the required accuracy of each task, we conducted an extensive survey in late 2023. The two key variables obtained from the survey are: required accuracy and random-guess accuracy for each task, which we then feed into a prediction function~\eqref{eq:entropy_accuracy} that assesses the corresponding cross-entropy loss for the computer vision classification tasks.

\subsection{Occupation and Task Characteristics Survey}
\label{sec:results_Survey}
To evaluate whether a task is exposed to computer vision technology, it is essential to differentiate between vision tasks and non-vision tasks within the economy. Our analysis relies on data from the O*NET Database 27.1 \citet{onet_center_2023}, which provides standardized information about jobs and workers in the United States. The dataset includes descriptions for 1,016 occupations and a total of 18,796 unique tasks. Several recent studies, including \citet{webb2019impact}, \citet{eloundou2024gpts}, and \citet{brynjolfsson2018can}, have utilized the O*NET task framework to analyze the impact of AI. \citet{svanberg2024beyond} also works within this framework and identifies 420 tasks as vision related. In our analysis, we use the set of vision-related tasks identified by \citet{svanberg2024beyond}.

We collected answers from 3,778 respondents. Each of them chose one occupation that they have familiarity with within the 263 occupations that has suitable vision tasks. At least 5, and on average 10 data points are collected for each of the 461 computer vision tasks, and we use the average values for the variables.\footnote{Our analysis focused on 420 out of the 461 surveyed tasks, as certain occupations were excluded due to incomplete employment or wage data.} More details of the survey could be found in Appendix~\ref{app:survey}. The required accuracy variable comes from the question: ``What is a typical error rate for employees currently doing this task? (i.e. how likely are workers who perform this task to make mistakes?)'' The random accuracy variable comes from the question: ``What would be the error rate on this task for a worker that had to guess their answer without any information? (e.g. while blindfolded).'' 

The survey also examines task frequency and time allocation to capture the prevalence and importance of visual tasks within the respondents' occupations. Respondents indicate how often they perform the task, with frequencies ranging from rare (less than once per year) to extremely frequent (over 3,000 times a day). They also report the percentage of their overall work time dedicated to performing the task, offering insight into its significance in their day-to-day responsibilities. This information is critical for understanding the role of these tasks in professional settings and for assessing the feasibility of automating tasks that are both prevalent and time intensive. Appendix~\ref{app:scaling_entropy_acc} provides the equation for estimating cross-entropy loss as a function of required accuracy and number of classes.

\subsection{Quantifying Task Complexity and Visual Intensity with ChatGPT-4o}
\label{sec:Quantifying Task Complexity}
The complexity of applying computer vision systems varies significantly across O*NET job tasks, leading to heterogeneous costs in developing suitable AI models for automation. To systematically measure these costs, we characterize the complexity of the computer vision task associated with each O*NET task along two key dimensions: the number of computer vision tasks required (henceforth, number of tasks), i.e. $m_i$ as in Equation~\eqref{eq:accuracy_production}; and the number of possible outcome categories within each task (henceforth, number of classes), i.e. $n_i$ as in Equation~\eqref{eq:accuracy_production}.

The number of tasks captures the extent to which an O*NET task decomposes into multiple distinct computer vision classification sub-tasks. A higher value indicates that completing the O*NET task necessitates a system composed of multiple specialized models. The number of classes, in turn, reflects the granularity of classification required within each sub-task, i.e., the number of distinct categories among which the AI system must differentiate. For instance, the O*NET task ``inspect motor vehicles'' for light truck drivers comprises sub-tasks, such as determining ``whether the gas system is operational,'' ``whether the oil level is adequate,'' ``whether the washer fluid level is sufficient,'' among others---amounting to approximately 20 distinct sub-tasks. In this case, the number of tasks is 20. Each sub-task requires a binary classification---e.g., ``working'' vs. ``not working''---implying a number of classes of two per sub-task.

By contrast, a computer vision system assisting zoologists and wildlife biologists in analyzing animal characteristics to classify species may involve a single overarching classification task (number of tasks = 1) yet requires distinguishing among 500 species (number of classes = 500).

A complementary dimension in characterizing O*NET tasks is the vision proportion $- \delta_i$ in Equation ~\eqref{eq:ai_cost_coef} -- that is, the fraction of the task that is inherently reliant on visual information rather than other sensory modalities or cognitive processes. Many computer-vision-exposed O*NET tasks involve additional non-visual components. For example, in the light truck driver case, workers may need to listen to engine sounds to detect mechanical issues or manually assess tire pressure through tactile feedback. To avoid overestimating the potential automation of such tasks via computer vision systems, we scale task-level estimates by the proportion of time that workers allocate specifically to vision-dependent activities.

To quantify these measures across over 420 O*NET tasks with computer vision exposure, we employed ChatGPT-4o to generate initial estimates for number of tasks, number of classes, and vision proportion. These outputs were subsequently validated by a team of five researchers through a structured manual review process to ensure alignment with domain knowledge and practical feasibility. The prompts used to generate these estimates are provided in Appendix~\ref{app:prompt}.

In addition to these original data sources, we incorporate several external datasets to support the simulation implementation of the model. For task and occupation definitions, we use the Occupational Information Network (O*NET), a comprehensive database maintained by the U.S. Department of Labor that provides standardized descriptors of job tasks across occupations. From O*NET, we also extract the share of time that workers in each occupation allocate to specific tasks and the distribution of occupations in different industries/sectors. Employment and wage data come from the 2024 Occupational Employment and Wage Statistics (OEWS) published by the U.S. Bureau of Labor Statistics (BLS).\footnote{To convert wages into total employer-side labor costs, we apply the BLS-reported wage-to-compensation ratio for civilian workers, which is 1.4498.} Since scale effects play a critical role in firms' automation decisions, it is essential to obtain estimates of the firm size distribution in the economy. We use data from the 2022 Statistics of U.S. Businesses (SUSB), which reports the number of establishments and firm size distributions by six-digit NAICS industry codes. Regarding the engineers and domain experts required for AI training and implementation, as well as their wage levels, the costs for training data and computational resources are based on the findings of \citet{svanberg2024beyond}.

\subsection{Computer Vision Tasks Identification Based on O*NET}
While it is difficult to directly measure task-level output $Y_i$, we estimated $\tau_i Y_i$---the product of labor time allocated to the task and output---using the following strategy, which is sufficient for the purposes of our analysis:

\begin{equation}
    w_o \tau_i Y_i = w_{oi} \rho_{oi} \Rightarrow \tau_i Y_i = l_o \rho_{oi}
\end{equation}

where $w_o$ is the annual wage for occupation $o$, which can be obtained from Occupational Employment and Wage Statistics (OEWS). $l_o$ represents the total number of employees in each business entity (firms or industry subgroups). The most granular employment data we obtain come from the 4-digit NAICS level in the 2024 Occupational Employment and Wage Statistics (OEWS). While firm-level employment data are available from the 2022 Statistics of U.S. Businesses (SUSB), they require adjustments to be usable in our analysis. To address this, we follow the imputation strategy outlined in \citet{svanberg2024beyond}, as detailed in Appendix~\ref{app:firm_size}. $\rho_{oi}$ represents the time proportion that a human employee with occupation $o$ would spend on task $i$. We calculated $\rho_{oi}$ by weighting each task according to its score on the O*NET task importance scale, following the methodology of \citet{brynjolfsson2018can} and \citet{webb2019impact}.

Recall that improved performance of AI systems corresponds to reduced cross entropy loss which in turn is a function of available data, model size, and training steps along with the number of classes per subtask (Equation~\eqref{eq:scalinglaw}). Consequently, the optimization problem is to minimize cost subject to the constraints of required accuracy and cost for required GPU time to produce tasks incurred during deployment (See Equation~\eqref{eq:define_cost_minimization}).

The solution seeks the optimal combination of data, model size, and training steps relative to the cost of labor to perform the same task or subtasks. We discuss the results of this experiment, optimal combination with respect to the reported required task accuracy, over the next subsections. Labor substitutability must consider not only the required accuracy but also the task or subtask complexity.

\subsection{Scaling Law}
\label{sec:results_scalinglaw}
With the parameter estimates in Table~\ref{tab:third_compute2_params}, the implementation of Equation~\eqref{eq:parametric_scaling_law} will vary according to the number of classes (n). The estimation of scaling law described in Equation~\eqref{eq:scalinglaw}, is summarized in Table~\ref{tab:scalinglaw_params}. The table shows the relationship between task complexity and cross-entropy loss, which in turn is a determinant of accuracy. As the number of classes increase, task complexity increases. As each of the classes increase, $\alpha$, $\beta$, and $\sigma$ increase as well, contributing to increased cross-entropy loss. With greater complexity, the models have more difficulty in executing tasks with the required accuracy. In addition, the denominators of the first two terms of Equation~\eqref{eq:scalinglaw} have coefficient values for a and b that increase as the number of classes increase, finding that as the volume of data increase and the number of training steps increase, the cross-entropy loss decreases. With greater complexity, the models reduce loss and improve accuracy with more data and training. Conversely, as the number of classes increase, the coefficient c decreases, finding that as model sizes increase, the cross-entropy loss increases -- suggesting there are diminishing marginal returns to model size.

Our estimated scaling-law coefficients are broadly consistent in magnitude with prior work on data and model scaling. \citet{rosenfeld2019constructive} report fitted power-law exponents on data size and model size in the range $\alpha\approx 0.40$--$1.10$ and $\beta\approx 0.51$--$1.16$ across vision datasets, suggesting our estimates are within the range documented in established work. The positive returns to data and compute are directionally consistent with \citet{hoffmann2022training}, and the diminishing returns to model size as task complexity grows are conceptually aligned with \citet{sharma2022scaling}, who show theoretically that the parameter-scaling exponent decreases as the intrinsic dimension of the data manifold increases, though their notion of complexity is intrinsic dimension rather than class count. We emphasize that our scaling law is empirical and calibrated to our computer vision setting; it is intended to describe behavior within the observed data regime rather than to claim a universal law.

\renewcommand{\arraystretch}{1.5}
\begin{table}[h]
    \centering
    \footnotesize
    \caption{Scaling Law Parameters under Varying Number of Classes (n) }
    \begin{tabular}{ccccccccc}
        \toprule
        \textbf{Number of Classes} & \textbf{$\alpha$} & \textbf{a} & \textbf{$\beta$} & \textbf{b} & \textbf{$\sigma$} & \textbf{c} & \textbf{G} & \textbf{k} \\
        \midrule
        2    & 0.40  & 0.02 & 9.03    & 0.40 & 71.35  & 0.59  & -0.30 & -0.10 \\
        5    & 0.79  & 0.09 & 23.62   & 0.42 & 94.61  & 0.55  & -0.30 & -0.24 \\
        10   & 1.33  & 0.14 & 48.87   & 0.43 & 117.13 & 0.52  & -0.30 & -0.35 \\
        50   & 4.46  & 0.27 & 264.34  & 0.46 & 192.26 & 0.46  & -0.30 & -0.59 \\
        100  & 7.52  & 0.32 & 546.91  & 0.48 & 238.01 & 0.43  & -0.30 & -0.69 \\
        500  & 25.24 & 0.44 & 2958.46 & 0.51 & 390.69 & 0.36  & -0.30 & -0.93 \\
        1000 & 42.52 & 0.50 & 6120.92 & 0.52 & 483.64 & 0.33  & -0.30 & -1.04 \\
        \bottomrule
    \end{tabular}
    \label{tab:scalinglaw_params}
\end{table}

\subsubsection{Performance Elasticity}
Combining Equation~\eqref{eq:r} and Equation~\eqref{eq:scalinglaw}, we can obtain the labor substitution ratio as a function of the fine-tuning inputs:

\begin{equation}
r_i(D_i,T_i,M_i)
=
\min\left(
\frac{
H_{\text{task},i}
-
e^{k}\!\left(
\frac{\alpha}{D_i^{a}}
+
\frac{\beta}{T_i^{b}}
+
\frac{\sigma}{M_i^{c}}
+
G
\right)
}
{
H_{\text{task},i}
-
\tilde{H}_{\text{req},i}
},
\,1\right).
\label{eq:performance_elasticity_substitution_ratio}
\end{equation}

In the context of this paper, the extent to which labor can be substituted depends jointly on task complexity, the required accuracy, and the achieved accuracy of the model. For convenience, we define the \textit{performance elasticity} of input $X$ as

\begin{equation}
\epsilon_X = \frac{\partial r}{\partial X} \cdot \frac{X}{r},
\label{eq:performance_elasticity}
\end{equation}
capturing the percentage change in performance r (defined as the labor substitution ratio) in response to a 1\% change in input factor X. We obtain the formulas for performance elasticities for data, training steps, and model size, which are reported in Appendix \ref{sec:substitution_elasticity}. However, it is important to keep in mind that this elasticity is itself influenced by the intrinsic difficulty of the task.

To facilitate a more intuitive analysis, we examine four representative cases based on the parameter values reported in Table~\ref{tab:scalinglaw_params}. Specifically, we consider two levels of task complexity, proxied by the number of classes equal to $2$ and $500$, respectively, with corresponding task entropy set to $\ln 2$ and $\ln 500$ under the assumption of a uniform distribution (i.e., entropy equals $\ln(n)$). For each level, we evaluate the performance elasticity under two input bundles: a small-scale configuration with $D = 25{,}000$, $T = 200{,}000$, and $M = 250{,}000$; and a medium-scale configuration with $D = 100{,}000$, $T = 1{,}000{,}000$, and $M = 5{,}000{,}000$. \footnote{In large-scale computer vision projects, the input bundle can comprise up to a billion images, with training often requiring tens of millions of training steps and models containing hundreds of millions of parameters. However, we deliberately abstract away from such scenarios. The primary reason is that our analysis focuses on fine-tuning a model to perform a specific task, typically a classification task. In contrast, large-scale foundation models are often designed to be general-purpose, capable of handling a wide range of tasks---including generative ones---rather than being tailored to a single narrowly defined use case. In practical applications, especially in industrial settings, the vast majority of tasks are relatively simple binary classification problems---for example, detecting the presence or absence of a defect. As such, the scale of models considered in our framework is aligned with the practical requirements of these tasks. Notably, in our setting, excessively large models would trivially achieve full automation (the corner solution) at reasonable task complexity levels. In these corner solutions, the performance elasticity becomes irrelevant.} Table \ref{tab:performance_elasticity_analysis} reports the results.

\begin{table}[h]
\centering
\caption{Performance Elasticity}
\label{tab:performance_elasticity}
\footnotesize
\begin{tabular}{l>{\bfseries}c*{3}{p{2.2cm}}c}
\toprule
\textbf{Scenario} & \textbf{Performance} & \multicolumn{1}{c}{\textbf{Data}} & \multicolumn{1}{c}{\textbf{Training Steps}} & \multicolumn{1}{c}{\textbf{Model Size}} & \textbf{Total} \\
 & \textbf{($r$)} & \multicolumn{1}{c}{\textbf{($\epsilon_D$)}} & \multicolumn{1}{c}{\textbf{($\epsilon_T$)}} & \multicolumn{1}{c}{\textbf{($\epsilon_M$)}} & \textbf{Elasticity} \\
\midrule
\multicolumn{6}{c}{\textbf{2-Class Classification Task}} \\
\midrule
Small Scale (I)   & 0.804 & \multicolumn{1}{c}{0.010} & \multicolumn{1}{c}{0.046} & \multicolumn{1}{c}{0.046} & 0.102 \\
Medium Scale (II) & 0.911 & \multicolumn{1}{c}{0.009} & \multicolumn{1}{c}{0.021} & \multicolumn{1}{c}{0.007} & 0.037 \\
\midrule
\multicolumn{6}{c}{\textbf{500-Class Classification Task}} \\
\midrule
Small Scale (I)   & 0.351 & \multicolumn{1}{c}{0.023} & \multicolumn{1}{c}{0.544} & \multicolumn{1}{c}{0.283} & 0.849 \\
Medium Scale (II) & 0.751 & \multicolumn{1}{c}{0.006} & \multicolumn{1}{c}{0.112} & \multicolumn{1}{c}{0.045} & 0.163 \\
\bottomrule
\end{tabular}

\begin{tablenotes}
\footnotesize
\item Notes: \(r\) denotes the labor-substitution ratio.  
Scenario I (Small Scale): \(D = 25{,}000\), \(T = 200{,}000\), \(M = 250{,}000\).  
Scenario II (Medium Scale): \(D = 100{,}000\), \(T = 1{,}000{,}000\), \(M = 5{,}000{,}000\).  
\(D\) is data size, \(T\) training steps, \(M\) model size.  
Elasticities are evaluated at the respective input bundles.
\end{tablenotes}
\end{table}

Table ~\ref{tab:performance_elasticity} shows the labor substitution ratios and the performance elasticities -- the responsiveness of labor to data, training steps, and model size. When the performance elasticity approaches one (e.g. 0.80), increases in model elements (data, training steps, and model size) are nearly fully reflected in labor substitution increases. Conversely, when the performance elasticity approaches zero (e.g. 0.10), increases in model elements show limited response in labor substitution. However, total elasticities remain below one in every scenario, signaling decreasing returns to scale in AI fine-tuning.

First, task complexity clearly shapes both the labor-substitution ratio and the elasticity patterns. Compare the top panel of Table 4 (less complexity) with the bottom of Table 3 (more complexity). For the simple binary-classification problem (n = 2), a small-scale configuration delivers a substitution ratio of 80.4\%, while moving to medium scale lifts the substitution ratio to 91.1\%. By contrast, the 500-class task attains only 35.1\% under small-scale training yet rises sharply to 75.1\% at medium scale.

Second, elasticities systematically decline as training scale expands. In the 500-class task, training-steps elasticity falls from 0.544 to 0.112, and model-size elasticity drops from 0.283 to 0.045 when moving from small to medium scale. These reductions confirm diminishing marginal gains: each additional unit of input yields progressively smaller performance improvements at larger scales.

The ranking of elasticities varies with task complexity and resource constraints. Under small-scale conditions the 2-class task shows $\epsilon_M > \epsilon_T> \epsilon_D$, whereas the 500-class task reverses to $\epsilon_T > \epsilon_M> \epsilon_D$. Organizations should therefore adapt their investment strategy to both the complexity of the classification problem and their budget. Simple tasks benefit most from enlarging the model, while complex tasks gain more from extending training steps, at least until diminishing returns set in.

\subsubsection{Input Substitutability}
\label{sec:substitution_results}
\begin{figure}[]
    \centering
    \includegraphics[width=\linewidth]{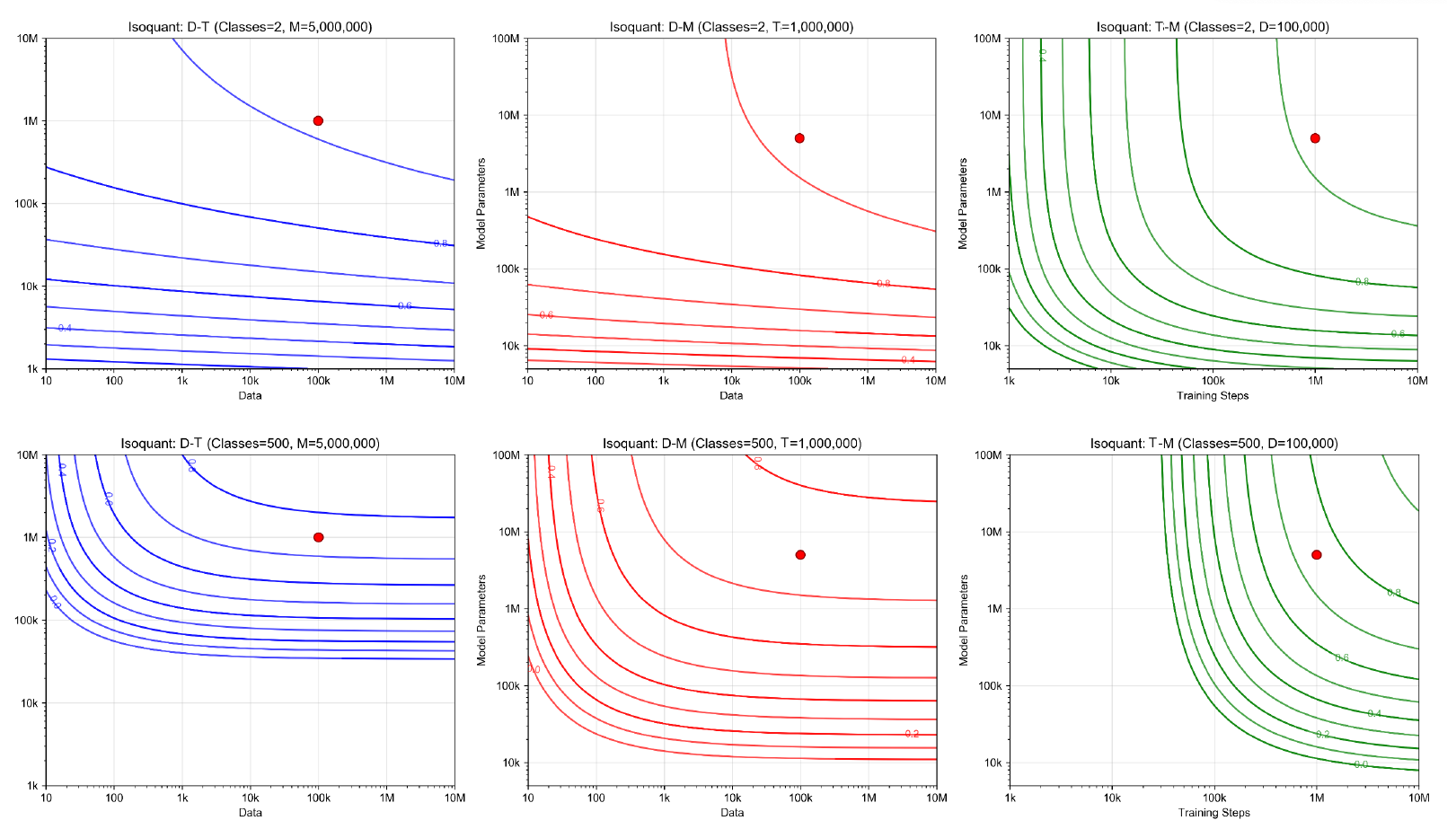}
    \caption{Pairwise Isoquant Contours Illustrating Substitutability among Data, Training Steps, and Model Size across Task Complexity (2-Class vs. 500-Class Classification Tasks)}
    \label{fig:pairwise_isoquants}
\end{figure}

This section quantifies how the three key inputs to model performance---training data ($D$), training steps ($T$), and model size ($M$)---substitute for one another as task complexity grows. Table~\ref{tab:elasticities} reports local elasticities of substitution, each evaluated at the same baseline input bundle \(\bigl(D,T,M\bigr)=\bigl(100{,}000,\;1{,}000{,}000,\;5{,}000{,}000\bigr)\).\footnote{
    We obtain the formulas for pairwise elasticities of substitution, which are reported in Appendix \ref{sec:substitution_elasticity}. Because the underlying production function is not CES, elasticities in principle vary with the evaluation point.  Sensitivity checks using alternative bundles change the levels only marginally, whereas task complexity drives pronounced variation; hence we omit additional bundles for brevity.} Figure~\ref{fig:pairwise_isoquants} visualises the corresponding isoquant contours for a simple (2-class) and a complex (500-class) task, each drawn with one input held at its baseline level while the other two vary; the elasticities reported in Table~\ref{tab:elasticities} are measured \emph{along} these constant-performance isoquants.

We begin with a few observations that align closely with basic intuition. At low input levels the contours are tightly packed, so a modest increase in data, steps, or parameters quickly boosts performance; farther from the origin the contours spread out, illustrating diminishing marginal product (second derivative of output with respect to the input). None of the panels reach the $r=1$ isoquant (perfect performance), which lies well outside the plotted range. The $r=0$ isoquant (performance no better than random guessing) sits noticeably higher in the 500-class task than in the 2-class task, confirming the intuition that more complex problems require a larger ``entry fee'' of resources before the model becomes useful at all.

\textbf{Near--Cobb--Douglas behaviour at low complexity. }
For the 2-class task, both \(\sigma_{DT}\) and \(\sigma_{DM}\) are essentially unity (\(0.982\)), placing the technology at the Cobb--Douglas production function boundary.  In practical terms, a 1\% reduction in training steps can be offset by roughly a 1\% increase in either data volume or model parameters.  The left and middle isoquants in the top row of Figure~\ref{fig:pairwise_isoquants} confirm this: contours are smooth and almost hyperbolic, indicating that the marginal rate of technical substitution changes proportionally with the input ratio.

\begin{table}[H]
    \centering
    \footnotesize
    \caption{Pairwise elasticities of substitution by task complexity}
    \renewcommand{\arraystretch}{1.3}
    \begin{tabular}{cccc}
        \toprule
        \multirow{2}{*}{Number of classes} & \multicolumn{3}{c}{Elasticity of substitution}\\
        \cmidrule(lr){2-4}
        & $\sigma_{DT}$ & $\sigma_{DM}$ & $\sigma_{TM}$ \\
        \midrule
          2     & 0.982 & 0.982 & 0.716 \\
          5     & 0.918 & 0.918 & 0.706 \\
         10     & 0.875 & 0.875 & 0.698 \\
         50     & 0.790 & 0.790 & 0.685 \\
        100     & 0.758 & 0.758 & 0.694 \\
        500     & 0.693 & 0.695 & 0.732 \\
       1\,000 & 0.668 & 0.682 & 0.749 \\
        \bottomrule
    \end{tabular}
    
    \vspace{0.35em}
    \raggedright\footnotesize
    \textit{Notes}:  Elasticities are evaluated at the baseline bundle \(\bigl(D,T,M\bigr)=\bigl(100{,}000,\;1{,}000{,}000,\;5{,}000{,}000\bigr)\).  
    \label{tab:elasticities}
\end{table}

\textbf{Rising complementarity with task complexity.}  
The functional form we estimate mechanically keeps each pairwise elasticity between 0 and 1, so some degree of complementarity is always present. Even within that band, however, clear patterns emerge. As the number of classes grows to 50, 100, and 500, \(\sigma_{DT}\) and \(\sigma_{DM}\) fall monotonically to about \(0.69\). In the isoquant panels this shows up as contours that bend ever more sharply toward the axes, meaning that adding just one type of input delivers rapidly diminishing returns unless the other input is expanded in tandem. Put differently, rising task complexity pushes the technology from ``almost substitutable'' toward ``strongly complementary,'' with the curves becoming progressively more convex toward the origin.

\textbf{Stable but low substitutability between steps and model size.}  
The elasticity between training steps and model parameters, \(\sigma_{TM}\), remains in a narrow band (\(0.68\)--\(0.75\)) across all complexity levels.  This reflects hardware and optimisation constraints---e.g.\ memory limits, gradient stability---that restrict the extent to which longer training can compensate for a smaller model (or vice versa).

\textbf{Implications for resource allocation.}  
For low-complexity tasks, practitioners enjoy a high degree of freedom in trading off data collection, compute time, and parameter count, enabling cost-efficient rapid prototyping.  For more complex tasks, the diminishing elasticities mandate a balanced expansion of all three resources: doubling data alone or scaling parameters alone is insufficient to maintain performance. 

\subsection{Labor Compensation Saving from Full and Partial Automation by Computer Vision}
\label{sec:Labor Compensation Saving}
This section addresses two central research questions: 
(1)~How much of total labor compensation can be fully or partially automated by computer vision (CV) systems? 
(2)~How do automation outcomes differ when AI deployment occurs at the firm level versus AI-as-a-Service? 
To evaluate these questions, we define an automation rate that quantifies the proportion of labor compensation optimally automated under our model. Specifically, we measure this automation rate from two complementary perspectives: first, by considering \emph{all tasks} within a given deployment scale (whether at the firm level or under an industry-wide AI-as-a-Service arrangement), and second, by \emph{restricting attention to computer vision tasks} within that same deployment scale to isolate the contribution of CV automation specifically.

Let $E$ denote the number of distinct entities that require separate AI systems under each deployment assumption. So far, we have considered the scope of deployment at the firm level, hence, different entities denote different firms. Now, under an industry-wide AI agent or AI-as-a-Service arrangement, we can broaden our deployment scale, namely, changing entities to industry groups (4-digit NAICS), subsectors (3-digit NAICS), sectors (2-digit NAICS), or the entire economy. As the deployment scale expands from individual firms to any of these deployment scales, $E$ decreases monotonically. We index entities by $e$, and solving the optimization problem for any given deployment scale yields the optimal AI automation decisions $(s_{ie}^*, a_{ie}^*)$ for task $i$ within entity $e$. We then define the normalized automation rate $R$ as the proportion of labor compensation that is optimally automated conditional on tasks being technically automatable by computer vision (i.e., tasks whose $\delta_i \neq 0$). That is, $R$ captures the optimal automation share within the subset of tasks feasible to be performed by computer vision. Formally,

\begin{equation}
    R = 
    \frac{\sum_{i,e} \delta_i\,w_{i}\tau_{ie}Y_{ie}\,\mathcal{I}[s_{ie}^*=1]}
         {\sum_{i,e} \delta_i\,w_{i}\tau_{ie}Y_{ie}}r_i(a_{ie}^*).
    \label{eq:agg_auto_rate_normalized}
\end{equation}

\noindent where $\mathcal{I}[\cdot]$ is an indicator function, which equals one if the statement within its brackets is true, and zero otherwise. The numerator represents the total labor compensation of CV-related tasks that are economically automated, that is, tasks with $s_{ie}^*$ and optimal substitution rate $r_i(a_{ie}^*)>0$. The denominator represents the total compensation of tasks that are technically automatable by CV ($\delta_i \neq 0$). Hence, $R$ measures the optimal automation rate within automatable activities. Tasks with $r_i(a_{ie}^*)=1$. are fully automated, meaning AI systems completely replace human performance, while tasks with $0<r_i(a_{ie}^*)<1$ are partially automated, indicating human--AI complementarity.

The aggregate automation rate in Eq.~(\ref{eq:agg_auto_rate_normalized}) can be decomposed into contributions from fully automated and partially automated tasks:
\begin{equation}
\label{eq:R_full_partial}
\begin{split}
    R &= R_{\text{full}} + R_{\text{partial}}, \\
    R_{\text{full}} &= 
    \dfrac{\sum_{i,e} \delta_i w_{i} \tau_{ie} Y_{ie} \mathcal{I}[s_{ie}^*=1, r_i(a_{ie}^*)=1]}
    {\sum_{i,e} \delta_i w_{i} \tau_{ie} Y_{ie}}r_i(a_{ie}^*), \\
    R_{\text{partial}} &= 
    \dfrac{\sum_{i,e} \delta_i w_{i} \tau_{ie} Y_{ie} \mathcal{I}[s_{ie}^*=1, 0<r_i(a_{ie}^*)<1]}
    {\sum_{i,e} \delta_i w_{i} \tau_{ie} Y_{ie}} r_i(a_{ie}^*).
\end{split}
\end{equation}
\noindent This decomposition directly aligns with the entropy-based definitions in Section \ref{sec:auto_rate_collab}, where $r_i(a_{ie}^*)=1$ corresponds to complete substitution (zero residual entropy, compared to required entropy) and $0<r_i(a_{ie}^*)<1$ corresponds to partial substitution (positive residual entropy, compared to required entropy). Equation~\eqref{eq:R_full_partial} also makes explicit that the aggregate automation rate $R$ is an aggregation of task-level automation intensities $r_i(a_{ie}^*)$ across firms or industries.

We distinguish between two modes of AI agent or AI-as-a-Service deployment. First, AI systems are shared within industry groups, for example, across firms operating within the same 2-, 3-, or 4-digit NAICS category the model is trained once and used by firms performing comparable visual tasks. Second, a unified AI agent or AI-as-a-Service platform serves across the entire economy, allowing a single model to be deployed wherever equivalent visual recognition tasks appear, regardless of industry. The within-industry case reflects realistic near-term diffusion, while the economy-wide case represents an upper bound on attainable efficiency gains from shared deployment.

\begin{figure}[h]
    \centering
    \includegraphics[width=0.6\textwidth]{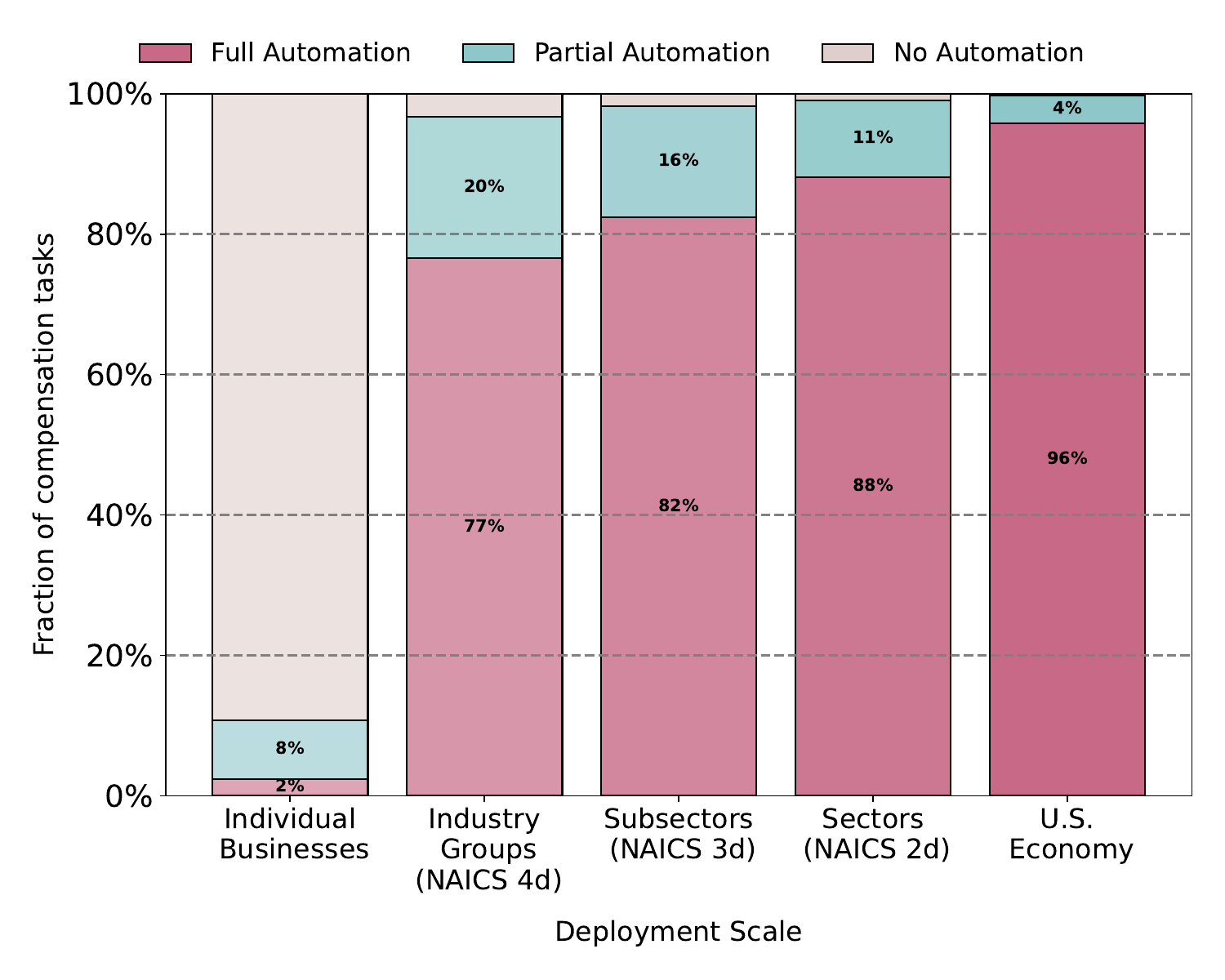}
    \caption{Composition of Full and Partial Automation of Vision-Task Labor Compensation Across Deployment Scales}
    \label{fig:fraction}
\end{figure}

In Figure~\ref{fig:fraction}, we vary assumptions about the extent to which a single AI system can be shared across users performing the same task. The leftmost bar represents the most restrictive case, where we assume that an AI system developed for a specific computer vision task is used exclusively by workers in a single occupation within a single firm. The second to fourth bars reflect progressively broader assumptions under which AI is provided as an agent or as a service, enabling the same system to be deployed by all workers in the same occupation performing the same task across progressively larger economic aggregates---namely, within a NAICS 4-digit, 3-digit, 2-digit industry, and eventually the entire U.S. economy.

The red segments of Figure~\ref{fig:fraction} represent the share of CV-automatable labor compensation for which full automation is optimal. At the firm deployment scale, only 2.4\% of labor compensation is fully automated. Under economy-wide deployment, however, this share rises to 95.8\%. The remaining automated portion shown in blue corresponds to tasks where partial automation and human--AI complementarity are optimal. At the firm level, total automation accounts for 10.8\% of labor compensation: 8.4\% through partial automation and 2.4\% through full automation. Thus, 89.2\% of technically feasible CV tasks remain economically unattractive to automate at firm scale, as fixed development costs outweigh potential labor savings when systems cannot be shared across firms. As deployment scope expands, automation potential increases sharply: 96.7\% at the 4-digit NAICS level; 98.3\% at the 3-digit level; 99.1\% at the 2-digit level; and 99.6\% under economy-wide deployment. This monotonic increase reflects a fundamental economic mechanism: broader deployment amortizes AI development costs across larger user bases, making tasks that are unprofitable to automate in isolation increasingly attractive at scale.

To place these results in the context of the overall economy, we compute an unnormalized automation rate $R'$, defined relative to total labor compensation across all activities, including those not feasible for CV automation:
\begin{equation}
\label{eq:agg_auto_rate}
    R' = 
    \frac{\sum_{i,k}
\delta_i\,w_{i}\tau_{ie}Y_{ie}\,\mathcal{I}[s_{ie}^*=1]}
{\sum_{i,k} w_{i}Y_{ie}}r_i(a_{ie}^*).
\end{equation}
$R$ and $R'$ provide different perspectives on AI automation. While $R$ describes the economic attractiveness within technically feasible computer vision tasks only, $R'$ describes the economic attractiveness to AI automation considering all tasks in the economy.

\vspace{2mm}

\begin{figure}[]
    \centering
    \includegraphics[width=0.6\textwidth]{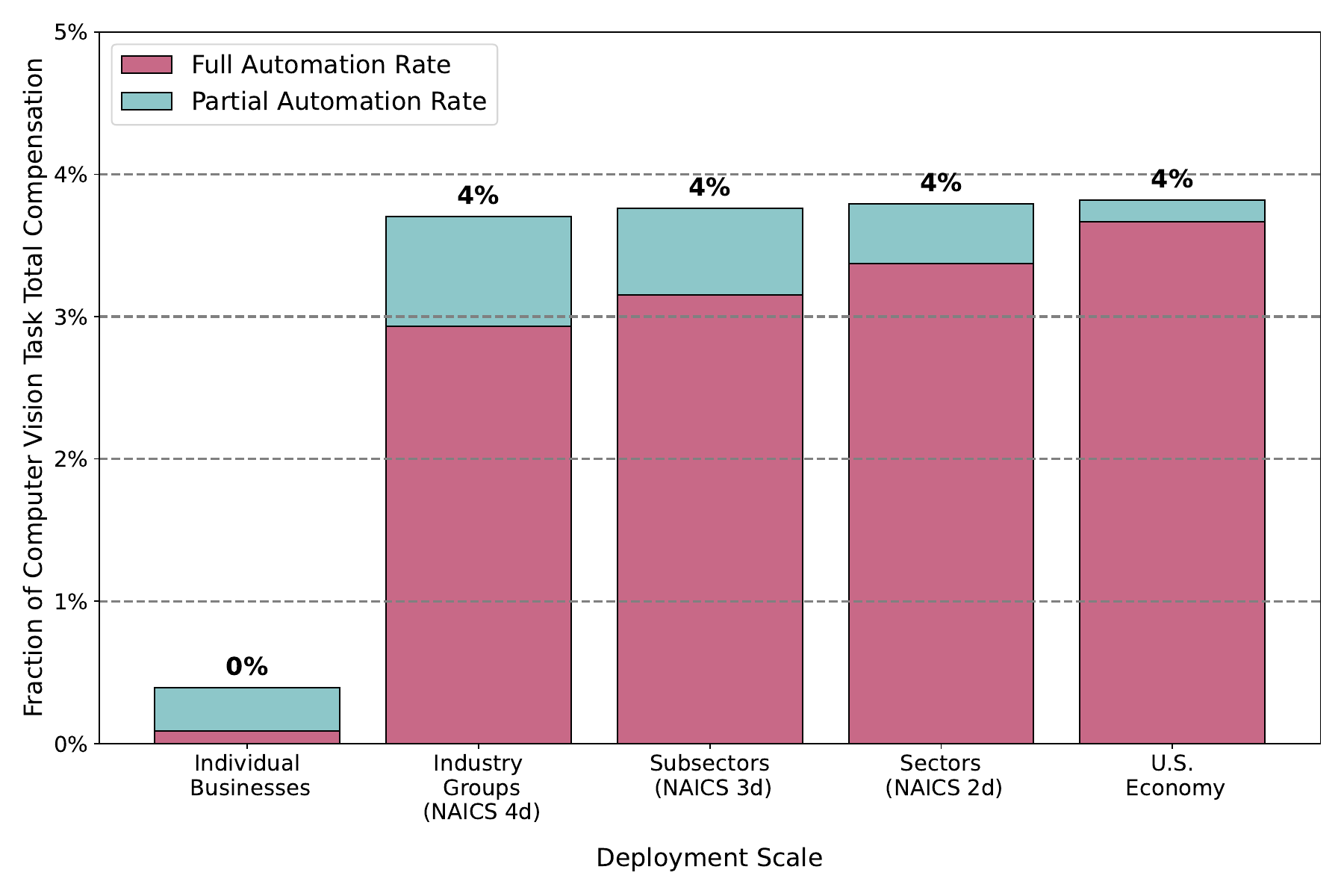}
    \caption{Fraction of Computer Vision Task Compensation Economically Attractive to Automate.} %
    \label{fig:fraction_all}
\end{figure}

While Figure~\ref{fig:fraction} considers the percentage of the automation fraction among all AI-exposed vision tasks, Figure~\ref{fig:fraction_all} presents its proportion across all economic activities. As shown in Figure~\ref{fig:fraction}, the percentage of the automation fraction among all AI-exposed vision tasks as automation within automatable CV activities approaches completeness at higher aggregation levels, reflecting near-total substitution potential when shared AI systems are deployed widely. 

In contrast, Figure~\ref{fig:fraction_all} indicates that when considered by all economic activities, the overall automation rate remains below 4\% of total labor compensation, underscoring that only a limited share of the U.S. economy is currently automatable by existing CV technologies. Taken together, these results demonstrate that broader aggregation magnifies attainable benefits and increases automation potential, yet the overall share of total labor compensation affected remains small when evaluated across the entire economy. This dual perspective, spanning firm-level and AI agent or AI-as-a-Service deployment, as well as CV- specific and economy-wide scopes clarifies both the magnitude and the limits of current automation potential.

\subsection{Characteristics of Computer Vision System Adoption}
\begin{figure}[b]
    \centering
    \includegraphics[width=0.6\textwidth]{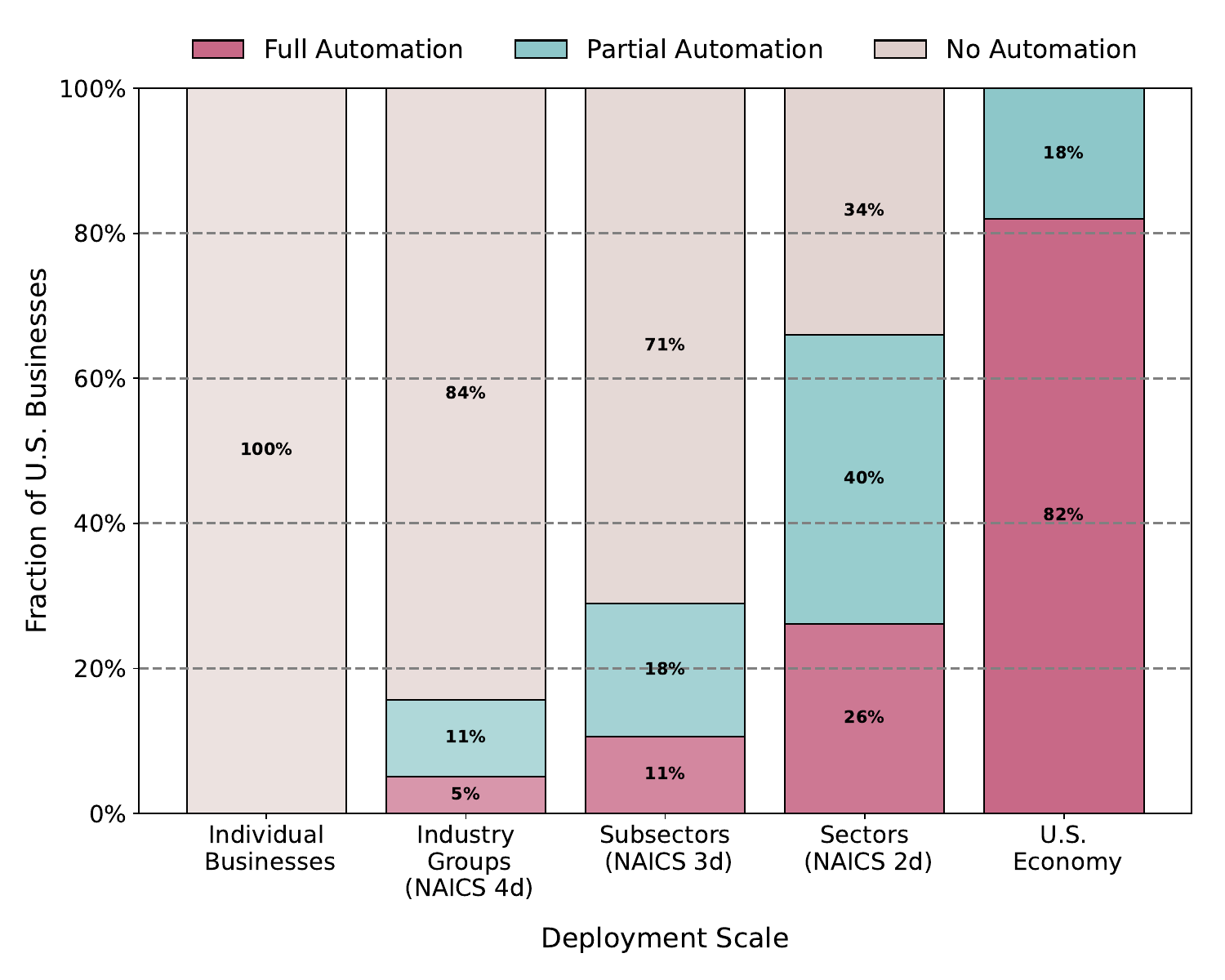}
    \caption{Fraction of the U.S. businesses (by business counts) over different deployment scales.}
    \label{fig:count_businesses}        
\end{figure}

Figure~\ref{fig:count_businesses} presents the distribution of firms by automation level choice. Notably, the proportion of firms opting for automation is significantly lower than the corresponding share observed in Figure~\ref{fig:fraction}. For instance, at the firm level, over 99.99\% of firms do not adopt any form of automation. This outcome can be attributed to the predominance of small firms in the economy, as automation tends to be implemented first by larger firms. Scale plays a crucial role in AI-driven automation. However, despite only a small subset of large firms initially adopting automation, the total labor compensation affected remains substantial (see Figure~\ref{fig:fraction}). As we move from the left to the right side of the figure, assuming that an AI platform can serve an entire industry group, subsector, or even sector, the share of deployment entities capable of adopting AI models increases significantly.

\begin{figure}[]
    \centering
    \includegraphics[width=0.6\textwidth]{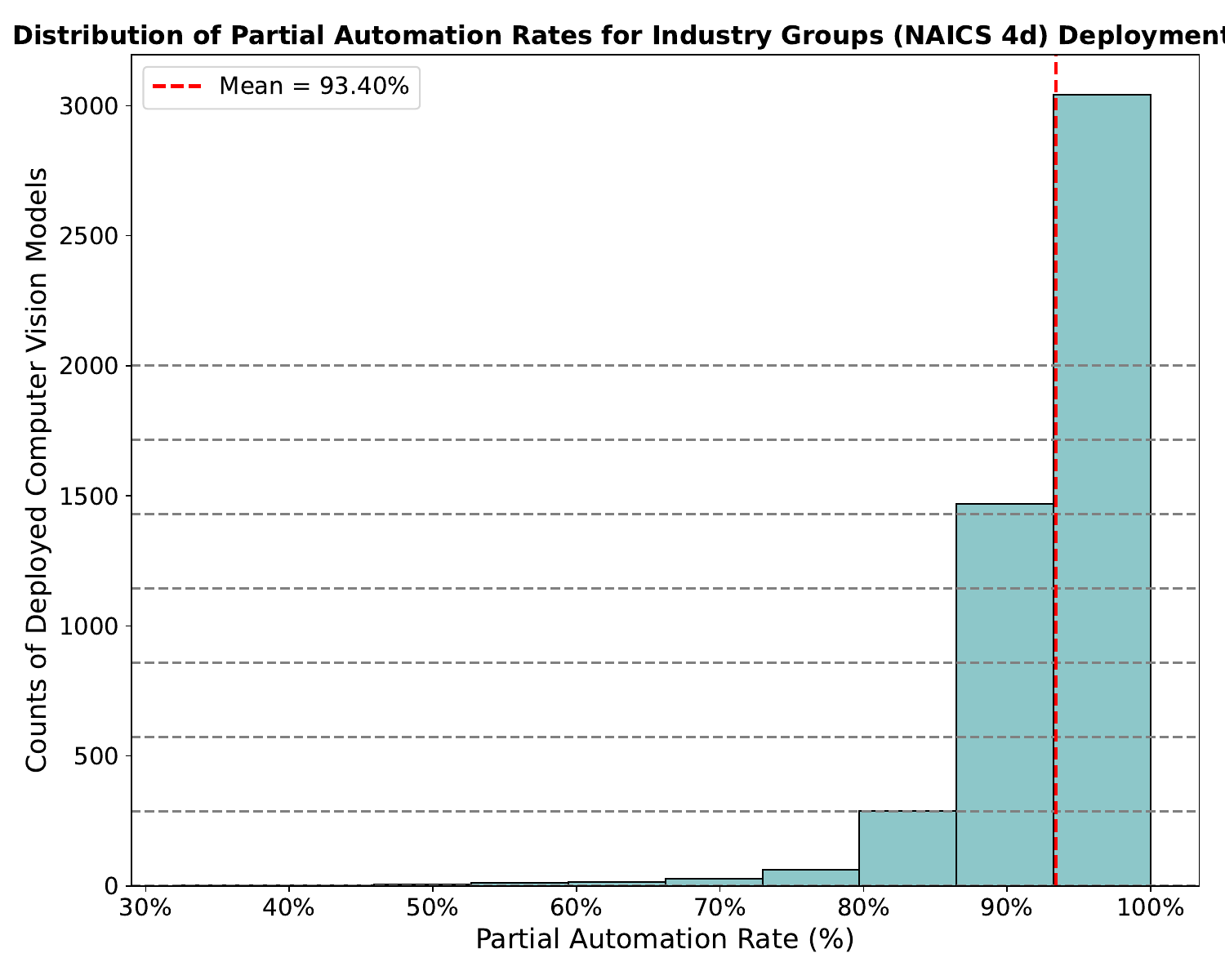}
    \caption{Counts of deployed computer vision models at industry group (4d) deployment scale by partial automation rates.}
    \label{fig:count_cv_4d}    
\end{figure}

Figure ~\ref{fig:count_cv_4d} displays the distribution of partial-automation intensities at the NAICS 4-digit deployment scale, conditional on adoption. As shown in Figure 7, only 10.6\% of industry groups adopt partial automation at this scale, but those that do tend to implement systems with very high automation intensity. The histogram in Figure~\ref{fig:count_cv_4d} is heavily right-skewed: nearly all systems automate between 90\% and close to 100\% of the task, and the mean partial automation rate is 93.4\%. This indicates that, conditional on adoption, firms typically deploy computer-vision systems that perform almost the entire visual component of the task, leaving only a small residual share to human workers. The left tail of the distribution reflects heterogeneity in task requirements and the diminishing returns to automating small or idiosyncratic subtasks, but the concentration of mass near full automation underscores that partial automation---when chosen---is generally high-intensity rather than marginal. Taken together, the results show that partial automation at the 4-digit industry level operates predominantly as a high-coverage technology: although only a subset of industry groups may adopt it, those that do tend to automate nearly all CV-feasible components of the task.

\subsection{Automation Across Occupation}

\begin{figure}[]
    \centering
    \includegraphics[width = 0.8\linewidth]{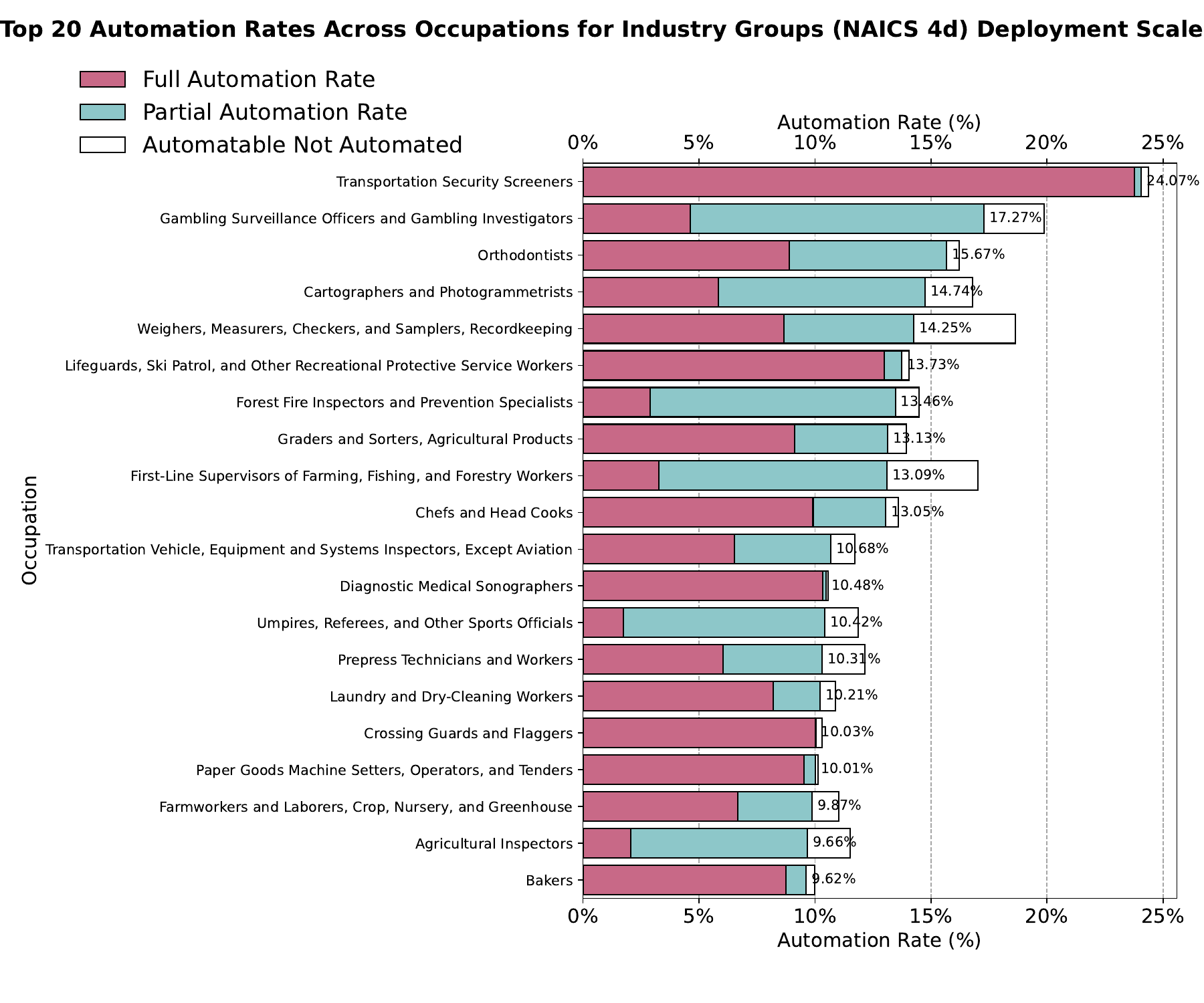}
    \caption{Top 20 Automation Rates across Occupations at Industry Group (4d) Deployment Scale. See Appendix Figure~\ref{fig:occupationrate_new} for all occupations.}
    \label{fig:occupationrate_top20}
\end{figure}

Figure~\ref{fig:occupationrate_top20} presents automation rates for the top 20 occupations at the NAICS 4-digit deployment scale; Appendix Figure~\ref{fig:occupationrate_new} provides the complete occupation-level breakdown. The x-axis represents the proportion of an occupation's total tasks---both computer vision -exposed and non-exposed---that can be substituted by computer vision systems. The automation rate is influenced by two key factors: (i) the share of an occupation's tasks that are computer vision-exposed and (ii) the extent to which these exposed tasks are suitable for full or partial automation.

In the figure, the red segments indicate tasks that can be fully automated, while the blue segments correspond to tasks that can be partially automated. Some occupations have bars containing both red and blue segments, signifying that multiple tasks within the occupation are automatable, with some being more suited for full automation and others for partial automation.

According to our findings, occupations with the highest potential for computer vision system adoption include: Transportation Security Screeners (24\% of work time automated), Orthodontists (16\%), Lifeguards, Ski Patrol, and Other Recreational Protective Service Workers (14\%), Cartographers and Photogrammetrists (15\%), and Gambling Surveillance Officers (17\%), among others.

It is important to note that the figure only displays occupations that contain at least one computer vision-exposed task. Many occupations that lack tasks suitable for computer vision applications are therefore not included. As a result, occupations appearing at the bottom of the figure should not be interpreted as the least affected by AI. However, we also observe that several occupations at the lower end of the figure exhibit an automation rate of 0\%. This indicates that while some tasks within these occupations are technically computer vision-exposed, further economic feasibility analysis suggests that none of them are viable candidates for automation through computer vision AI systems.

To illustrate these aggregate patterns, Table~\ref{tab:occupation_tasks} provides a task-level breakdown for two representative occupations: Transportation Security Screeners, and Zoologists and Wildlife Biologists. These examples highlight how technical feasibility (e.g., number of tasks and classes) and economic feasibility jointly shape realized automation outcomes.

\begin{table}[htp]
\centering
\scriptsize
\caption{Task-Level Automation Profiles}
\label{tab:occupation_tasks}
\renewcommand{\arraystretch}{1.15}
\setlength{\tabcolsep}{2pt}
\begin{tabularx}{\textwidth}{>{\raggedright\arraybackslash}p{7.4cm} >{\raggedleft\arraybackslash}p{1.3cm} >{\raggedleft\arraybackslash}p{1cm} >{\raggedleft\arraybackslash}p{1cm} >{\raggedleft\arraybackslash}p{1.3cm} >{\raggedleft\arraybackslash}p{0.5cm} >{\raggedleft\arraybackslash}p{0.5cm}}
\toprule
\textbf{Occupation / Task / DWA} & \multicolumn{1}{c}{\textbf{Task Automation}} & \multicolumn{1}{c}{\textbf{Full}} & \multicolumn{1}{c}{\textbf{Partial}} & \multicolumn{1}{c}{\textbf{Time Proportion}} & \multicolumn{1}{c}{\textbf{\#Task}} & \multicolumn{1}{c}{\textbf{\#Class}} \\
\midrule
\textbf{Transportation Security Screeners}\newline
\textit{Task:} Inspect carry-on items using x-ray equipment to determine\newline whether items contain objects that warrant further investigation.\newline
\textit{DWA:} Inspect cargo to identify potential hazards.
& 2.80\% & 2.72\% & 0.08\% & 2.81\% & 2 & 10 \\
\addlinespace[2pt]
\textbf{Transportation Security Screeners}\newline
\textit{Task:} Inspect checked baggage for signs of tampering.\newline
\textit{DWA:} Inspect cargo to identify potential hazards.
& 3.24\% & 3.24\% & 0.00\% & 3.24\% & 2 & 7 \\
\addlinespace[2pt]
\textbf{Transportation Security Screeners}\newline
\textit{Task:} Locate suspicious bags pictured in printouts sent from remote\newline monitoring areas, and set these bags aside for inspection.\newline
\textit{DWA:} Locate suspicious objects or vehicles.
& 4.09\% & 4.09\% & 0.00\% & 4.09\% & 2 & 3 \\
\addlinespace[2pt]
\textbf{Transportation Security Screeners}\newline
\textit{Task:} Monitor passenger flow through screening checkpoints to\newline ensure order and efficiency.\newline
\textit{DWA:} Monitor access or flow of people to prevent problems.
& 3.05\% & 3.05\% & 0.00\% & 3.05\% & 2 & 3 \\
\addlinespace[2pt]
\textbf{Transportation Security Screeners}\newline
\textit{Task:} Record information about any baggage that sets off alarms in\newline monitoring equipment.\newline
\textit{DWA:} Record information about suspicious objects.
& 2.79\% & 2.71\% & 0.08\% & 2.79\% & 13 & 3 \\
\addlinespace[2pt]
\textbf{Transportation Security Screeners}\newline
\textit{Task:} Send checked baggage through automated screening machines,\newline and set bags aside for searching or rescreening as indicated by equipment.\newline
\textit{DWA:} Inspect cargo to identify potential hazards.
& 3.08\% & 2.99\% & 0.09\% & 3.09\% & 3 & 7 \\
\addlinespace[2pt]
\textbf{Transportation Security Screeners}\newline
\textit{Task:} View images of checked bags and cargo using remote screening\newline equipment and alert baggage screeners or handlers to possible problems.\newline
\textit{DWA:} Inspect cargo to identify potential hazards.
& 2.10\% & 2.04\% & 0.06\% & 2.11\% & 2 & 15 \\
\addlinespace[2pt]
\textbf{Transportation Security Screeners}\newline
\textit{Task:} Watch for potentially dangerous persons whose pictures are\newline posted at checkpoints.\newline
\textit{DWA:} Maintain surveillance of individuals or establishments.
& 3.21\% & 3.21\% & 0.00\% & 3.21\% & 2 & 1 \\
\midrule
\textbf{Transportation Security Screeners (Agg Results)} & 24.07\% & 23.77\% & 0.30\% & 24.38\% & - & -\\
\midrule
\midrule
\textbf{Zoologists and Wildlife Biologists}\newline
\textit{Task:} Inventory or estimate plant and wildlife populations.\newline
\textit{DWA:} Measure environmental characteristics.
& 0.75\% & 0.00\% & 0.75\% & 6.08\% & 10 & 251 \\
\addlinespace[2pt]
\textbf{Zoologists and Wildlife Biologists}\newline
\textit{Task:} Analyze characteristics of animals to identify and classify them.\newline
\textit{DWA:} Examine characteristics or behavior of living organisms.
& 5.13\% & 0.00\% & 5.13\% & 7.13\% & 1 & 260 \\
\midrule
\textbf{Zoologists and Wildlife Biologists (Agg Results)} & 5.87\% & 0.00\% & 5.87\% & 13.21\% & - & -\\
\bottomrule
\end{tabularx}
\end{table}

Transportation Security Screeners demonstrate both high technical feasibility and an optimal automation rate. Most daily activities involve standardized image inspection or object recognition, tasks that CV systems can replicate with high consistency and minimal contextual variation. The overall automation rate of 24\% closely mirrors their task time proportion (0.24), indicating that nearly all visually intensive activities are represented and economically meaningful. This reflects not only the dominance of visual inspection tasks but also their relatively low complexity, averaging fewer than ten visual classes per task. The small number of classes and subtasks enables high substitution rates (full automation exceeding 90\% of automatable portions), consistent with economies of scale achieved through AI agents or AI-as-a-Service deployment. In this domain, automation effectively reduces labor time while maintaining reliability standards.

By contrast, zoologists illustrate the opposite pattern, high task complexity and low optimal automation rate. Despite only two automatable tasks, each accounts for a substantial time proportion (0.13 overall), nearly double the aggregate automation rate (7\%). These tasks, such as identifying or classifying animal species, involve extremely large classification spaces (over 250 classes per task) and require reasoning under uncertainty. The resulting automation rate, entirely partial, indicates that while technical feasibility exists, full automation is economically infeasible due to high model development costs and the necessity of human interpretive oversight. These results show how the combination of high complexity and large time shares amplifies both economic and technical barriers to automation.

Across occupations, two factors jointly determine automation potential. The first is technical feasibility---the share of total work time and subtasks that can be automated, influenced by task standardization and complexity (e.g., number of visual classes). The second is economic feasibility---the profitability of implementing automation given the costs of model development, integration, and required human oversight. The interaction between these two is mediated by time proportion: tasks occupying a larger share of total work time have greater economic leverage, yet only when technical feasibility is sufficiently high. Occupations with low complexity and high standardization, such as Transportation Security Screeners, exhibit both strong technical feasibility and high time alignment, making automation economically viable. By contrast, occupations such as Zoologists demonstrate that even when tasks are visually intensive or consume substantial time, high variability, tacit knowledge, and extensive classification complexity constrain economic attractiveness. These findings reinforce that partial automation, where computer vision systems assist rather than replace human judgment remains the most prevalent and economically rational trajectory for adoption in practice.

\section{Conclusion}
\label{sec:conclusion}
This paper develops a microeconomic framework to determine not just \textit{if} a task can be automated, but \textit{to what extent} automation is optimal. By integrating the technical realities of AI scaling laws with an entropy-based measure of task complexity, we bridge the gap between computer science metrics (such as cross-entropy loss) and economic outcomes (labor substitution). This unified approach moves the analysis of automation beyond binary exposure measures, allowing for a rigorous evaluation of the trade-off between full automation and partial, human-in-the-loop collaboration.

Our results highlight a sharp divergence between technical feasibility and economic viability. While computer vision systems are technically capable of performing a wide range of visual tasks, the convex cost structure of achieving human-level accuracy---driven by diminishing returns to data, model size and training steps---often renders full automation economically suboptimal. Instead, we find that partial automation is frequently the cost-minimizing equilibrium. In our calibrated model, the optimal strategy for most viable tasks involves AI systems reducing vision-related labor with human workers retaining the high-entropy share of the workload. This finding provides a micro-foundation rationale for why AI deployment has thus far been characterized more by augmentation than by wholesale displacement.

Furthermore, our analysis demonstrates that the scale of deployment is a fundamental determinant of the automation frontier. Under a firm-level deployment model, high fixed development costs act as a barrier to entry, making automation feasible only for the largest firms or the most standardized tasks. However, when costs are distributed through AI agents or an AI-as-a-Service model, the range of economically viable tasks expands significantly, shifting the optimal choice toward higher-quality models and higher automation rates. Despite this expansion, we estimate that even under optimistic economy-wide deployment assumptions, the total share of U.S. labor compensation attractive for computer-vision automation remains \textbf{below 4\%}. This suggests that while AI will be transformative for specific occupations---particularly those with standardized, low-complexity visual components---its aggregate labor market impact may be more gradual than purely exposure-based estimates suggest.

While our empirical application focuses on computer vision, the logic of our framework is modality-agnostic. The interplay among scaling laws, task entropy, and cost components applies equally to Large Language Models (LLMs) and multimodal systems. As these technologies continue to mature, our framework provides a tractable tool for predicting which cognitive tasks will remain in the human domain and which will be ceded to machines. Ultimately, our findings suggest that the future of work will not be defined by a simple race against the machine, but by a complex optimization of human-AI collaboration, governed as much by the economics of model training and its life cycle overall as by the capabilities of the models themselves.

The framework opens multiple directions for future work. Extending the model to include quality-adjusted output or error-sensitive payoff functions would capture settings where exceeding human-level accuracy yields additional economic value. Incorporating endogenous task creation would enable richer macroeconomic implications, including how AI reshapes occupations over time. Finally, applying the approach to multimodal or language-based AI systems would broaden the analysis beyond computer vision. This work contributes to a deeper understanding of how advances in AI technology propagate through firms, tasks, and labor markets, and provides a foundation for evaluating the economic consequences of human--AI collaboration.

\newpage
\bibliographystyle{apalike}
\bibliography{refs}

@misc{hestness2017deeplearningscalingpredictable,
      title={Deep Learning Scaling is Predictable, Empirically}, 
      author={Joel Hestness and Sharan Narang and Newsha Ardalani and Gregory Diamos and Heewoo Jun and Hassan Kianinejad and Md. Mostofa Ali Patwary and Yang Yang and Yanqi Zhou},
      year={2017},
      eprint={1712.00409},
      archivePrefix={arXiv},
      primaryClass={cs.LG},
      url={https://arxiv.org/abs/1712.00409} 
}

@misc{mikami2021scalinglawsynthetictorealtransfer,
      title={A Scaling Law for Synthetic-to-Real Transfer: How Much Is Your Pre-training Effective?}, 
      author={Hiroaki Mikami and Kenji Fukumizu and Shogo Murai and Shuji Suzuki and Yuta Kikuchi and Taiji Suzuki and Shin-ichi Maeda and Kohei Hayashi},
      year={2021},
      eprint={2108.11018},
      archivePrefix={arXiv},
      primaryClass={cs.LG},
      url={https://arxiv.org/abs/2108.11018}, 
}

@misc{svanberg2024beyond,
  title        = {Beyond {AI} Exposure: Which Tasks Are Cost-Effective to Automate with Computer Vision?},
  author={Svanberg, Maja and Li, Wensu and Fleming, Martin and Goehring, Brian and Thompson, Neil},
  year         = {2024},
  note         = {SSRN Working Paper No.\ 4700751},
  url          = {https://ssrn.com/abstract=4700751}
}

@article{hyman1953stimulus,
  title={Stimulus information as a determinant of reaction time},
  author={Hyman, Ray},
  journal={Journal of Experimental Psychology},
  volume={45},
  number={3},
  pages={188--196},
  year={1953},
  publisher={American Psychological Association}
}

@inproceedings{brynjolfsson2018can,
  title={What Can Machines Learn, and What Does It Mean for Occupations and the Economy?},
  author={Brynjolfsson, Erik and Mitchell, Tom and Rock, Daniel},
  organization={AEA papers and proceedings},
  volume={108},
  pages={43--47},
  year={2018}
}

@article{webb2019impact,
  title={{The Impact of Artificial Intelligence on the Labor Market}},
  author={Webb, Michael},
  journal={Available at SSRN 3482150},
  year={2019}
}

@techreport{NBERw33949,
 title = "Designing Human-{AI} Collaboration: A Sufficient-Statistic Approach",
 author = "Agarwal, Nikhil and Moehring, Alex and Wolitzky, Alexander",
 institution = "National Bureau of Economic Research",
 type = "Working Paper",
 series = "Working Paper Series",
 number = "33949",
 year = "2025",
 month = "June",
 doi = {10.3386/w33949},
 URL = "http://www.nber.org/papers/w33949",
}

@techreport{NBERw31422,
 title = "Combining Human Expertise with Artificial Intelligence: Experimental Evidence from Radiology",
 author = "Agarwal, Nikhil and Moehring, Alex and Rajpurkar, Pranav and Salz, Tobias",
 institution = "National Bureau of Economic Research",
 type = "Working Paper",
 series = "Working Paper Series",
 number = "31422",
 year = "2023",
 month = "July",
 doi = {10.3386/w31422},
 URL = "http://www.nber.org/papers/w31422",
}

@article{Brynjolfsson2021productivity,
Author = {Brynjolfsson, Erik and Rock, Daniel and Syverson, Chad},
Title = {The Productivity J-Curve: How Intangibles Complement General Purpose Technologies},
Journal = {American Economic Journal: Macroeconomics},
Volume = {13},
Number = {1},
Year = {2021},
Month = {January},
Pages = {333–72},
DOI = {10.1257/mac.20180386},
URL = {https://www.aeaweb.org/articles?id=10.1257/mac.20180386}}

@inproceedings{Zhang2023Optimizing,
  author    = {Zhang, Yang and Zong, Ruohan and Shang, Lanyu and Zeng, Huimin and Yue, Zhenrui and Wei, Na and Wang, Dong},
  title     = {On Optimizing Model Generality in AI‐based Disaster Damage Assessment: A Subjective Logic‐driven Crowd‐AI Hybrid Learning Approach},
  booktitle = {Proceedings of the Thirty-Second International Joint Conference on Artificial Intelligence (IJCAI)},
  pages     = {6317--6325},
  year      = {2023},
  doi       = {10.24963/ijcai.2023/701},
  url       = {https://www.ijcai.org/proceedings/2023/701/},
}

@misc{russakovsky2015imagenetlargescalevisual,
      title={ImageNet Large Scale Visual Recognition Challenge}, 
      author={Olga Russakovsky and Jia Deng and Hao Su and Jonathan Krause and Sanjeev Satheesh and Sean Ma and Zhiheng Huang and Andrej Karpathy and Aditya Khosla and Michael Bernstein and Alexander C. Berg and Li Fei-Fei},
      year={2015},
      eprint={1409.0575},
      archivePrefix={arXiv},
      primaryClass={cs.CV},
      url={https://arxiv.org/abs/1409.0575}, 
}

@misc{singh2024benchmarkingobjectdetectorscoco,
      title={Benchmarking Object Detectors with COCO: A New Path Forward}, 
      author={Shweta Singh and Aayan Yadav and Jitesh Jain and Humphrey Shi and Justin Johnson and Karan Desai},
      year={2024},
      eprint={2403.18819},
      archivePrefix={arXiv},
      primaryClass={cs.CV},
      url={https://arxiv.org/abs/2403.18819}, 
}

@ARTICLE{Auto-Pytorch,
  author={Zimmer, Lucas and Lindauer, Marius and Hutter, Frank},
  journal={IEEE Transactions on Pattern Analysis and Machine Intelligence}, 
  title={Auto-Pytorch: Multi-Fidelity MetaLearning for Efficient and Robust AutoDL}, 
  year={2021},
  volume={43},
  number={9},
  pages={3079-3090},
  doi={10.1109/TPAMI.2021.3067763}}

@inproceedings{
wistuba2022supervising,
title={Supervising the Multi-Fidelity Race of Hyperparameter Configurations},
author={Martin Wistuba and Arlind Kadra and Josif Grabocka},
booktitle={Advances in Neural Information Processing Systems},
editor={Alice H. Oh and Alekh Agarwal and Danielle Belgrave and Kyunghyun Cho},
year={2022},
url={https://openreview.net/forum?id=0Fe7bAWmJr}
}

@inproceedings{
kadra2023scaling,
title={Scaling Laws for Hyperparameter Optimization},
author={Arlind Kadra and Maciej Janowski and Martin Wistuba and Josif Grabocka},
booktitle={Thirty-seventh Conference on Neural Information Processing Systems},
year={2023},
url={https://openreview.net/forum?id=ghzEUGfRMD}
}

@misc{lin2015microsoftcococommonobjects,
      title={Microsoft COCO: Common Objects in Context}, 
      author={Tsung-Yi Lin and Michael Maire and Serge Belongie and Lubomir Bourdev and Ross Girshick and James Hays and Pietro Perona and Deva Ramanan and C. Lawrence Zitnick and Piotr Dollár},
      year={2015},
      eprint={1405.0312},
      archivePrefix={arXiv},
      primaryClass={cs.CV},
      url={https://arxiv.org/abs/1405.0312}, 
}

@misc{onet_center_2023, title={{O*NET}}, note={\url{https://www.onetcenter.org/overview.html}, Accessed: 2023-04-04}, journal={O*NET Resource Center}, author={{U.S. Department of Labor}}, year={2023}, month={Feb}}

@article{acemoglu2024task,
  title={A task-based approach to inequality},
  author={Acemoglu, Daron and Restrepo, Pascual},
  journal={Oxford Open Economics},
  volume={3},
  number={Supplement\_1},
  pages={i906--i929},
  year={2024},
  publisher={Oxford University Press}
}

@incollection{acemoglu2011skills,
  title={Skills, tasks and technologies: Implications for employment and earnings},
  author={Acemoglu, Daron and Autor, David},
  booktitle={Handbook of labor economics},
  volume={4},
  pages={1043--1171},
  year={2011},
  publisher={Elsevier}
}

@misc{thompson2022computationallimitsdeeplearning,
      title={The Computational Limits of Deep Learning}, 
      author={Neil C. Thompson and Kristjan Greenewald and Keeheon Lee and Gabriel F. Manso},
      year={2022},
      eprint={2007.05558},
      archivePrefix={arXiv},
      primaryClass={cs.LG},
      url={https://arxiv.org/abs/2007.05558}, 
}

@article{frey2017future,
  title={The future of employment: How susceptible are jobs to computerisfation?},
  author={Frey, Carl Benedikt and Osborne, Michael A},
  journal={Technological forecasting and social change},
  volume={114},
  pages={254--280},
  year={2017},
  publisher={Elsevier}
}

@article{eloundou2024gpts,
  title={{GPTs are GPTs}: Labor market impact potential of LLMs},
  author={Eloundou, Tyna and Manning, Sam and Mishkin, Pamela and Rock, Daniel},
  journal={Science},
  volume={384},
  number={6702},
  pages={1306--1308},
  year={2024},
  publisher={American Association for the Advancement of Science}
}

@article{handa2025economic,
  title={Which Economic Tasks are Performed with AI? Evidence from Millions of Claude Conversations},
  author={Handa, Kunal and Tamkin, Alex and McCain, Miles and Huang, Saffron and Durmus, Esin and Heck, Sarah and Mueller, Jared and Hong, Jerry and Ritchie, Stuart and Belonax, Tim and others},
  journal={arXiv preprint arXiv:2503.04761},
  year={2025}
}

@article{rosenfeld2019constructive,
  title={A constructive prediction of the generalization error across scales},
  author={Rosenfeld, Jonathan S and Rosenfeld, Amir and Belinkov, Yonatan and Shavit, Nir},
  journal={arXiv preprint arXiv:1909.12673},
  year={2019}
}

@article{acemoglu2018race,
  title={The Race Between Man and Machine: Implications of Technology for Growth, Factor Shares, and Employment},
  author={Acemoglu, Daron and Restrepo, Pascual},
  journal={American Economic Review},
  volume={108},
  number={6},
  pages={1488--1542},
  year={2018},
  publisher={American Economic Association}
}

@inproceedings{thompson2024economic,
  title={A Model for Estimating the Economic Costs of Computer Vision Systems That Use Deep Learning},
  author={Thompson, Neil and Fleming, Michael and Tang, Brandon J. and Pastwa, Andrew M. and Borge, Nicholas and Goehring, Benjamin C. and Das, Saurabh},
  booktitle={Proceedings of the AAAI Conference on Artificial Intelligence},
  volume={38},
  number={21},
  pages={23012--23018},
  year={2024},
  publisher={AAAI Press},
  doi={10.1609/aaai.v38i21.30343}
}

@article{lowder2018lexical,
  title={Lexical predictability during natural reading: Effects of surprisal and entropy reduction},
  author={Lowder, Matthew W and Choi, Wonil and Ferreira, Fernanda and Henderson, John M},
  journal={Cognitive science},
  volume={42},
  pages={1166--1183},
  year={2018},
  publisher={Wiley Online Library}
}

@article{hu2022human,
  title={Human decision time in uncertain binary choice},
  author={Hu, Lunhu and Pan, Xing and Ding, Song and Kang, Rui},
  journal={Symmetry},
  volume={14},
  number={2},
  pages={201},
  year={2022},
  publisher={MDPI}
}

@article{hobbhahn2022trends,
  title={Trends in GPU price-performance},
  author={Hobbhahn, Marius and Besiroglu, Tamay},
  journal={Epoch AI},
  year={2022}
}

@article{sullivan2023average,
  title={Average stock market return},
  author={Sullivan, Bob},
  journal={Forbes Advisor},
  volume={16},
  year={2023}
}

@article{assets2003consumer,
  title={Consumer Durable Goods in the United States, 1925-99},
  author={Assets, Fixed},
  journal={Washington, DC: US Government Printing Office, September},
  year={2003}
}

@article{kaplan2020scaling,
  title={Scaling laws for neural language models},
  author={Kaplan, Jared and McCandlish, Sam and Henighan, Tom and Brown, Tom B and Chess, Benjamin and Child, Rewon and Gray, Scott and Radford, Alec and Wu, Jeffrey and Amodei, Dario},
  journal={arXiv preprint arXiv:2001.08361},
  year={2020}
}

@misc{megatronlm2024,
  author       = {{NVIDIA}},
  title        = {Megatron-LM: Training Multi-Billion Parameter Language Models Using Model Parallelism},
  year         = {2024},
  howpublished = {\url{https://raw.githubusercontent.com/NVIDIA/Megatron-LM/main/README.md}},
  note         = {Accessed: 2024-07-21}
}

@article{thompson2021demand,
  title={Demand Forecasting with A.I.: Building the Business Case},
  author={Thompson, Neil and Borge, Nicholas J and Pande, Ananya and Fleming, Martin},
  journal={MIT Sloan Research Brief},
  year={2021}
}

@article{rosenfeld2021scaling,
  title={Scaling laws for deep learning},
  author={Rosenfeld, Jonathan S},
  journal={arXiv preprint arXiv:2108.07686},
  year={2021}
}

@article{hoffmann2022training,
  title={Training compute-optimal large language models (2022)},
  author={Hoffmann, Jordan and Borgeaud, S and Mensch, A and Buchatskaya, E and Cai, T and Rutherford, E and de Las Casas, D and Hendricks, LA and Welbl, J and Clark, A and others},
  journal={arXiv preprint arXiv:2203.15556},
  year={2022}
}

@article{brynjolfsson2017machinelearning,
  title        = {What Can Machine Learning Do? Workforce Implications},
  author       = {Erik Brynjolfsson and Tom Mitchell},
  journal      = {Science},
  volume       = {358},
  number       = {6370},
  pages        = {1530--1534},
  year         = {2017},
  doi          = {10.1126/science.aap8062}
}

@techreport{autor2025expertise,
  title        = {Expertise},
  author       = {David Autor and Neil Thompson},
  institution  = {National Bureau of Economic Research},
  number       = {33941},
  type         = {Working Paper},
  year         = {2025},
  url          = {http://www.nber.org/papers/w33941}
}

@techreport{hampole2025ailabor,
  title        = {Artificial Intelligence and the Labor Market},
  author       = {Menaka Hampole and Dimitris Papanikolaou and Lawrence D.W. Schmidt and Bryan Seegmiller},
  institution  = {National Bureau of Economic Research},
  type         = {Working Paper},
  year         = {2025}
}

@misc{shao2025futurework,
  title        = {Future of Work with {AI} Agents: Auditing Automation and Augmentation Potential Across the {U.S.} Workforce},
  author       = {Yijia Shao and Humishka Zope and Yucheng Jiang and Jiaxin Pei and David Nguyen and Erik Brynjolfsson and Diyi Yang},
  year         = {2025},
  eprint       = {2506.06576},
  archivePrefix= {arXiv},
  primaryClass = {cs.CY},
  url          = {https://arxiv.org/abs/2506.06576}
}

@article{brynjolfsson2025generative,
  title        = {Generative AI at Work},
  author       = {Erik Brynjolfsson and Danielle Li and Lindsey Raymond},
  journal      = {The Quarterly Journal of Economics},
  year         = {2025},
  doi          = {10.1093/qje/qjae044}
}

@article{acemoglu2025simple,
  title        = {The Simple Macroeconomics of {AI}},
  author       = {Daron Acemoglu},
  journal      = {Economic Policy},
  volume       = {40},
  number       = {121},
  pages        = {13--58},
  year         = {2025}
}

@misc{henighan2020generative,
  title        = {Scaling Laws for Autoregressive Generative Modeling},
  author       = {Tom Henighan and Jared Kaplan and Maxwell Katz and Mark Chen and others},
  year         = {2020},
  eprint       = {2010.14701},
  archivePrefix= {arXiv},
  primaryClass = {cs.LG},
  url          = {https://arxiv.org/abs/2010.14701}
}

@article{autor2003skill,
  title={The Skill Content of Recent Technological Change: An Empirical Exploration},
  author={Autor, David H. and Levy, Frank and Murnane, Richard J.},
  journal={The Quarterly Journal of Economics},
  volume={118},
  number={4},
  pages={1279--1333},
  year={2003}
}

@inproceedings{acemoglu2018modeling,
  title={Modeling Automation},
  author={Acemoglu, Daron and Restrepo, Pascual},
  booktitle={AEA Papers and Proceedings},
  volume={108},
  pages={48--53},
  year={2018}
}

@techreport{sheridan1978human,
  title={Human and Computer Control of Undersea Teleoperators},
  author={Sheridan, Thomas B. and Verplank, William L.},
  institution={MIT Man-Machine Systems Laboratory},
  year={1978}
}

@article{parasuraman2000model,
  title={A Model for Types and Levels of Human Interaction with Automation},
  author={Parasuraman, Raja and Sheridan, Thomas B. and Wickens, Christopher D.},
  journal={IEEE Transactions on Systems, Man, and Cybernetics---Part A: Systems and Humans},
  volume={30},
  number={3},
  pages={286--297},
  year={2000},
  publisher={IEEE}
}

@article{endsley1999level,
  title={Level of Automation Effects on Performance, Situation Awareness and Workload in a Dynamic Control Task},
  author={Endsley, Mica R. and Kaber, David B.},
  journal={Ergonomics},
  volume={42},
  number={3},
  pages={462--492},
  year={1999},
  publisher={Taylor \& Francis}
}

@article{zeira1998workers,
  title={Workers, Machines, and Economic Growth},
  author={Zeira, Joseph},
  journal={The Quarterly Journal of Economics},
  volume={113},
  number={4},
  pages={1091--1117},
  year={1998}
}

@article{hick1952rate,
  title={On the Rate of Gain of Information},
  author={Hick, W. E.},
  journal={Quarterly Journal of Experimental Psychology},
  volume={4},
  number={1},
  pages={11--26},
  year={1952},
  doi={10.1080/17470215208416600}
}

@INPROCEEDINGS{Liu2021swin,
  author={Liu, Ze and Lin, Yutong and Cao, Yue and Hu, Han and Wei, Yixuan and Zhang, Zheng and Lin, Stephen and Guo, Baining},
  booktitle={2021 IEEE/CVF International Conference on Computer Vision (ICCV)}, 
  title={Swin Transformer: Hierarchical Vision Transformer using Shifted Windows}, 
  year={2021},
  volume={},
  number={},
  pages={9992-10002},
  doi={10.1109/ICCV48922.2021.00986}}

@article{arrow1961capital,
 ISSN = {00346535, 15309142},
 URL = {http://www.jstor.org/stable/1927286},
 author = {K. J. Arrow and H. B. Chenery and B. S. Minhas and R. M. Solow},
 journal = {The Review of Economics and Statistics},
 number = {3},
 pages = {225--250},
 publisher = {The MIT Press},
 title = {Capital-Labor Substitution and Economic Efficiency},
 urldate = {2026-03-26},
 volume = {43},
 year = {1961}
}

@article{solow1956econ,
 ISSN = {00335533, 15314650},
 URL = {http://www.jstor.org/stable/1884513},
 author = {Robert M. Solow},
 journal = {The Quarterly Journal of Economics},
 number = {1},
 pages = {65--94},
 publisher = {Oxford University Press},
 title = {A Contribution to the Theory of Economic Growth},
 urldate = {2026-03-26},
 volume = {70},
 year = {1956}
}

@article{sharma2022scaling,
  title={Scaling Laws from the Data Manifold Dimension},
  author={Sharma, Utkarsh and Kaplan, Jared},
  journal={Journal of Machine Learning Research},
  volume={23},
  number={9},
  pages={1--34},
  year={2022}
}

\newpage
\begin{appendices}
\setcounter{section}{0}
\counterwithin{figure}{section}
\counterwithin{table}{section}
\counterwithin{equation}{section}
\setcounter{equation}{0}

\section{Occupation and Task Characteristics Survey}
\label{app:survey}

\subsection{Survey Participants} 
\leavevmode
The survey leveraged Prolific, a widely regarded online platform known for its diverse and reliable respondent pool, to recruit participants and direct them to the survey. The target population comprised U.S. adults aged 18 and older, with at least two years of work experience and familiarity with occupations involving the visual tasks under investigation. Screening and survey implementation were conducted via Qualtrics, a platform that enabled dynamic question flows, tailored question paths, and streamlined participant navigation.

The final sample exhibited substantial demographic diversity, encompassing respondents aged 18 to 85, with a median age of 35 years. Gender representation within the sample was approximately 44.34\% female and 41.21\% male, with the remainder identifying as non-binary, other genders, or preferring not to disclose their gender. The average annual household income of respondents was approximately \$94,000, with 8.44\% reporting incomes below \$25,000 and 9.26\% reporting incomes above \$150,000. 

\subsection{Survey Questions}
\label{sec:Survey Questions}
\leavevmode

The survey questions fall into following categories: 

\textbf{Task Navigation Questions. }These questions direct participants to identify their industry, occupation, and familiar computer vision tasks. Once an occupation is selected, the first computer vision task associated with that occupation is presented. Participants are then asked whether they are familiar with the task. If they indicate familiarity, they proceed to answer a series of detailed questions related to the task; otherwise, the survey bypasses the task and moves on to the next. Tasks within the same occupation are presented in random order. Among the 263 occupations included in the survey, 153 are associated with two computer vision tasks, with an overall average of 1.75 tasks per occupation. After completing the questions for a given task, participants are asked whether they are willing to continue with an additional task. By allowing participants to skip tasks or exit the survey at their discretion, we mitigate the risk of response quality deterioration due to cognitive limitation or participant fatigue.

\textbf{Questions on Task Characteristics.}
These questions aim to collect detailed information about the nature of visual tasks, focusing on their outcomes, error tolerance, and frequency, as well as their significance within the respondent's occupation.

Participants are guided to categorize the vision task into classification, visual generation (e.g., drawing or films), or other types of vision tasks according to the outcome of the task. For classification tasks, which is the focus of this paper, respondents further determine the specific type: Binary classification, e.g., whether there's defect in the product. Single-choice multi-class classification, involving the selection of one option from a predefined list. E.g., determining whether it is sunny, cloudy, or rainy based on weather images. Multi-choice classification, requiring the identification of multiple items from a set. E.g., determining which products at a grocery store need to be re-stocked.
Respondents are also asked to list the number of classes and potential classes relevant to the task.

The survey includes a series of error tolerance questions to evaluate performance standards for tasks. Respondents are asked to estimate the typical error rate under normal conditions for competent workers, providing a benchmark for the expected level of accuracy. Additionally, they are asked to identify the worst acceptable error rate, beyond which workers would no longer be considered qualified to perform the task. To contextualize the difficulty of the task, respondents also report the error rate that would occur under random guessing. These measures allow for a nuanced assessment of the precision requirements of each task and the potential challenges of automating them while maintaining acceptable performance levels.

The survey also examines task frequency and time allocation to capture the prevalence and importance of visual tasks within the respondents' occupations. Respondents indicate how often they perform the task, with frequencies ranging from rare (less than once per year) to extremely frequent (over 3,000 times a day). They also report the percentage of their overall work time dedicated to performing the task, offering insight into its significance in their day-to-day responsibilities. This information is critical for understanding the role of these tasks in professional settings and for assessing the feasibility of automating tasks that are both prevalent and time-intensive. 

\textbf{Questions on Data Availability and Data Collection Costs.} The survey examines the feasibility of collecting and recording data, a critical component for training AI systems and automating vision tasks. Participants were asked to determine whether the necessary task inputs---such as images, videos, and task outcomes---are already being recorded in their workflows. For tasks lacking sufficient data, the survey investigated the estimated costs of data collection, ranging from ``very cheap'' to ``very expensive.'' Additionally, respondents identified any adjustments required to facilitate data capture, including the need for specialized equipment, technical expertise, or changes in existing procedures.

\subsection{Survey Data Quality Control}
\leavevmode

To ensure the reliability and accuracy of the collected data, the survey implemented rigorous quality control measures throughout its design and administration. Key methods included:

\textbf{Attention Checks.} Specific criteria were embedded within the survey to identify inconsistent or illogical responses. For example, respondents were required to logically align their answers to questions on error rates. Workers were expected to make fewer errors than randomly guessing and less than the worst acceptable error rate. Responses failing these criteria were flagged and excluded from the dataset.

\textbf{Screening for Familiarity.} At the start of the survey and before each task, participants were asked to confirm their familiarity with the chosen occupation and the listed tasks. This ensured that only qualified respondents contributed data, enhancing the relevance and credibility of the results.

\textbf{Survey Duration Monitoring.} The average completion time for each wave was recorded to identify outliers. We excluded respondents who completed the survey too quickly, potentially indicating rushed or careless responses.

\textbf{Iterative Survey Refinement.}To ensure high-quality survey results, we conducted a trial run (Wave 1) in July 2023, which collected 451 valid responses, and followed it with a formal survey (Wave 2) in November 2023, yielding 3,327 valid responses. The trial run provided an initial dataset for analyzing survey data quality and identifying areas for improvement. Based on insights from Wave 1, we refined the survey design by rephrasing ambiguous questions, introducing additional response options to capture more nuanced perspectives, and enhancing the instructions on both the Prolific and Qualtrics platforms. Furthermore, the payment plan was revised to better align with respondent expectations and boost engagement. To preserve data integrity, we ensured that no respondents participated in both waves. After thorough post-survey processing, we confirmed that the results from the two waves could be seamlessly combined following appropriate adjustments. This iterative approach allowed us to address potential issues early and significantly enhance data clarity and accuracy in the formal survey. %

\section{Cost Function Specifications}
\label{app:cost}

To interpret the solution to the cost minimization problem, it is helpful to describe how the different terms in the objective function reflect the economic costs of developing and deploying an AI system. The reduced-form cost function $\kappa_i(a_i,Y_i)$ represents the minimum expenditure needed to achieve a system with accuracy level $a_i$ that can support a usage level of $Y_i$. Conceptually, these costs fall into two broad categories: fixed costs associated with building and training the model, and variable costs associated with using it at scale.

The fixed component corresponds to the one-time investment necessary to create a model capable of meeting the target accuracy level. This includes engineering and development labor involved in designing the system, setting up the training pipeline, and maintaining the model throughout its lifecycle. It also includes the cost of acquiring and preparing training data, as reflected in the term proportional to $D_i$, since higher accuracy generally requires a larger and more carefully curated datasets. In addition, training a more accurate model requires greater computational resources, which is captured by the component proportional to $M_i$ and $T_i$: larger models require more compute per update, and more training steps increase the total computational workload. Together, these fixed elements determine the minimum development cost necessary to produce a model of quality $a_i$.

A second part of the cost arises from deployment and scales with the amount of output the model produces. Once the model is trained, each inference run requires computational resources that depend on the model size. Because larger models involve more parameters and higher per-call compute requirements, the term proportional to $I_i$ captures the expenditure associated with running the system to produce the required $Y_i$ task outputs. This component therefore reflects the variable cost of using the model in practice and increases with both the model size chosen to achieve accuracy $a_i$ and the usage level $Y_i$ that the firm must support.

These two components---development costs that depend on achieving accuracy $a_i$, and deployment costs that depend on supporting usage $Y_i$---together constitute the overall cost structure summarized in $\kappa_i(a_i,Y_i)$. The structure implies two useful properties. First, the cost is increasing in $a_i$, because achieving higher accuracy requires more data, more computation, or larger models. Second, the cost is increasing in $Y_i$, since each additional model call requires inference compute that grows with the model size. These properties clarify how AI-system costs enter the firm's optimization problem in Section \ref{sec:ai_cost_min}, with fixed costs governing whether AI adoption is economically viable at all, and variable costs determining the marginal trade-off between AI usage and human labor for task $i$.

Our approach to deriving the cost function $\kappa(D,T,M,I; m)$ is primarily based on the cost framework established by \cite{svanberg2024beyond}. Specifically, $\kappa(D,T,M,I; m)$ is decomposed as
\begin{equation}
    \kappa(D,T,M,I; m) = \kappa^{\text{eng}} + m (\kappa^{\text{data}}(D) + \kappa^{\text{train}}(T, M) + \kappa^{\text{inf}}(M, I)).
    \label{eq:cost}
\end{equation}
Here, $\kappa^{eng}$ refers to the cost of hiring a group of engineering team and subject matter experts to write the training and inference code. We assume that the coding effort holds constant regardless of the change in other input factors (which mostly just involves changing the hyperparameters in the code). $\kappa^{data}$ refers to the cost of collecting and manually labeling the data, which scales linearly with $D$. $\kappa^{train}$ refers to the GPU cost of training the AI system, which scales linearly with $T$ and $M$. $\kappa^{inf}$ refers to the GPU cost of deploying the AI system to produce the task output, which scales linearly with $M$ and $I$. 

In the case where there are multiple classification tasks, i.e., $m>1$, we assume that a distinct CV system is developed for each classification task, each undergoing respective data collection, training, and inference processes. Hence, the corresponding cost terms in Equation~\eqref{eq:cost} is proportional to $m$. On the other hand, we assume that the engineering effort remains constant regardless of $m$. Since an AI system, once developed, can be used over a finite operational lifespan, let $L$ denotes the number of years the system remains in use.\footnote{U.S. Bureau of Economic Analysis specifies a 5-year service life for custom software in the context of fixed assets. We use this number as the operational lifespan of a deployed AI system.} Each of the cost terms above should thus be interpreted as the present value of the discounted stream of costs incurred over the $L$-year period with a discount rate of $d$.In the following subsections, we detail the specifications for each term.

\subsection{Data Collection Cost}

We assume that data collection is a continuous effort. Specifically, to collect and maintain a training data set of size $D$, we assume there is an initial round of collecting $D$ images, and $\kappa_\text{recur}$ rounds of data renewal per year for $L$ years. Each data renewal round collects a new set of $D$ data. Therefore, the data collection cost is specified as follows
\begin{equation}
    \kappa^{\text{data}}(D) = p_\text{data}D + \sum_{t=1}^L \frac{\kappa_\text{recur}p_\text{data} D}{(1+d)^t},
    \label{eq:B2}
\end{equation}
where $p_\text{data}$ represents the cost of collecting and labeling one datum. $d$ represents the discount rate.

We assume that the cost of collecting and labeling one datum for a visual-judgment task equals the opportunity cost of the worker performing that judgment within their regular occupation. 
In other words, this cost is proportional to the worker's wage and the share of time spent on the task's vision-related component, and is normalized by the frequency with which such judgments are performed within the occupation. 
Formally, for each task $i$, we define the per-datum data-collection cost as
\begin{equation}
p_{\text{data}}
\,=\, 
\frac{w_o \, \rho_{oi} \, \delta_i}{F_i} \, \Phi_i,
\end{equation}
where $w_o$, $\rho_{oi}$, and $\delta_i$ follow the definitions introduced in Equation~\eqref{eq:ai_cost_coef}. 
$F_i$ denotes the annualized frequency of visual-judgment decisions for task $i$, derived directly from survey questions detailed in Appendix \ref{sec:Survey Questions}. 
$\Phi_i$ represents the multi-annotator reliability factor, indicating the number of independent judgments required per datum to achieve consensus-quality labels. We adopt $\Phi_i = 8$ consistent with established standards for multi-annotator labeling practices in canonical computer vision datasets \citep{russakovsky2015imagenetlargescalevisual,lin2015microsoftcococommonobjects,singh2024benchmarkingobjectdetectorscoco} to ensure inter-rater reliability in visual annotation across diverse task settings. To prevent unrealistically low cost estimates when judgment frequency is exceptionally high or wage data are missing, we apply a conservative minimum threshold of \$0.05 per datum based on \cite{Zhang2023Optimizing}, ensuring that our cost model remains both stable and realistic.

Thus, $p_{\text{data},i}$ captures the marginal opportunity cost of labeling one datum for task $i$, expressed in wage-time per decision. 
This measure forms the micro-foundation for the aggregate data-collection cost component employed in the automation-rate analysis in Section \ref{sec:results}.

\subsection{Training Cost}

Similar to data collection, training a CV system is also modelled as a continuous effort, consisting of $\kappa_\text{init}$ rounds of initial training (for algorithmic development) and $\kappa_\text{recur}$ rounds of re-training per year. Each training step involves approximately $F_\text{GPU} M \cdot Z_\text{input}$ GPU FLOPs, where $F_\text{GPU}$ denotes the number of GPU FLOPs per floating point operation, $Z_\text{input}$ denotes the number of pixels of the input image. Hence each training round, with $T$ training steps in total, requires $F_\text{GPU}Z_\text{input}MT$ GPU FLOPs.

The training cost involves the GPU cost, which is paid by GPU hours. Denote $p_{GPU}$ as the price for one GPU hour, and we assume that the price will drop by $1 + d_\text{GPU}$ each year. Denote $r_\text{FLOP}$ as the number of FLOPs the GPU can perform per hour, and $U_\text{GPU}$ as GPU utilization, which is the percentage of time when the GPU is being utilized. Then the total training cost, including the initial and recurring training rounds, amounts to
\begin{equation}
    \kappa^{\text{train}}(T, M) = \bigg(\kappa_\text{init} + \sum_{t=1}^L \frac{\kappa_\text{recur}}{(1 + d_\text{GPU})^t(1+d)^t}\bigg) \cdot \frac{p_\text{GPU}}{r_\text{FLOP}U_\text{GPU}} \cdot F_\text{GPU}Z_\text{input}MT.
\label{eq:B4}
\end{equation}

\subsection{Inference Cost}

The inference cost involves the GPU cost to perform inference. The number of GPU hours needed for inference per year is $I$. In the main text we assume that adopting AI does not change the total quantity of output---that is, the number of times the task is completed remains the same as it would under human labor. However, it is very difficult to estimate how many times a human worker performs a given task over the course of a year. Given that the inference cost term is relatively small compared to other cost components in computer vision applications, we adopt a conservative principle to avoid underestimating it. Specifically, we assume an extremely efficient human worker who can make one classification decision per second (recall that the share of time a worker spends on a given task varies across tasks). This assumption gives us an upper bound on the number of task completions, which in turn provides a reasonable approximation for $Y_i$. Using this, and considering the inference speed of typical neural networks such as VGG-Net, we estimate
\[
\frac{\tau}{\tau_{\mathrm{GPU}}} = 2.01138 \times 10^{13},
\]
which, combined with Equation~\eqref{eq:D6}, allows us to eliminate the unknown $Y_i$ in the cost minimization process as in Equation~\eqref{eq:cost_min}.

Taking into account the annual price drop of the GPU and the discount rate, the inference cost of $L$ years can be computed as
\begin{equation}
    \kappa^{\text{inf}}(M, Y) = \sum_{t=1}^L \frac{1}{(1+d_\text{GPU})^t(1+d)^t} \cdot \frac{p_\text{GPU}}{U_{GPU}}I.
\label{eq:B5}
\end{equation}

\subsection{Engineering Cost}

The engineering cost consists of two terms, one is the implementation cost, $C_\text{impl}$, incurred during the initial CV system development stage, and the other is the maintenance cost, $C_\text{maint}$, incurred annually. So the total engineering cost amounts to
\begin{equation}
    \kappa^{\text{eng}} = C_\text{impl} + \sum_{t=1}^L \frac{C_\text{maint}}{(1+d)^t}.
\end{equation}

We used the original implementation and maintenance costs reported in \cite{thompson2021demand}. The original case study, based on a real-world IBM deep learning deployment project for time series forecasting, reports an upfront implementation cost of \$1,765,000 and an annual maintenance cost of \$242,840, primarily reflecting the compensation of IBM engineers, client engineers, and subject matter experts. To account for changes in labor costs over time, we adjust the original costs reported to 2024 dollars. Given that these expenditures are driven by high-skilled technical labor, we apply an inflation adjustment using the U.S. Employment Cost Index (ECI) for Private Industry Workers in Professional and Technical Services. According to the BLS ECI data, the index value was approximately 130.2 in Q1 2018 and rose to approximately 158.3 by Q4 2024, reflecting a cumulative increase of about 21.5\% over this period.\footnote{U.S. Bureau of Labor Statistics, Employment Cost Index (ECI) for Private Industry Workers in Professional and Technical Services, available at: \url{https://www.bls.gov/news.release/eci.t01.htm}} Applying this adjustment, the inflation-adjusted 2024 values are approximately \$2.14 million for the implementation cost and \$295,000 for the annual maintenance cost.

\subsection{Cost Function Parameters}

The values of the cost function parameters are listed below.

\begin{table}[H]
\centering
\scriptsize
\renewcommand{\arraystretch}{1.1}
\setlength{\tabcolsep}{4.5pt}
\caption{Cost Function Parameters}
\begin{tabular}{|l|l|p{6.5cm}|p{3.6cm}|}
\hline
\textbf{Symbol} & \textbf{Value} & \textbf{Description} & \textbf{Source} \\ \hline
$\kappa_{\text{init}}$ & 1000 & Number of initial fine-tuning rounds for algorithmic development & \cite{kadra2023scaling}, \cite{wistuba2022supervising}, \cite{Auto-Pytorch} \\ \hline
$\kappa_{\text{recur}}$ & 6 & Retraining frequency per year & \cite{thompson2024economic}\\ \hline
$d_{\text{GPU}}$ & 0.22 & GPU-specific annual price discount rate & \cite{hobbhahn2022trends} \\ \hline
$d$ & 0.05 &  Discount rate applied across the economy  & \cite{sullivan2023average} \\ \hline
$L$ & 5 & System operational lifespan in years & \cite{assets2003consumer} \\ \hline
$p_{\text{data}}$ & dynamic & Cost of collecting and labeling one datum & \cite{thompson2024economic}, \cite{singh2024benchmarkingobjectdetectorscoco}, \cite{russakovsky2015imagenetlargescalevisual}, \cite{lin2015microsoftcococommonobjects} \\ \hline
$F_{\text{GPU}}$ & 6 & FLOPs multiplier for training per token per parameter (accounts for forward + backward passes) & \cite{kaplan2020scaling} \\ \hline
$r_{\text{FLOP}}$ & $4 \times 10^{12}$ & FLOPs per GPU-hour & \cite{svanberg2024beyond} \\ \hline
$p_{\text{GPU}}$ & 0.34 & GPU hour cost for 4 FP-32 TFLOPS GPU on cloud & \cite{svanberg2024beyond} \\ \hline
$U_{\text{GPU}}$ & 0.4 & Average GPU utilization rate during training & \cite{megatronlm2024} \\ \hline
$Z_{\text{input}}$ & $256^2$ & Input image size in pixels & ImageNet image size and standard CV model input size \\ \hline
$C_{\text{impl}}$ & 2,144,475 & Initial implementation cost for system development & \cite{thompson2021demand} \\ \hline
$C_{\text{maint}}$ & 295,123 & Annual system maintenance cost & \cite{thompson2021demand} \\ \hline
\end{tabular}
\label{tab:cost_hyperparams}
\end{table}

\begin{table}[h]
\centering
\caption{Input prices under default parameter values (2024 USD).}
\label{tab:coef_values}
\vspace{0.5em}
\small
\renewcommand{\arraystretch}{2.2}
\setlength{\tabcolsep}{8pt}
\begin{tabular}{c|p{8.5cm}|c}
\toprule
\textbf{Coeff.} & \textbf{Expression} & \textbf{Value (USD)} \\ 
\midrule
$c_F$ &
$\displaystyle C_{\text{impl}} + \frac{1-\left(1/(1+d)\right)^{L}}{1-1/(1+d)}\,C_{\text{maint}}$
& $3{,}486{,}090$ \\ 
\addlinespace[0.5em]
$c_D$ &
$\displaystyle m\left[1+\kappa_{\text{recur}} \frac{1-\left(1/(1+d)\right)^{L}}{1-1/(1+d)}\right] p_{\text{data}}$
& $6.19$ \\ 
\addlinespace[0.5em]
$c_T$ &
$\displaystyle 6\,m\left[ \kappa_{\text{init}} + \kappa_{\text{recur}} \frac{1-\left(\frac{1}{(1+d)(1+d_{\text{GPU}})}\right)^{L}}{1-\frac{1}{(1+d)(1+d_{\text{GPU}})}}\right] \frac{p_{\text{GPU}}\,Z_{\text{input}}}{r_{\text{FLOP}}\,U_{\text{GPU}}}$
& $3.83\times 10^{-6}$ \\ 
\addlinespace[0.5em]
$c_I$ &
$\displaystyle m\left[ \frac{1-\left(\frac{1}{(1+d)(1+d_{\text{GPU}})}\right)^{L}}{1-\frac{1}{(1+d)(1+d_{\text{GPU}})}}\right] \frac{p_{\text{GPU}}}{U_{\text{GPU}}} \frac{40\times50\,\delta_i N_i}{\tau/\tau_{\mathrm{GPU}}}$
& $1.29\times 10^{-8}$ \\ 
\bottomrule
\end{tabular}
\end{table}

\section{Cost Minimization First-Order Conditions}
\label{app:cost_minimization_first}
To find the first-order conditions for the AI cost minimization problem in Equation~\eqref{eq:define_cost_minimization}, the Lagrangian is:

\begin{equation*}
\mathcal{L} = \kappa_i(D_i, T_i, M_i, I_i) 
+ \lambda_1 \left[ a_i - Q(D_i, T_i, M_i; m_i, n_i) \right]
+ \lambda_2 \left[ Y_i - \frac{I_i}{\tau_{\text{GPU}} M_i} \right]
\end{equation*}

\noindent where $\lambda_1 \geq 0$ is the Lagrange multiplier on the quality constraint and $\lambda_2 \geq 0$ is the Lagrange multiplier on the inference capacity constraint.

\subsection{Partial Derivatives of the Lagrangian}

The first-order condition with respect to $D_i$ (Data) is:
\begin{equation*}
\frac{\partial \mathcal{L}}{\partial D_i} = \frac{\partial \kappa_i}{\partial D_i} - \lambda_1 \frac{\partial Q}{\partial D_i} = 0
\end{equation*}
and rearranging:
\begin{equation*}
\frac{\partial \kappa_i}{\partial D_i} = \lambda_1 \frac{\partial Q}{\partial D_i}
\end{equation*}
The marginal cost of data is the shadow price of quality times the marginal product of data.

The first-order condition with respect to $T_i$ (Training Steps) is:
\begin{equation*}
\frac{\partial \mathcal{L}}{\partial T_i} = \frac{\partial \kappa_i}{\partial T_i} - \lambda_1 \frac{\partial Q}{\partial T_i} = 0
\end{equation*}
and rearranging:
\begin{equation*}
\frac{\partial \kappa_i}{\partial T_i} = \lambda_1 \frac{\partial Q}{\partial T_i}
\end{equation*}
The marginal cost of training steps is the shadow price of quality times the marginal product of training steps.

The first-order condition with respect to $M_i$ (Model Size) is:
\begin{equation*}
\frac{\partial \mathcal{L}}{\partial M_i} = \frac{\partial \kappa_i}{\partial M_i} - \lambda_1 \frac{\partial Q}{\partial M_i} + \lambda_2 \frac{I_i}{\tau_{\text{GPU}} M_i^2} = 0
\end{equation*}
and rearranging:
\begin{equation*}
\frac{\partial \kappa_i}{\partial M_i} = \lambda_1 \frac{\partial Q}{\partial M_i} - \lambda_2 \frac{I_i}{\tau_{\text{GPU}} M_i^2}
\end{equation*}
The marginal cost of model size is the shadow price of quality times the marginal product of model size minus the shadow price of inference capacity times the effect of model size on inference capacity. Larger models reduce inference capacity per GPU hour, hence the negative term.

The first-order condition with respect to $I_i$ (Inference Compute) is:
\begin{equation*}
\frac{\partial \mathcal{L}}{\partial I_i} = \frac{\partial \kappa_i}{\partial I_i} - \frac{\lambda_2}{\tau_{\text{GPU}} M_i} = 0
\end{equation*}
and rearranging:
\begin{equation*}
\frac{\partial \kappa_i}{\partial I_i} = \frac{\lambda_2}{\tau_{\text{GPU}} M_i}
\end{equation*}
The marginal cost of inference compute is the shadow value of inference capacity times the marginal contribution of inference to capacity.

\subsection{Partial Derivatives of the Cost Function}

From Equation~\eqref{eq:cost} in Appendix~\ref{app:cost}, compute partial derivatives of $\kappa$:
\begin{equation*}
\kappa(D, T, M, I; m) = \kappa^{\text{eng}} + m\left(\kappa^{\text{data}}(D) + \kappa^{\text{train}}(T, M) + \kappa^{\text{inf}}(M, I)\right)
\end{equation*}

From Equation~\eqref{eq:B2}:
\begin{equation*}
\kappa^{\text{data}} = p_{\text{data}} D + \sum_{t=1}^{L} \frac{\kappa_{\text{recur}}\, p_{\text{data}}\, D}{(1+d)^t}
\end{equation*}
\begin{equation*}
\frac{\partial \kappa_i}{\partial D_i} = m\, p_{\text{data}} \left[1 + \sum_{t=1}^{L} \frac{\kappa_{\text{recur}}}{(1+d)^t} \right]
\end{equation*}

From Equation~\eqref{eq:B4}:
\begin{equation*}
\kappa^{\text{train}}(T, M) = \left(\kappa^{\text{int}} + \sum_{t=1}^{L} \frac{\kappa_{\text{recur}}}{(1+d_{\text{GPU}})^t (1+d)^t}\right) \cdot \frac{p_{\text{GPU}}}{r_{\text{FLOP}}\, U_{\text{GPU}}} \cdot F_{\text{GPU}}\, Z_{\text{input}}\, M T
\end{equation*}
\begin{equation*}
\frac{\partial \kappa_i}{\partial T} = m \left(\kappa^{\text{int}} + \sum_{t=1}^{L} \frac{\kappa_{\text{recur}}}{(1+d_{\text{GPU}})^t (1+d)^t}\right) \cdot \frac{p_{\text{GPU}}}{r_{\text{FLOP}}\, U_{\text{GPU}}} \cdot F_{\text{GPU}}\, Z_{\text{input}}\, M
\end{equation*}
\begin{equation*}
\frac{\partial \kappa_i}{\partial T} 
= m \cdot c_T \cdot M
\end{equation*}
where $c_T$ is the effective training cost coefficient. Model size appears in both training and inference costs:
\begin{equation*}
\frac{\partial \kappa_i}{\partial M} = m \left[\frac{\partial \kappa^{\text{train}}}{\partial M} + \frac{\partial \kappa^{\text{inf}}}{\partial M}\right]
\end{equation*}

From the training component:
\begin{equation*}
\frac{\partial \kappa^{\text{train}}}{\partial M} = c_T \cdot T
\end{equation*}

From the inference component (Equation~\eqref{eq:B5}):
\begin{equation*}
\kappa^{\text{inf}}(M, Y) = \sum_{t=1}^{L} \frac{1}{(1+d_{\text{GPU}})^t (1+d)^t} \cdot \frac{p_{\text{GPU}}}{U_{\text{GPU}}} \cdot I
\end{equation*}
But $I = \tau_{\text{GPU}} M Y$ (from the binding constraint), so:
\begin{equation*}
\frac{\partial \kappa^{\text{inf}}}{\partial M} = \sum_{t=1}^{L} \frac{1}{(1+d_{\text{GPU}})^t (1+d)^t} \cdot \frac{p_{\text{GPU}}}{U_{\text{GPU}}} \cdot \tau_{\text{GPU}}\, M Y
= c_I \cdot Y
\end{equation*}
where $c_I$ is the inference cost coefficient. Therefore:
\begin{equation*}
\frac{\partial \kappa_i}{\partial M} = m\left[c_T \cdot T + c_I \cdot Y\right]
\end{equation*}

From Equation~\eqref{eq:B5}:
\begin{equation*}
\kappa^{\text{inf}}(M, Y) = \sum_{t=1}^{L} \frac{1}{(1+d_{\text{GPU}})^t (1+d)^t} \cdot \frac{p_{\text{GPU}}}{U_{\text{GPU}}} \cdot I
\end{equation*}

\begin{equation*}
\frac{\partial \kappa^{\text{inf}}}{\partial I} = \sum_{t=1}^{L} \frac{1}{(1+d_{\text{GPU}})^t (1+d)^t} \cdot \frac{p_{\text{GPU}}}{U_{\text{GPU}}}
\end{equation*}

\begin{equation*}
\frac{\partial \kappa^{\text{inf}}}{\partial I} = m \cdot c_I
\end{equation*}

\subsection{Partial Derivatives of $Q$}

From the scaling law (Equation~\eqref{eq:scalinglaw}):
\begin{equation*}
\ln(\tilde{H}_{\text{AI}}) = \ln\!\left(\frac{\alpha}{D^a} + \frac{\beta}{T^b} + \frac{\sigma}{M^c} + G\right) + k
\end{equation*}
and $Q$ is related to accuracy through Equation~\eqref{eq:q_acc}:
\begin{equation*}
Q(D_i, T_i, M_i, m_i, n_i) = F^{-1}\!\left(\tilde{H}_{\text{AI}}(D, T, M, n);\, m_i, n_i\right)
\end{equation*}
Using the chain rule:
\begin{equation*}
\frac{\partial Q}{\partial D} = \frac{\partial F^{-1}}{\partial \tilde{H}_{\text{AI}}} \cdot \frac{\partial \tilde{H}_{\text{AI}}}{\partial D}
\end{equation*}
From the scaling law:
\begin{equation*}
\frac{\partial \tilde{H}_{\text{AI}}}{\partial D} = \tilde{H}_{\text{AI}} \cdot \frac{-\alpha a / D^{a+1}}{\alpha/D^a + \beta/T^b + \sigma/M^c}
\end{equation*}
Since $F$ is decreasing in accuracy (higher accuracy results from lower entropy):
\begin{equation*}
\frac{\partial F^{-1}}{\partial \tilde{H}_{\text{AI}}} < 0
\end{equation*}
Therefore 
\begin{equation*}
\partial Q / \partial D > 0
\end{equation*}
Increased data results in higher accuracy. 
By similar logic, 
\begin{equation*}
\partial Q / \partial T > 0
\end{equation*}
More training steps also results in higher accuracy. 
And 
\begin{equation*}
\partial Q / \partial M > 0
\end{equation*}
larger models also provide higher accuracy. 

\subsection{Solve for Shadow Prices}

Assuming both constraints bind, the system of first-order conditions gives:

\noindent For data:
\begin{equation*}
m\, p_{\text{data}} \left[1 + \sum_{t=1}^{L} \frac{\kappa_{\text{recur}}}{(1+d)^t}\right] = \lambda_1 \frac{\partial Q}{\partial D}
\end{equation*}

\noindent For training steps:
\begin{equation*}
m\, c_T\, M = \lambda_2 \frac{\partial Q}{\partial T}
\end{equation*}

\noindent For model size:
\begin{equation*}
m\left[c_T \cdot T + c_I \cdot Y\right] = \lambda_1 \frac{\partial Q}{\partial M_i} - \lambda_2 \frac{I_i}{\tau_{\text{GPU}} M_i^2}
\end{equation*}
With $I = \tau_{\text{GPU}} M Y$:
\begin{equation*}
m\left[c_T \cdot T + c_I \cdot Y\right] = \lambda_1 \frac{\partial Q}{\partial M_i} - \lambda_2 \frac{Y}{M}
\end{equation*}

\noindent For inference:
\begin{equation*}
m \cdot c_I = \frac{\lambda_2}{\tau_{\text{GPU}} M}
\end{equation*}
From the inference first-order condition:
\begin{equation*}
\lambda_2 = m \cdot c_I \cdot \tau_{\text{GPU}} M
\end{equation*}
The shadow price of requiring higher inference capacity, $\lambda_2 = \partial\kappa/\partial Y$, is the cost of producing one more unit of task output. Substituting into the model size first-order condition and solving for $\lambda_1$:
\begin{equation*}
\lambda_1 = \frac{m}{\partial Q/\partial M_i} \left\{ \left[c_T \cdot T + c_I \cdot Y\right] + c_I \cdot \tau_{\text{GPU}} M \cdot \frac{I_i}{\tau_{\text{GPU}} M_i^2} \right\}
\end{equation*}
The shadow price of achieving higher accuracy, $\lambda_1 = \partial\kappa/\partial a_i$, is the increase in total cost per unit increase in required accuracy --- the ``marginal cost'' that appears in the Stage~2 optimization (Equation~\eqref{eq:D6}).

The optimality conditions for input ratios combining first-order conditions yields optimal input ratios. From the first-order conditions for data and training steps:

\begin{equation*}
\frac{\partial Q/\partial D}{\partial Q/\partial T} = \frac{p_{\text{data}}D\left[1 + \sum_{t=1}^{L} \dfrac{\kappa_{\text{recur}}}{(1+d)^t}\right]}{c_T\, M}
\end{equation*}

\noindent Marginal rate of technical substitution between data and training steps equals their cost ratio. From the first order conditions for training steps and model size. More complex due to inference constraint interaction, but principle is same: equalize marginal product per dollar across inputs.

\subsection{Complete System of First-Order Conditions}

The complete system of first-order conditions is:
\begin{equation*}
\frac{\partial \kappa}{\partial D} = \lambda_1 \frac{\partial Q}{\partial D}
\end{equation*}
\begin{equation*}
\frac{\partial \kappa}{\partial T} = \lambda_1 \frac{\partial Q}{\partial T}
\end{equation*}
\begin{equation*}
\frac{\partial \kappa}{\partial M} = \lambda_1 \frac{\partial Q}{\partial M} - \lambda_2 \frac{Y}{M}
\end{equation*}
\begin{equation*}
\frac{\partial \kappa}{\partial I} = \frac{\lambda_2}{\tau_{\text{GPU}}\, M}
\end{equation*}
\begin{equation*}
Q(D, T, M, m, n) = a_i
\end{equation*}
\begin{equation*}
I = \tau_{\text{GPU}} \cdot M \cdot Y
\end{equation*}
\begin{equation*}
\lambda_1,\, \lambda_2 \geq 0
\end{equation*}
These seven equations (plus non-negativity) determine the six unknowns $\{D^*, T^*, M^*, I^*, \lambda_1, \lambda_2\}$. The solution characterizes the cost-minimizing way to build an AI system that achieves accuracy $a_i$ and can process $Y_i$ task outputs.

\section{Task Level Optimization in AI Adoption}
\label{app:firm_backward_induction}

Before proceeding, we introduce four terms that will be used throughout the analysis. Total Benefit (TB) refers to the labor cost savings generated by automation. Total Cost (TC) denotes the cost of deploying an AI system with accuracy $a_i$. Marginal Benefit (MB) refers to the incremental labor-saving gain from a marginal improvement in AI performance, and Marginal Cost (MC) denotes the incremental increase in AI-system cost associated with such an improvement. These definitions will allow us to characterize the firm's optimization problem using standard cost-benefit analysis.

\subsection{Stage Two Optimization}

We begin by analyzing the firm's choice of the optimal AI quality $a_i$ conditional on adopting AI ($s_i = 1$) in stage one. To elucidate the structure of the firm's cost minimization problem, we rewrite Equation~\eqref{eq:ai_cost_coef} as:

\begin{equation}
C_i(1, a_i, Y_i) = w_i \tau_i Y_i - \text{TB}_i(a_i, Y_i) + \text{TC}_i(a_i, Y_i)
\label{eq:D1}
\end{equation}

\noindent where total cost consists of the baseline labor cost $w_i \tau_i Y_i$, net of automation benefits and plus automation costs. The term

\begin{equation}
\text{TB}_i(a_i, Y_i) = \delta_i\, r_i(a_i)\, w_i \tau_i Y_i
\label{eq:D2}
\end{equation}

\noindent captures the labor cost savings enabled by the AI system---interpreted as the automation benefit---where $\delta_i$ denotes the proportion of the task technically feasible to be automated and $r_i(a_i)$ the labor substitution ratio at AI quality level $a_i$. The cost of implementing AI, denoted by

\begin{equation}
\text{TC}_i(a_i, Y_i) = \kappa_i(a_i, Y_i)
\label{eq:D3}
\end{equation}

\noindent represents the total expenditure associated with deploying an AI system of quality $a_i$ to produce output $Y_i$. The firm's optimization problem in Stage two is therefore:

\begin{equation}
\max \left[ \text{TB}_i(a_i, Y_i) - \text{TC}_i(a_i, Y_i) \right]
\label{eq:D4}
\end{equation}

Equation~\eqref{eq:entropy_accuracy} establishes a monotonic transformation between accuracy $a$ and cross-entropy loss $\tilde{H}$. Since cross-entropy loss aligns more linearly with labor time, we recast the optimization problem using cross-entropy loss as the choice variable. The reformulated problem is:

\begin{equation}
\max_{a_i} \left[ \text{TB}_i\!\left(a_i(\tilde{H}_{\text{AI},i}), Y_i\right) - \text{TC}_i\!\left(a_i(\tilde{H}_{\text{AI},i}), Y_i\right) \right]
\label{eq:D5}
\end{equation}

\noindent The first-order condition over $\tilde{H}_{\text{AI},i}$ is:

\begin{equation}
\text{MB}_{-\tilde{H}} = \text{MC}_{-\tilde{H}}, \qquad \text{where} \quad
\text{MB}_{-\tilde{H}} = \frac{\partial\, \text{TB}_i}{\partial(-\tilde{H}_{\text{AI},i})}, \quad
\text{MC}_{-\tilde{H}} = \frac{\partial\, \text{TC}_i}{\partial(-\tilde{H}_{\text{AI},i})}
\label{eq:D6}
\end{equation}

\noindent For ease of interpretation, we introduce a negative sign to $\tilde{H}_{\text{AI},i}$ such that increases in the AI system's quality correspond to decreases in the AI model's cross-entropy loss $\tilde{H}_{\text{AI},i}$.

The sufficient conditions for an interior solution are:
\begin{equation*}
\frac{\partial\, \text{TB}_i}{\partial(-\tilde{H}_{\text{AI},i})} = \frac{\partial\, \text{TC}_i}{\partial(-\tilde{H}_{\text{AI},i})}
\qquad \text{and} \qquad
\frac{\partial^2\, \text{TB}_i}{\partial(-\tilde{H}_{\text{AI},i})^2} < \frac{\partial^2\, \text{TC}_i}{\partial(-\tilde{H}_{\text{AI},i})^2}
\end{equation*}
Because TB is piecewise linear in entropy and MB is piecewise constant while TC is convex due to AI scaling laws and MC is increasing, an interior solution arises when the MB curve intersects the rising MC curve before reaching $a_{\text{req}}$. See Figure \ref{fig:partial_equilibrium}, panel~2 (partial automation).

Conversely, the sufficient conditions for a corner solution at $a = a_{\text{req}}$ are MB $\geq$ MC throughout the interval. Equivalently:
\begin{equation*}
\frac{\partial\, \text{TB}_i}{\partial(-\tilde{H}_{\text{AI},i})} \geq \frac{\partial\, \text{TC}_i}{\partial(-\tilde{H}_{\text{AI},i})} \qquad \forall\; a_i \leq a_{\text{req}}
\end{equation*}
When the marginal benefit of improving accuracy exceeds the marginal cost throughout, it is optimal to fully automate to the required accuracy level. See Figure \ref{fig:partial_equilibrium}, panel~1 (full automation).

There is also a corner solution for no automation when at $a = a_{\text{req}}$, MB $\leq$ MC. Even at the lowest feasible accuracy, the cost of minimal AI exceeds benefits. See Figure \ref{fig:partial_equilibrium}, panel~3 (no automation).

\subsection{Stage One Optimization}

The stage one optimization problem is straightforward. The firm compares two options: (i) adopting the AI system and incurring the minimized cost from the stage two optimization; or (ii) not adopting AI and bearing the baseline labor cost. The firm chooses the option that yields the lower total cost. Denote $\tilde{H}_i^*$ and $a_i^*$ as the optimal cross-entropy loss and optimal accuracy of the AI system, respectively. The second step of the backward induction is to solve for the optimal $s_i$, formally:

\begin{equation}
s_i^* = \begin{cases} 1 & \text{if } C_i(1, a_i, Y_i) < C_i(0, Y_i) \\ 0 & \text{otherwise} \end{cases}
\label{eq:D7}
\end{equation}

To support the subsequent analysis, we formally define a set of terms related to the feasibility and optimality of task automation. These definitions clarify under what conditions automation may be considered possible or desirable from a cost-minimization perspective.

\begin{itemize}
    \item \textbf{Full automation is feasible} if the cost of deploying an AI system that delivers performance at the required level is less than or equal to the baseline cost of performing the task entirely with human labor:
    \begin{equation*}
        C_i(1, a_{\text{req},i}, Y_i) \leq C_i(0, Y_i)
    \end{equation*}

    \item \textbf{Partial automation is feasible} if there exists an AI system accuracy level $a_i \leq a_{\text{req},i}$ such that the total cost of performing the task through collaboration between human workers and an AI system $a_i$ is less than or equal to the baseline cost of performing the task entirely with human labor:
    \begin{equation*}
        \exists\; a_i \leq a_{\text{req},i} \quad \text{such that} \quad C_i(1, a_i, Y_i) \leq C_i(0, Y_i)
    \end{equation*}

    \item \textbf{Full automation is optimal} if full automation is feasible and the cost-minimizing accuracy $a_i^*$ satisfies:
    \begin{equation*}
        a_i^* \geq a_{\text{req},i}
    \end{equation*}

    \item \textbf{Partial automation is optimal} if partial automation is feasible and the cost-minimizing accuracy $a_i^*$ satisfies:
    \begin{equation*}
        a_i^* < a_{\text{req},i}
    \end{equation*}
\end{itemize}

Figure \ref{fig:partial_equilibrium} presents illustrative marginal benefit--marginal cost (MB--MC) and total cost--total benefit (TC--TB) curves under three distinct automation scenarios. The horizontal axis represents improvements in AI system performance, corresponding to reductions in cross-entropy loss. Moving to the right on the axis thus indicates better AI model accuracy and performance.

To the left of the required performance threshold, according to Equation~\eqref{eq:r} and Equation~\eqref{eq:D2}, $\text{TB}_i$ is piecewise linear with respect to $H_{\text{AI},i}$, and hence the MB curves are piecewise constant. Panel~1 illustrates the full automation case. In this scenario, the marginal cost (MC) of improving AI performance lies below the MB curve throughout the entire interval between random-guess cross-entropy loss and the required threshold. The two curves may intersect only beyond the required level of performance. Hence, it is optimal to adopt full automation, and the AI system achieves performance equal to or exceeding that of human workers.

Panel~2 represents the case of partial automation. In this case, the MB and MC curves intersect before reaching the required performance threshold. The total task workload is captured by the horizontal distance between the random-guess and required cross-entropy loss levels (denoted as segment $b$). The portion of the task completed by the AI system corresponds to the distance between the random-guess level and the AI system's optimal cross-entropy loss (segment $a$). The remaining portion, from the AI's optimal performance to the required threshold, is completed by humans. The labor substitution ratio is therefore given by the ratio $a/b$.

Panel~3 shows the no automation scenario. In this case, the total cost curve remains above the total benefit curve across the entire performance spectrum, rendering AI adoption infeasible at any level of model performance. This outcome reflects the presence of substantial costs in developing AI systems. Even if MB were to exceed MC marginally at certain levels, the fixed cost may still make adoption suboptimal. The fixed cost primarily arises from assembling and maintaining an engineering team to develop domain-specific AI models. Additional details on the cost parameters used in our analysis---including fixed costs of developing and deploying AI systems---can be found in Appendix~\ref{app:cost}. For related modeling assumptions and empirical estimates of such costs, see also \citet{svanberg2024beyond}.

\subsection{Aggregating Occupation-Level Benefits from Labor Saving }
We define the benefits of labor savings at the occupation level by aggregating benefits at the task level in specific automation scenarios. Suppose that occupation $j$ comprises a set of computer vision tasks indexed by $i$. For each task $i$, we calculate the task-level benefit $TB_i$ under its optimal automation choice.
Let $Z$ denote a particular automation condition (e.g., full automation is optimal). We identify the subset of tasks within occupation $j$ that satisfy condition $Z$. The aggregate benefit under condition $Z$ is defined as:
\begin{equation*}
\text{TB}_j^Z = \sum_{i \in Z} \text{TB}_i
\end{equation*}

\noindent This formulation allows us to calculate, for each occupation, the total benefit from labor saving across all computer vision--related tasks under various automation conditions.

To facilitate comparative analysis, we report the following three occupation-level labor-saving scenarios in the results section:

\begin{itemize}
    \item Full automation is both feasible and optimal;
    \item Both full and partial automation are feasible, but partial automation is optimal;
    \item Full automation is not feasible, but partial automation is both feasible and optimal.
\end{itemize}

Figure~\ref{fig:partial_equilibrium} also illustrates how labor compensation associated with all computer vision--related tasks in a given occupation is distributed across different automation outcomes. The 100\% reference point represents the total labor compensation that would be incurred if no AI-based automation were applied.

The red segment corresponds to the share of compensation attributable to tasks that should be fully automated. The green segment reflects the labor cost savings achieved through partial automation---i.e., tasks for which AI systems substitute for part, but not all, of the human input.

The grey segment represents the portion of compensation that is not substituted by AI. This residual can be further decomposed into three components: (i) the proportion of a task that is technically non-automatable (e.g., the $(1 - \delta_i)$ portion); (ii) tasks for which the optimal decision is not to automate, despite being technically feasible; and (iii) tasks that are optimally partially automated but still require human input (e.g., the residual $(1 - r_i(a_i^*))$ portion). Taken together, components (ii) and (iii) account for the portion of the automatable share that is not replaced by AI, which can be summarized compactly as $(1 - r_i(a_i^*))\,\delta_i$; that is, within the automatable part of a task, the residual share corresponds to $1 - r_i(a_i^*)$.

\section{Functional Mapping from Accuracy to Cross-Entropy Loss}
\label{app:scaling_entropy_acc}

We utilize the data from the scaling law experiment described in Section \ref{sec:data} to estimate the relationship between cross-entropy loss and accuracy (Equation~\eqref{eq:entropy_accuracy}). 

The function form is specified as
\begin{equation}
\begin{aligned}
\tilde{H}_{AI} =\ & \beta_0 + \beta_1\,a + \beta_2\,a^2 + \beta_3\,a^3 \\
&+ \gamma_1\,a\,\log(a ) + \gamma_2\,(1-a)\,\log(1-a ) \\
&+ \delta_1\,a\,\log(n) + \delta_2\,\log(n) + \delta_3\,\frac{1}{n}
\end{aligned}
\label{eq:entropy_acc_func}
\end{equation}
which is then estimated via OLS. Note that, similar to the case in the scaling law function, we allow this relationship to be adjusted based on task complexity $n$.

Table~\ref{tab:entropy_params} presents the fitted parameters, highlighting the quantitative relationship between the two metrics in our setting.

\renewcommand{\arraystretch}{1.5}

\begin{table}[htbp]
\centering
\caption{Fitted Coefficients for Mapping Accuracy to Cross-Entropy Loss.}
\label{tab:entropy_params}
\footnotesize
\begin{tabular}{l|ccccccccc|c}
\toprule
& $\beta_0$ & $\beta_1$ & $\beta_2$ & $\beta_3$ & $\gamma_1$ & $\gamma_2$ & $\delta_1$ & $\delta_2$ & $\delta_3$ & $R^2_{\mathrm{fitted}}$ \\
\midrule
\textbf{Coefficient} & 2.86 & 14.22 & -27.10 & 10.48 & 11.99 & -1.82 & -0.70 & 0.61 & -0.62 & 0.98 \\
\textbf{Std. Error} & 0.06 & 0.82 & 1.74 & 0.96 & 0.52 & 0.29 & 0.01 & 0.01 & 0.05 &  \\
\bottomrule
\end{tabular}
\end{table}

The parameters were estimated using ordinary least squares linear regression on a feature matrix constructed from the accuracy 
and the number of classes, with additional polynomial and logarithmic transformations to capture nonlinear relationships. After fitting the model, the covariance matrix of the estimated coefficients was computed as 
\[
\mathrm{Cov}(\hat{\beta}) = \sigma^2 (X^\top X)^{-1},
\]
where \(\sigma^2\) is the error variance, i.e., estimated by the residual sum of squares divided by the degrees of freedom. The standard errors of the coefficients were then derived as the square roots of the diagonal elements of this covariance matrix, providing a quantification of the uncertainty in each parameter estimate. The model achieved an \(R^2\) score of 0.98, indicating an excellent fit between the predicted and observed cross-entropy loss values.

\section{Required Accuracy in Multiple Classification Tasks}
\label{app:a_and_m}

When $m > 1$, i.e., an O*NET task involves multiple vision classification tasks, the required accuracy for each classification task, $a_{req,i}$ should be discounted from the O*NET task-level required accuracy, denoted as $a'_{req,i}$, as follows
\begin{equation}
    1-a_{req, i} =  \frac{1-a'_{req,i}}{m_i}.
\end{equation}
This is because, assuming the probability of making an error for each classification task is independent, if each classification task has an accuracy of $a$, then the O*NET task would have an accuracy of $a^{m_i}$. If $a$ is close to one, then we can apply the first-order Taylor approximation: $a^{m_i} = (1 - (1-a))^{m_i} \approx 1 - m_i(1-a)$. In other words, the aggregated error rate of all the classification tasks of an O*NET task is $m_i$ times that of each classification task.

\section{Firm Size Estimation}
\label{app:firm_size}

The firm-level analysis requires estimating the occupation-specific firm size distributions, i.e., for each occupation, what is the distribution of the number of employees with this distribution across all US firms. Formally, denote $q^{occ}_o(l)$ as the firm size distribution for occupation $o$: $q^{occ}_o(3) = 500$ means there are 500 firms with 3 employees with occupation $o$.

The detailed data for $q^{occ}_o(l)$ are not available. However, the data of firm size distributions for each NAICS-4-digit sub-sector can be obtained from Business Dynamic Statistics\footnote{\url{https://www.census.gov/programs-surveys/bds.html}}. In what follows, we will describe how we estimated the $q^{occ}_o(l)$ from sub-sector firm size distributions.

\subsection{Estimating Occupation-Specific Firm Size Distributions}

Denote the subsector-level firm size distribution as $q^{4d}_i(l)$: $q^{4d}_i(100) = 500$ means there are 500 firms with 100 employees working in sub-sector $i$. We made an assumption that the same NAICS-4-digit sector shares the same proportion of employees working in each occupation. Hence, the proportion of occupation $o$ within sub-sector $i$ can be estimated by $l_{oi} / l_i$, where $l_{oi}$ denotes the total number of employees in sub-sector $i$ and occupation $o$, and $l_i$ denotes the total number of employees in sub-sector $i$. Accordingly, $q^{occ}_o(l)$ can be estimated as
\begin{equation}
    q^{occ}_o(l) = \sum_i q^{4d}_i\left(\frac{l_i}{l_{oi}} \cdot l\right).
\end{equation}

\subsection{Firm Size Imputation}

BDS only provides the firm size distribution at coarse bins: 1-4, 5-9, 10-19, 20-99, 100-499, 500-999, 1,000-2,499, 2,500-4,999, 5,000-9,999, and 10,000+ employees, respectively. To obtain finer-grained firm-size distributions, we adopt the firm size imputation method in \citep{svanberg2024beyond}. Except for the 10,000+ bin, each bin is divided into 10 sub-bins uniformly on the logarithmic scale. The number of firms within each coarse bin is evenly divided into its 10 sub-bins. For the 10,000+ bin, we perform the Zipf law extrapolation --- where we assume a linear decay, in the log-log scale, in $q^{4d}_i(l)$ with a slope of $-1$. The extrapolation is cut off at the largest firm size in the US.

\section{Prompt for Computer Vision Task Characteristics}
\label{app:prompt}

\begin{lstlisting}


You will be given a job, a specific task of the job, and a direct working 
activity (DWA) involved in the specific task. The definitions of task, job, and DWA come from the ONET dataset. The given activity involves or partially involves an image classification problem. Your goal is to identify the classification problem, as well as the number of classes and classification tasks involved in each decision in the given problem.

### Requirements:
1. **Show your reasoning process** and provide examples of your identified classes and tasks before giving your final answer.
2. **Output must contain exactly 6 fields**:
   - "thought_process"
   - "example_tasks"
   - "example_classes"
   - "number_of_classification_tasks"
   - "number_of_classes_per_task"
   - "proportion_of_vision_tasks"
3. **In the 'thought_process' field**, define each classification task and derive the number of tasks/classes from the definition. Rate the importance score of these vision tasks, from 1-10.  
   - **If the DWA includes tasks that require other cognitive or motor skills to achieve the same goal, list these as alternative tasks.**  
   - **Ensure all major non-vision-based skills involved in this DWA are considered.**  
   - **Do not stop at one alternative task - consider multiple reasonable ways the DWA is performed without vision.**  
   - **If no alternative tasks are directly relevant to the DWA, do not list any.**  
4. **In the 'example_classes' and 'example_tasks' fields**, enumerate a full or partial list:
   - If there are **fewer than 20**, list all.
   - If there are **more than 20**, list a subset of 20.
5. **For 'number_of_classification_tasks' and 'number_of_classes_per_task' fields**:
   - Always output a **list of two integers** representing the minimum and maximum estimates.
   - If the estimate has no uncertainty, the minimum and maximum can be the same.
   - If greater than 20, provide an incomplete list of 20 examples.
6. **For 'proportion_of_vision_tasks'**:
   - Compute the ratio of the importance score of the vision tasks over the sum of the importance scores of all the listed tasks under this DWA in the thought process.
   - **Alternative tasks must be directly related to the DWA.** Do not include unrelated activities.
   - **Alternative tasks must primarily involve different skills applied to the same DWA goal.**
   - **Ensure a complete list of relevant alternative tasks is provided.**
   - **If no strong alternative tasks exist, vision tasks should take up 100%

---

### Examples:

#### Example 1:
**Job:** Light truck drivers  
**Task:** Inspect and maintain vehicle supplies and equipment, such as gas, oil, water, tires, lights, or brakes, to ensure that vehicles are in proper working condition.  
**DWA:** Inspect motor vehicles.

**Expected JSON output:**
```json
{
 "thought_process": ["This activity involves checking multiple parts of the vehicle and deciding whether each part is functioning or not. Therefore, it can be cast as multiple binary classification tasks. Each binary decision involves deciding whether a specific part is working or not. The number of tasks is equal to the number of vehicle parts to be inspected. Since the number of vehicle parts ranges from 6 to 20, so does the number of tasks. The importance score for these vision tasks is **8**. Alternative tasks include: (1) Listening to engine sounds to detect issues without a visual inspection (importance score: 1), (2) Feeling tire pressure manually without relying on visual readings (importance score: 1)."],
 "example_tasks": ["Whether gas is in working condition", "Whether oil is in working condition", "...", "Whether washer fluid level is sufficient"],
 "example_classes": ["working", "not working"],
 "number_of_classification_tasks": [6, 20],
 "number_of_classes_per_task": [2, 2],
 "proportion_of_vision_tasks": [8/10]  // Vision tasks take 8 out of 10 importance points
}

---

#### Example 2:
**Job:** Zoologists and Wildlife Biologists  
**Task:** Analyze the characteristics of animals to identify and classify them.  
**DWA:** Examine characteristics or behavior of living organisms.

**Expected JSON output:**
{
 "thought_process": ["The classification task is to classify animals based on their observed characteristics. This requires visual examination of multiple features, such as species type, coloration, body shape, and behavior. Therefore, it contains only one classification task. The number of classes should equal the number of distinct species in the inventory, which typically ranges between 20 to 500. The importance score for these vision tasks is **10**, as the entire DWA consists of examining and classifying organisms. Since there are no significant alternative tasks that do not involve vision, the proportion of vision tasks is 100%
 "example_tasks": ["What type of animal the target animal belongs to."],
 "example_classes": ["Lion", "Tiger", "Elephant", "Giraffe", "Zebra", "Kangaroo", "Panda", "Cheetah", "Leopard", "Rhinoceros", "Hippopotamus", "Crocodile", "Alligator", "Gorilla", "Orangutan", "Koala", "Sloth", "Wolf", "Fox", "Bald Eagle"],
 "number_of_classification_tasks": [1,1],
 "number_of_classes_per_task": [20,500],
 "proportion_of_vision_tasks": [1.0]  // 100%
}

---

#### Example 3:
**Job:** Teachers, Secondary School  
**Task:** Evaluate and grade students' work in class.  
**DWA:** Assess educational performance of students.

**Expected JSON output:**
{
 "thought_process": ["The classification task involves assessing students' performance, but only some of these assessments involve vision-based classification. Specifically, grading image-based problems - such as diagrams, geometric drawings, or labeled scientific illustrations - requires visual recognition and classification. Other grading activities, such as evaluating numerical problems or essay responses, rely on reasoning or language comprehension instead of visual classification and should be treated as alternative tasks. The importance score for grading image-based problems is **4**. Alternative grading tasks include: (1) Evaluating numerical responses using reasoning skills (importance score: 5), (2) Assessing essays based on language comprehension (importance score: 3), (3) Entering grades into a digital system without requiring classification (importance score: 3)."],
 "example_tasks": ["Classify the accuracy of a student's labeled diagram in Biology", "Classify a student's hand-drawn geometric proof in Mathematics", "Classify a student's artistic composition in Art class"],
 "example_classes": ["Excellent", "Good", "Satisfactory", "Needs Improvement", "Fail"],
 "number_of_classification_tasks": [1,3],
 "number_of_classes_per_task": [5,10],
 "proportion_of_vision_tasks": [4/15]  // Vision tasks take 4 out of 15 importance points
}

---

Below is the actual task you need to label:
Job: {JOB}
Task: {TASK}
DWA: {DWA}


Provide the output in strict JSON format as follows:
{{
 "thought_process": ["Your detailed reasoning here."],
 "example_tasks": ["Example task 1", "Example task 2", "...", "Example task 20"],
 "example_classes": ["Example class 1", "Example class 2", "...", "Example class 20"],
 "number_of_classification_tasks": [minimum number of tasks, maximum number of tasks],
 "number_of_classes_per_task": [minimum number of classes per task,maximum number of classes per task],
 "proportion_of_vision_tasks": ["..."]
}}


1. The JSON object must contain exactly these 5 fields:
  - "thought_process"
  - "example_tasks"
  - "example_classes"
  - "number_of_classification_tasks"
  - "number_of_classes_per_task"


2. Do not include any text or explanations outside of the JSON object.

\end{lstlisting}

\section{Detailed Analysis of the Scaling Law Production Function}

\subsection{Output Elasticity}

Following Equation~\eqref{eq:performance_elasticity} We calculate the performance elasticities for data, training steps, and model size, which are given by:

\begin{equation}
\epsilon_D = \frac{a\alpha \cdot e^k}{D^a \cdot \left(H_{task,i} - e^k \cdot \left(\frac{\alpha}{D^a} + \frac{\beta}{T^b} + \frac{\sigma}{M^c} + G\right)\right)}
\end{equation}

\begin{equation}
\epsilon_S = \frac{b\beta \cdot e^k}{S^b \cdot \left(H_{task,i} - e^k \cdot \left(\frac{\alpha}{D^a} + \frac{\beta}{T^b} + \frac{\sigma}{M^c} + G\right)\right)}
\end{equation}

\begin{equation}
\epsilon_M = \frac{c\sigma \cdot e^k}{M^c \cdot \left(H_{task,i} - e^k \cdot \left(\frac{\alpha}{D^a} + \frac{\beta}{T^b} + \frac{\sigma}{M^c} + G\right)\right)}
\end{equation}

\begin{table}[h]
\centering
\caption{Performance Elasticity Analysis Results}
\label{tab:performance_elasticity_analysis}
\footnotesize
\begin{tabular}{l>{\bfseries}c*{3}{p{2.2cm}}c}
\toprule
\textbf{Scenario} & \textbf{Performance} & \multicolumn{1}{c}{\textbf{Data}} & \multicolumn{1}{c}{\textbf{Training Steps}} & \multicolumn{1}{c}{\textbf{Model Size}} & \textbf{Total} \\
 & \textbf{($r$)} & \multicolumn{1}{c}{\textbf{($\epsilon_D$)}} & \multicolumn{1}{c}{\textbf{($\epsilon_S$)}} & \multicolumn{1}{c}{\textbf{($\epsilon_M$)}} & \textbf{Elasticity} \\
\midrule
\multicolumn{6}{c}{\textbf{2-Class Classification Task}} \\ \midrule
Small Scale (I)   & 0.804 & \multicolumn{1}{c}{0.010} & \multicolumn{1}{c}{0.046} & \multicolumn{1}{c}{0.046} & 0.102 \\
Medium Scale (II) & 0.911 & \multicolumn{1}{c}{0.009} & \multicolumn{1}{c}{0.021} & \multicolumn{1}{c}{0.007} & 0.037 \\
\midrule
\multicolumn{6}{c}{\textbf{5-Class Classification Task}} \\ \midrule
Small Scale (I)   & 0.866 & \multicolumn{1}{c}{0.016} & \multicolumn{1}{c}{0.035} & \multicolumn{1}{c}{0.032} & 0.083 \\
Medium Scale (II) & 0.960 & \multicolumn{1}{c}{0.013} & \multicolumn{1}{c}{0.016} & \multicolumn{1}{c}{0.006} & 0.034 \\
\midrule
\multicolumn{6}{c}{\textbf{10-Class Classification Task}} \\ \midrule
Small Scale (I)   & 0.860 & \multicolumn{1}{c}{0.016} & \multicolumn{1}{c}{0.040} & \multicolumn{1}{c}{0.034} & 0.089 \\
Medium Scale (II) & 0.961 & \multicolumn{1}{c}{0.012} & \multicolumn{1}{c}{0.018} & \multicolumn{1}{c}{0.006} & 0.036 \\
\midrule
\multicolumn{6}{c}{\textbf{50-Class Classification Task}} \\ \midrule
Small Scale (I)   & 0.771 & \multicolumn{1}{c}{0.015} & \multicolumn{1}{c}{0.080} & \multicolumn{1}{c}{0.056} & 0.151 \\
Medium Scale (II) & 0.925 & \multicolumn{1}{c}{0.009} & \multicolumn{1}{c}{0.032} & \multicolumn{1}{c}{0.012} & 0.052 \\
\midrule
\multicolumn{6}{c}{\textbf{100-Class Classification Task}} \\ \midrule
Small Scale (I)   & 0.695 & \multicolumn{1}{c}{0.015} & \multicolumn{1}{c}{0.122} & \multicolumn{1}{c}{0.078} & 0.215 \\
Medium Scale (II) & 0.894 & \multicolumn{1}{c}{0.007} & \multicolumn{1}{c}{0.044} & \multicolumn{1}{c}{0.017} & 0.068 \\
\midrule
\multicolumn{6}{c}{\textbf{500-Class Classification Task}} \\ \midrule
Small Scale (I)   & 0.351 & \multicolumn{1}{c}{0.023} & \multicolumn{1}{c}{0.544} & \multicolumn{1}{c}{0.283} & 0.849 \\
Medium Scale (II) & 0.751 & \multicolumn{1}{c}{0.006} & \multicolumn{1}{c}{0.112} & \multicolumn{1}{c}{0.045} & 0.163 \\
\midrule
\multicolumn{6}{c}{\textbf{1000-Class Classification Task}} \\ \midrule
Small Scale (I)   & 0.080 & \multicolumn{1}{c}{0.088} & \multicolumn{1}{c}{3.459} & \multicolumn{1}{c}{1.629} & 5.176 \\
Medium Scale (II) & 0.636 & \multicolumn{1}{c}{0.006} & \multicolumn{1}{c}{0.188} & \multicolumn{1}{c}{0.075} & 0.269 \\
\bottomrule
\end{tabular}

\begin{tablenotes}
\footnotesize
\item Notes: \(r\) represents the labor-substitution ratio.  
Scenario I (Small Scale): \(D = 25{,}000\), \(T = 200{,}000\), \(M = 250{,}000\).  
Scenario II (Medium Scale): \(D = 100{,}000\), \(T = 1{,}000{,}000\), \(M = 5{,}000{,}000\).  
\(D\) denotes data size, \(T\) training steps, and \(M\) model size.  
Performance elasticities are calculated at the respective input bundles.
\end{tablenotes}
\label{table:performance_full}
\end{table}

Table \ref{table:performance_full} reveals that, for a given bundle of inputs---data, compute, and model parameters---the attainable labor-substitution ratio $r$ differs systematically with task complexity, proxied here by the number of output classes ($n$). The direction of the relationship is shaped by two opposing forces.

First, greater task complexity raises the technical difficulty of the prediction problem. Holding the input bundle fixed, a more complex task requires a model with higher effective capacity to achieve the same accuracy. Because the bundle is not scaled up accordingly, model performance deteriorates as the number of classes $n$ rises, exerting a downward pressure on $r$.

Second, as $n$ increases, humans must spend more time or cognitive effort to complete the task unaided. This lengthens the benchmark completion time against which the AI's output is compared, enlarging the scope for automation. The result is an upward pressure on $r$.

Between the binary (2-class) and the modestly complex (10-class) problems, the second effect marginally outweighs the first: the potential time savings for humans rises faster than model accuracy falls, so $r$ edges upward. Beyond 10 classes, however---especially in the jump to 1,000 classes---the accuracy penalty dominates. Without commensurate increases in data, training steps, or parameter count, the labor-substitution ratio declines sharply with further increases in task complexity.

Across the small- and medium-sized computer-vision bundles we analyse, the overwhelming majority of configurations already lie in the region of diminishing returns to scale: the sum of input elasticities is less than one. Because more complex tasks require larger quantities of data, compute, and parameters to yield comparable performance, the slide into diminishing returns occurs more gradually as task complexity rises---the total elasticity falls toward unity at a slower pace for harder problems. The single exception is the 1 000-class task at the small-scale bundle, where total elasticity exceeds unity, indicating increasing returns. Once that inputs are upgraded to the medium-scale bundle, however, total elasticity falls below one, and the system quickly enters the diminishing-returns region.

\subsection{Elasticity of Substitution}
\label{sec:substitution_elasticity}

The pairwise (Allen--Uzawa) elasticities of substitution evaluated at any point \((D,T,M)\) holding the performance level fixed are

\begin{equation}\label{eq:elas_DS}
\sigma_{DT} \;=\;
\frac{(a+1)\,b\beta\,T^{b} \;+\; (b+1)\,a\alpha\,D^{a}}
     {b\beta\,T^{b} \;+\; a\alpha\,D^{a}},
\end{equation}

\begin{equation}\label{eq:elas_SM}
\sigma_{TM} \;=\;
\frac{(b+1)\,c\sigma\,M^{c} \;+\; (c+1)\,b\beta\,T^{b}}
     {c\sigma\,M^{c} \;+\; b\beta\,T^{b}},
\end{equation}

\begin{equation}\label{eq:elas_DM}
\sigma_{DM} \;=\;
\frac{(a+1)\,c\sigma\,M^{c} \;+\; (c+1)\,a\alpha\,D^{a}}
     {c\sigma\,M^{c} \;+\; a\alpha\,D^{a}}.
\end{equation}

Substituting the baseline bundle into Eqs.~\eqref{eq:elas_DS}--\eqref{eq:elas_DM} yields the values listed in Table~\ref{tab:elasticities}.

\subsection{Equation~\eqref{eq:performance_elasticity_substitution_ratio} Implementation}

\subsection*{Step 1: Data Loading (\texttt{main()})}

Three data sources are merged:
\begin{itemize}
    \item \texttt{BeyondAIExposure(Average\_industry)\_rev.xlsx} --- survey data with task characteristics (error rates, judgment frequency, etc.)
    \item \texttt{prompt\_engineering\_output.csv} --- task complexity information ($n_{\text{class}}$, $\text{num\_tasks}$)
    \item \texttt{oesmnaics\{granularity\}.csv} --- BLS employment and wage data at each industry level
\end{itemize}

\subsubsection*{Step 2: Merging (\texttt{map\_soc\_to\_occ()})}

Creates one row per (SOC Code $\times$ Task ID $\times$ DWA $\times$ NAICS) combination by joining the three sources. Also computes:
\begin{itemize}
    \item $n_{\text{class}}$, $\text{num\_tasks}$ --- from prompt engineering output
    \item $\tau_i = \text{dwa}/\text{occ} \times \text{importance\_score}$ --- fraction of work time spent on this visual task
\end{itemize}

\subsubsection*{Step 3: Row-level Extraction (\texttt{extract\_scenario\_data()})}

For each row, extracts:
\begin{itemize}
    \item $\text{accuracy} = 1 - \text{Q5 error rate}$ $\rightarrow$ required human-level accuracy
    \item $\tilde{\varepsilon}_{\text{rand}} = \text{Q7 random guess error rate}$
    \item $w_i$, $\text{num\_emplyee}$, $\tau_i$, $n_{\text{class}}$, $\text{num\_tasks}$
\end{itemize}

\subsubsection*{Step 4: Loss Computation}

Converts accuracy/error to cross-entropy loss space via \texttt{entropy\_fn()}:
\begin{align*}
\tilde{H}_{\text{req},i} &= \texttt{entropy\_fn}(\text{req\_err} / \text{num\_tasks},\; n_{\text{class}}) \\
H_{\text{task},i} &= \min\!\left(\texttt{entropy\_fn}(\tilde{\varepsilon}_{\text{rand}},\; n_{\text{class}}),\; \texttt{entropy\_fn}(1 - 1/n_{\text{class}},\; n_{\text{class}})\right)
\end{align*}

\subsubsection*{Step 5: Cost Structure Setup (\texttt{optim\_fns.\_\_init\_\_()})}

Initializes all cost components:
\begin{itemize}
    \item \texttt{data\_cost\_coef} --- annotation cost over model lifetime with retraining
    \item \texttt{compute\_cost\_coef\_val} --- GPU training cost
    \item \texttt{infer\_cost\_coef} --- inference cost scaled by employees
    \item \texttt{fixed\_cost} --- implementation and maintenance (NPV)
    \item $p_{\text{data}}$ --- row-specific data cost from Q8 (judgment frequency $\times$ wage)
\end{itemize}

\subsubsection*{Step 6: Scaling Law (\texttt{cross\_entropy\_fn()})}

Maps fine-tuning inputs $(D_i, T_i, M_i)$ to predicted cross-entropy loss using a fitted scaling law:
\begin{equation*}
\tilde{H}_{\text{AI}} = \exp\!\left(\frac{A}{D^a} + \frac{B}{T^b} + \frac{C}{M^c} + d\right) \times n_{\text{class}}^e
\end{equation*}
This is the $e^{k(\cdot)}$ term inside Equation~\eqref{eq:performance_elasticity_substitution_ratio} .

\subsubsection*{Step 7: Cost Minimization (\texttt{cost\_min()})}

For a given target loss, finds the cheapest $(D_i, T_i, M_i)$ combination:
\begin{equation*}
\min_{D_i,\, T_i,\, M_i} \quad \kappa^{\text{data}}(D_i) + \kappa^{\text{train}}(T_i, M_i) + \kappa^{\text{inf}}(M_i)
\end{equation*}
\begin{equation*}
\text{s.t.} \quad \tilde{H}_{\text{AI}}(D_i, T_i, M_i) \leq \tilde{H}_{\text{req},i}, \qquad T_i \geq D_i \times \text{complx} \times 1000
\end{equation*}
Solved via \texttt{scipy} \texttt{trust-constr}.

\subsubsection*{Step 8: Marginal Cost and Benefit}

\begin{itemize}
    \item $\text{MB} = \dfrac{\text{TB}_i}{H_{\text{task},i} - \tilde{H}_{\text{req},i}}$, \quad where $\text{TB}_i = n_{\text{empl}} \times w_i \times \tau_i \times \text{NPV\_factor} \times \text{comp\_ratio}$
    \item $\text{MC} = $ Lagrange multiplier from \texttt{cost\_min()}, representing $\partial \kappa / \partial \tilde{H}$
\end{itemize}

\subsubsection*{Step 9: Profit Maximization (\texttt{profit\_max()})}

Three cases arise:
\begin{enumerate}
    \item If $\text{MB}(H_{\text{task},i}) \leq 0$ $\rightarrow$ no automation, $r_i = 0$
    \item If $\text{MB}(\tilde{H}_{\text{req},i}) \geq 0$ $\rightarrow$ full automation, $r_i = 1$
    \item Otherwise $\rightarrow$ partial automation; find optimal loss via \texttt{inverse\_marginal\_cost()}:
    \begin{itemize}
        \item Solves $\min_{\tilde{H}} \left[\tilde{H} + \kappa(\tilde{H}) / \text{MB}\right]$ to find where $\text{MC} = \text{MB}$
        \item $r_i = \dfrac{H_{\text{task},i} - \tilde{H}^*}{H_{\text{task},i} - \tilde{H}_{\text{req},i}}$
    \end{itemize}
\end{enumerate}
Then checks profitability: if $\text{TB}_i - \kappa^{\text{var}}_i - \kappa^{\text{fix}}_i \leq 0$, sets $r_i = 0$. This implements Equation~\eqref{eq:performance_elasticity_substitution_ratio}.

\subsubsection*{Step 10: Output and Aggregation}

Results are saved per row to \texttt{profit\_maximization\_results\_\{granularity\}\_individual\_invmc.csv} with fields: Replace Ratio, Optimal Accuracy, Optimal Data/Model Size, Optimal Training Steps, Total Benefit, Variable Cost, and Fixed Cost.

The plotting script reads these CSVs and computes weighted automation rates across occupations and industries at each granularity level:

\begin{equation*}
\text{Automation Rate} = \frac{\sum_i r_i \times n_{\text{empl},i} \times w_i \times \tau_i}{\sum_i n_{\text{empl},i} \times w_i}
\end{equation*}
\clearpage
\pdfpageheight=27in
\enlargethispage{17in}
\thispagestyle{empty}
\section{Automation Rates across All Occupations at Industry Group (4d) Deployment Scale}
\label{app:Automation Rates across All Occupations}
\begin{figure}[H]
    \centering
    \includegraphics[width = \linewidth]{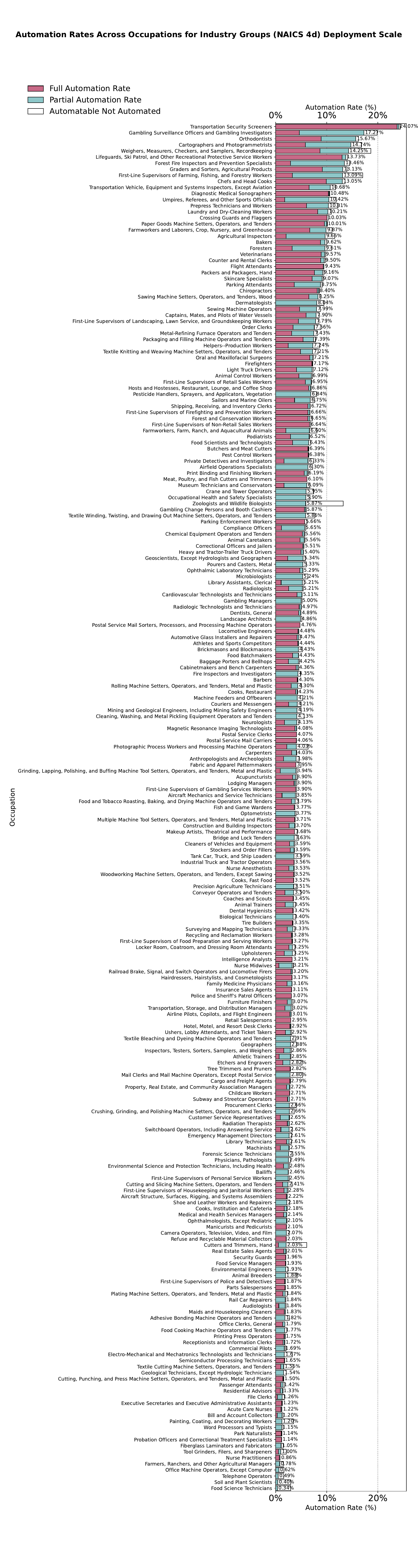}
    \caption{Automation rates across occupations at industry group (4d) deployment scale.}
    \label{fig:occupationrate_new}
\end{figure}
\begin{center}\thepage\end{center}
\clearpage
\pdfpageheight=11in %
\end{appendices}

\end{document}